\documentclass[fleqn,usenatbib]{mnras}
\usepackage{newtxtext,newtxmath}

\usepackage[T1]{fontenc}
\usepackage{ae,aecompl}
\usepackage{xcolor,xspace}
\usepackage{graphicx}	% Including figure files
\usepackage{subcaption}
\usepackage{amsmath}	% Advanced maths commands
\usepackage{amssymb}	% Extra maths symbols
\usepackage{cases}
\usepackage{wasysym}
\usepackage{xfrac}
\usepackage{gensymb}

\usepackage[noend]{algpseudocode}
\usepackage{algorithm}
\usepackage{enumitem}
\setlist[itemize,1]{leftmargin=\dimexpr 26pt-0.5cm}
\renewcommand{\algorithmicrequire}{\textbf{Input:}}

\usepackage{comment}

\newcommand{\stlight}{{\!\star}}
\newcommand{\cirsig}{{\!\odot}}
\newcommand{\convOp}{\ast}
\newcommand{\rotop}{\cl R}
\newcommand{\rotopmat}{\bs R}
\newcommand{\maskop}{\cl M}

\newcommand{\Hsvd}[1]{ \mathcal{H}_{#1}^{\text{SVD}}  }
\newcommand{\Lmodel}{\bar{\bs L}}
\newcommand{\xmodel}{\bar{\bs x}}
\newcommand{\inddisk}{{\rm d}}
\newcommand{\indpla}{{\rm p}}
\newcommand{\Diskmodel}{\bar{\bs x}_\inddisk}
\newcommand{\Planetmodel}{\bar{\bs x}_\indpla}
\newcommand{\NoiseDetector}{\bs N_{\rm{det}}}
\newcommand{\NoiseSpeckles}{\bs N_{\rm ns}}
\newcommand{\NoiseTotal}{\bs N}
\newcommand{\Smodel}{\bar{\bs S}}
\newcommand{\Hnorm}[1]{\|#1\|_{\delta,\bs \Xi}}
\newcommand{\gdsest}[1]{\hat{#1}}

\newcommand{\scp}[2]{\langle #1, #2 \rangle}

\newcommand{\inv}[1]{\frac{1}{#1}}
\newcommand{\supp}{{\rm supp}\,}

\renewcommand{\leq}{\leqslant}
\renewcommand{\geq}{\geqslant}

\DeclareMathOperator{\st}{{s.\!t.}\xspace}

\DeclareMathOperator{\MR}{MR}

\DeclareMathOperator{\iid}{iid}

\DeclareMathOperator{\rank}{rank}
\DeclareMathOperator{\Span}{span}

\DeclareMathOperator*{\argmin}{arg\,min}

\newcommand{\ts}{\textstyle}
\newcommand{\bb}{\mathbb}
\newcommand{\bs}{\boldsymbol}

\newcommand{\cl}{\mathcal}
\newcommand{\ie}{i.e.,\xspace}
\newcommand{\eg}{e.g.,\xspace}

\newcommand{\psf}{\bs \varphi}

\title[Circumstellar disks and exoplanets imaging]{MAYONNAISE: a morphological components analysis pipeline for circumstellar disks and exoplanets imaging in the near infrared}
\author[B. Pairet et al.]{
Beno\^it Pairet,$^{1}$\thanks{Email: benoit.pairet@uclouvain.be}\thanks{BP and LJ are funded by the Belgian F.R.S.-FNRS.}
Faustine Cantalloube$^2$,  Laurent Jacques$^1$\textcolor{blue}{\footnotemark[2]}\\
$^1$ ISPGroup, INMA-ELEN, ICTEAM, Belgium\\
$^2$ Max Planck Institute for Astronomy, Germany
}

\date{Accepted XXX. Received YYY; in original form ZZZ}

\pubyear{2020}

\begin{document}
  
\label{firstpage}
\pagerange{\pageref{firstpage}--\pageref{lastpage}}
\maketitle

% Abstract of the paper
\begin{abstract}

Imaging circumstellar disks in the near-infrared provides unprecedented information about the formation and evolution of planetary systems. However, current post-processing techniques for high-contrast imaging using ground-based telescopes have a limited sensitivity to extended signals and their morphology is often plagued with strong morphological distortions. Moreover, it is challenging to disentangle planetary signals from the disk when the two components are close or intertwined. We propose a pipeline that is capable of detecting a wide variety of disks and preserving their shapes and flux distributions. By construction, our approach separates planets from disks. After analyzing the distortions induced by the current angular differential imaging (ADI) post-processing techniques, we establish a direct model of the different components constituting a temporal sequence of high-contrast images. In an inverse problem framework, we jointly estimate the starlight residuals and the potential extended sources and point sources hidden in the images, using low-complexity priors for each signal. To verify and estimate the performance of our approach, we tested it on VLT/SPHERE-IRDIS data, in which we injected synthetic disks and planets. We also applied our approach on observations containing real disks. Our technique makes it possible to detect disks from ADI datasets of a contrast above $3\times10^{-6}$ with respect to the host star. As no specific shape of the disks is assumed, we are capable of extracting a wide diversity of disks, including face-on disks. The intensity distribution of the detected disk is accurately preserved and point sources are distinguished, even close to the disk.

\end{abstract}

% Select between one and six entries from the list of approved keywords.
% Don't make up new ones.
\begin{keywords}
techniques: image processing -- techniques: high angular resolution -- planet-disc interactions -- (stars:) circumstellar matter -- protoplanetary discs
\end{keywords}

%%%%%%%%%%%%%%%%%%%%%%%%%%%%%%%%%%%%%%%%%%%%%%%%%%

%%%%%%%%%%%%%%%%% BODY OF PAPER %%%%%%%%%%%%%%%%%%

\section{Introduction}
\label{sec:introduction}
High-contrast imaging in the near-infrared enables to constrain the scenarii of planet formation and evolution, by offering a unique view of the birthplace of exoplanets, through the starlight scattered by the surface of young protoplanetary disks ($\lesssim 10~\mathrm{Myrs}$), and of the outcome of planetary formation, through the starlight scattered by debris disks ($\gtrsim 10~\mathrm{Myrs}$). 
Whatever the formation stage of the system, planetary perturbers can explain the ubiquitous morphology of the circumstellar disks resolved so far \citep[see \eg][]{Andrews2020rev}. There exist various theoretical studies attempting to link the morphology and substructures observed in disks with the potential presence of planetary perturbers. Features such as gaps/rings \citep{zhang2019systematic}, spirals \citep{bae2018planet} or vortexes \citep{li2014protostars}, can be theoretically explained by the presence of planets, but some could also be explained by other mechanisms. There is no final evidence that these structures are really created by the presence of planets, as shown in \cite{Andrews2020rev}.

In that context, it is essential to (1) increase the detection rate of disks (towards fainter disks and for any disk inclination), (2) accurately restore the morphology of the disks (spirals, gaps, cavities, streamers and dips) and its flux distribution (to extract the surface-brightness or scattering phase function for debris disks), and (3) separate point source signals from the disk to study the planet-disk interactions with precision.

However as of today, not only have very few disks been resolved compared to the expected rate, but it is also difficult to extract their morphology accurately. Ground-based telescopes of the 8 m class assisted by adaptive optics (AO) provide the necessary resolution and sensitivity to image circumstellar disks. Specific instruments, equipped with coronagraphic devices and high-quality optics in a stable environment, make it possible to image faint circumstellar material up to a raw contrast of $10^{-4}$ with respect to the host star, at only a few hundred milliarcseconds (mas).

VLT/SPHERE \citep{Beuzit2019}, Gemini/GPI \citep{Macintosh2008}, MagAO-X \citep{males2018magao}, Keck/KPIC \citep{mawet2016kpic}, LBT/LMIRCam \citep{skrutskie2010lmircam,defrere2014lbthci,kenworthy2010lbthci}, and Subaru/SCExAO \citep{Jovanovic2015} are the latest generation of high-contrast instruments dedicated to exoplanets and disks imaging. 
Under the contrast reached by these instruments, the effect of small instrumental aberrations and atmospheric turbulence residuals become visible in the coronagraphic image \citep{cantalloube2019msgr}. These starlight residuals limit the raw contrast performance of the instrument, and post-processing techniques are necessary to gain from one to three orders of magnitude in contrast. 

The residual starlight present in the image is usually called speckle field as, under very good observing conditions, the dominant residuals form speckles, which originate from non-common path aberrations between the AO arm and the science arm of the instrument. Post-processing techniques consist in estimating and removing these speckles that are quasi-static. To do so, observing strategies introduce a diversity between the speckle field to be removed and the circumstellar objects to be recovered. The baseline observing strategy for high-contrast imaging is to use the pupil tracking mode of the telescope, so that aberrations (\ie the speckles) remain at the same position with time, while the field of view (\ie the circumstellar objects) rotate along with the parallactic angles. Angular differential imaging \citep[ADI,][]{marois2006angular} exploits this diversity to estimate the speckle field in the temporal image cube and subtract it from each image. The subtracted images are then aligned to a common direction for circumstellar signals and combined (e.g.~ median averaged) to increase the signal-to-noise ratio (S/R) of the objects of interest. Several algorithms have been developed during the last decade to improve the speckle subtraction using this ADI concept. However, by construction, all the ADI-based techniques developed so far are not suitable for extended sources: the centro-symmetric signals are erased, distorting the shape of disks, and self-subtraction effects (the estimated speckle field contains parts of the disk signal) alter the flux distribution of the disks \citep{milli2012impact}. Besides, speckle residuals, particularly consequential at close angular separation to the star, may contaminate the circumstellar signals. In this paper we focus solely on this specific type of data set taken with ground-based telescopes in pupil tracking mode (also called ADI data set).

Three solutions are commonly used to alleviate these limitations. The first is to use conservative parameters to avoid subtracting too many modes and breaking-down the optimization regions, while enforcing positivity and/or sparsity \citep{pueyo2012dloci, ren2018nmf}. The second is to mask the signal, for instance, using a ray-tracing model of the disk and analyze the distortion generated by the ADI process \citep{milli2012impact,esposito2013fm,ren2020nmfmask}. The third is to iterate the ADI subtraction to remove the effect of self-subtraction at each iteration \citep{pairet2018reference}. Those solutions are suitable for bright disks and do not preserve face-on disk signals. We also mention the development of reference differential imaging~\citep[RDI, see, \eg ][]{xuan2018rdi,ruane2019rdi,bohn2020rdi} techniques, particularly beneficial for disk imaging. RDI uses a large library of images from the same instrument taken with a uniform observing mode. The processing consists in finding the metric to create the model of the speckle field from the images within the library. However, RDI suffers from over-subtraction effects (due to different profiles and gradients in the reference image) and does not perform a proper image restoration, including deconvolution as we propose here. 

To fully address the scientific questions evoked above and to alleviate the current post-processing limitations, we propose a source separation pipeline, the Morphological Analysis Yielding separated Objects iN Near infrAred usIng Sources Estimation (MAYONNAISE or MAYO for short). It leverages the morphological diversity between point-like sources and extended structures, allowing us to separate exoplanets and disks, respectively. Furthermore, MAYO attenuates the influence of the telescope optics by deconvolving (deblurring) the circumstellar signal, using the empirical response of the telescope to a point source. This is the first inclusion of a deconvolution method in the context of high-contrast imaging. The capabilities of MAYO are thoroughly demonstrated on semi-synthetic datasets. Finally,  applying our algorithm to VLT/SPHERE datasets of known circumstellar systems, our deconvolution and source separation approach provide a clear view of the disk and exoplanet signals, such as the protoplanet PDS~70~c~\citep{Haffert2019pds70,Mesa2019pds70}.

\paragraph*{Paper structure:} We first present in Sec.~\ref{sec:pca-induc-dist} the PCA speckle field subtraction (PCA-SFS) method, which is the most widely used ADI-based method, and identify why this method strongly distorts the signal of extended sources. 
In Sec.~\ref{sec:acqu-model-greedy}, we introduce the acquisition model used in this work, from which we derive a greedy algorithm to restore the disk signal. Then in Sec.~\ref{sec:unmixing_algo}, we introduce the core contribution of this paper, namely the source separation algorithm, MAYO, which is based on specific priors listed in Sec.~\ref{priors-identification}. In Sec.~\ref{sec:numercial-exp}, we validate our method numerically on synthetic injected disk signals, and then apply it on datasets containing real circumstellar disks and planets.

\paragraph*{Conventions and notations}
Matrices and vectors are written as upper- and lower-case bold symbols respectively. The cardinality of a set $\cl S$ is $|\cl S|$. For $d \in \bb N$ and $p\geq 1$, $[d] := \{1, \cdots, d\}$, 
$\bs 1_d := (1, \cdots, 1)^\top \in \bb R^d$, $\bs 0_d := (0, \cdots, 0)^\top \in \bb R^d$, the $\ell_p$-norm of $\bs u \in \bb R^d$ reads $\|\bs u\|_p := (\sum_i |u_i|^p)^{1/p}$ and the $\ell_0$-(not-a)-norm of $\bs u$ is $\|\bs u\|_0 := |\supp \bs u| = |\{i \in [d]: u_i \neq 0\}|$. Given $\bs u, \bs v \in \bb R^d$, $\scp{\bs u}{\bs v} = \sum_i u_i v_i$ is the inner product of $\bs u$ and $\bs v$. Abusing the notation, the square Frobenius norm of a matrix $\bs A \in \bb R^{d \times d'}$ reads $ \| \bs A \|^2_2 = \sum_{i,j} (A_{ij})^2$, and $ \| \bs A \|_1 = \sum_{i,j} |A_{ij}|$. The operator norm of a square matrix $\bs A$ is $\|\bs A\|_{\rm op} = \sup \{|\bs A \bs u|: \|\bs u\|_2 \leq 1\}$, with $\|\bs A \bs B\|_2 \leq \|\bs A\|_{\rm op} \|\bs B\|_2$ for any dimension-compatible matrix $\bs B$.
For a vector $\bs u$ (a matrix $\bs A$), $\bs u \geq 0$ (resp. $\bs A \geq 0$) means that all components of $\bs u$ (resp. all entries of $\bs A$) are nonnegative.
Given an index set $\Omega \subset [d']$, $\bs A_\Omega \in \bb R^{d \times |\Omega|}$ denotes the submatrix made of the columns of $\bs A$ indexed in $\Omega$. In particular, $\bs A_{[r]}$ for $r < \min(d,d') \in \mathbb N$, is the submatrix made of the first $r$ columns of $\bs A$. 
Hereafter, given an ADI dataset consisting of $T$ $n$-by-$n$ images (or frames), we represent it, for convenience, as a $T \times n^2$ matrix $\bs Y := (\bs y_1,\cdots, \bs y_T)^\top \in \bb R^{T \times n^2}$, each image being associated with a vector $\bs y_i \in \bb R^{n^2}$ for $i \in [T]$. To facilitate the comparison between figures, we use a common zero in the colorbar, except when explicitly stated otherwise.

\section{Limitations of the PCA speckle field subtraction for disk imaging}
\label{sec:pca-induc-dist}

In this section, we describe in detail the widely used ADI post-processing technique using a principal component analysis~(PCA) for speckle field subtraction~\citep[][]{soummer2012detection, amara2012pynpoint}. We then demonstrate and discuss how this PCA distorts the shape and flux distribution of circumstellar disks. Finally, we recall that PCA can be recast as a low-rank matrix approximation, a framework that is more adapted to the rest of the paper.

The classical speckle field subtraction (SFS) post-processing methods involve four steps: \emph{(SFS-1)} we estimate a model of the speckle field $\bs L  \in \bb R^{T \times n^2}$ directly from the $T$-frames image cube $\bs Y \in \bb R^{T \times n^2}$;  \emph{(SFS-2)} we subtract $\bs L$ from the image cube to form $\bs S = \bs Y - \bs L \in \bb R^{T \times n^2}$;  \emph{(SFS-3)} we align the $T$ frames of $\bs S$ along a common direction for the circumstellar signal, providing with an aligned dataset $\bs S' \in \bb R^{T \times n^2}$;  \emph{(SFS-4)} finally, we compute the temporal mean (or median) of $\bs S'$ to get the \emph{processed frame}, an image $\bs x = (\sum_{i=1}^T \bs s'_i)/T $.
The estimation of $\bs L$ is the most critical step of the speckle subtraction algorithms. 

\subsection{Distortions caused by a PCA speckle field subtraction}
\label{sec:dataset}
In the PCA-SFS approach, $\bs L$ is estimated as the (left) projection of $\bs Y$ on its first principal components (PCs), that is, on the main eigenvectors of $\bs Y \bs Y^\top \in \bb R^{T \times T}$~\citep[see, \eg][]{abdi2010principal}. The number of selected PCs depends on the data variability and typically ranges from 1 to 100 for high-contrast imaging applications in the near infrared. The rationale is that in pupil tracking mode, $\bs L$ is supposedly quasi-static (slow temporal evolution with no apparent motions) in the temporal cube and is therefore well represented by the first few PCs of $\bs Y$. On the other hand, the circumstellar signals rotate with the parallactic angles and therefore they are spread across a large number of PCs. Hence, by removing only a few PCs from $\bs Y$, a large proportion of the signal of circumstellar objects is still present in $\bs S$. However, although the circumstellar signal is spread into a large number of PCs, it is not absent from the first PCs. This implies that parts of the circumstellar signal are removed from $\bs S$, a phenomenon called self-absorption.
Because of self-absorption, the intensity of the circumstellar objects is underestimated on the processed frame and the morphology of extended sources is severely altered. 

\begin{figure*}
  \centering
  {\includegraphics[width=0.31\textwidth]{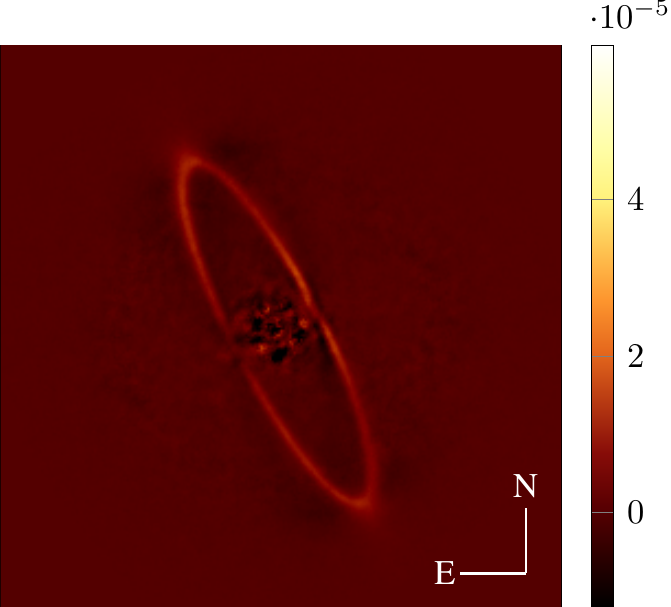}}
  {\includegraphics[width=0.31\textwidth]{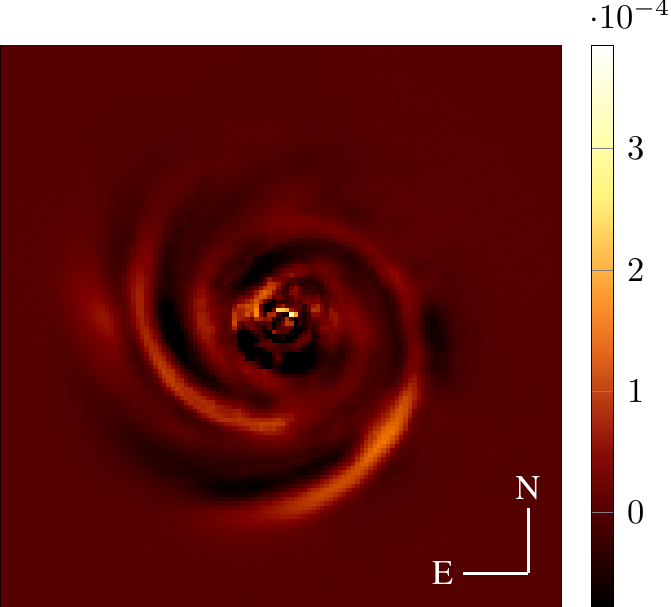}}
  {\includegraphics[width=0.31\textwidth]{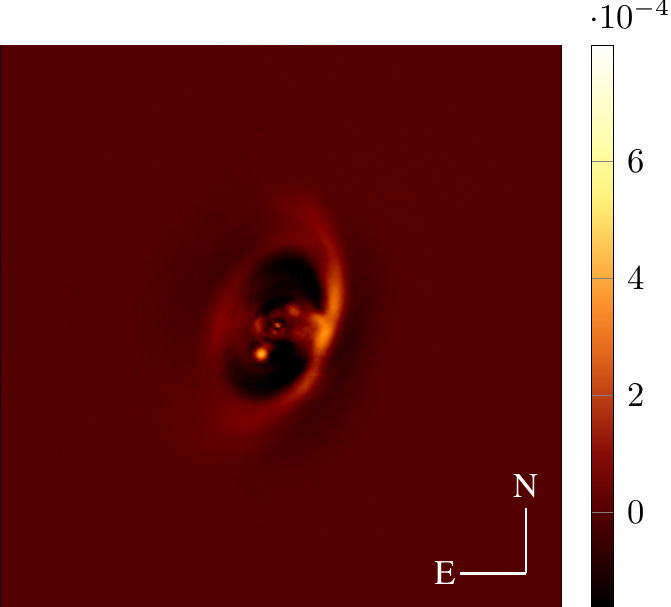}} \\
  {\includegraphics[width=0.31\textwidth]{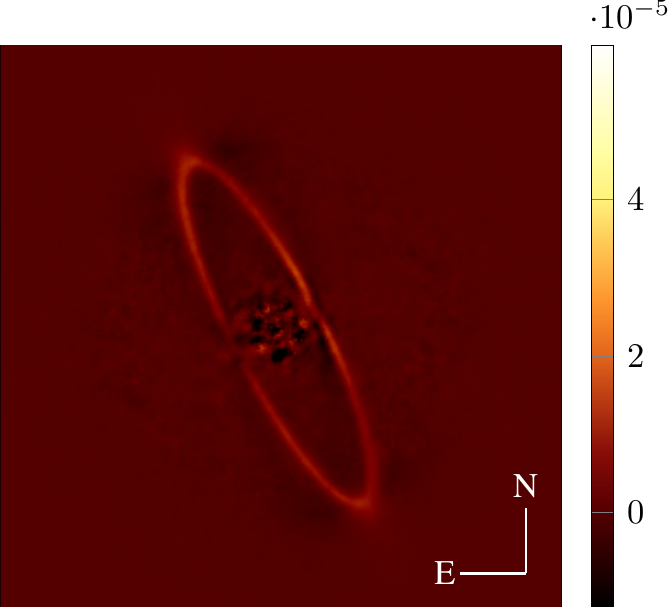}}
  {\includegraphics[width=0.31\textwidth]{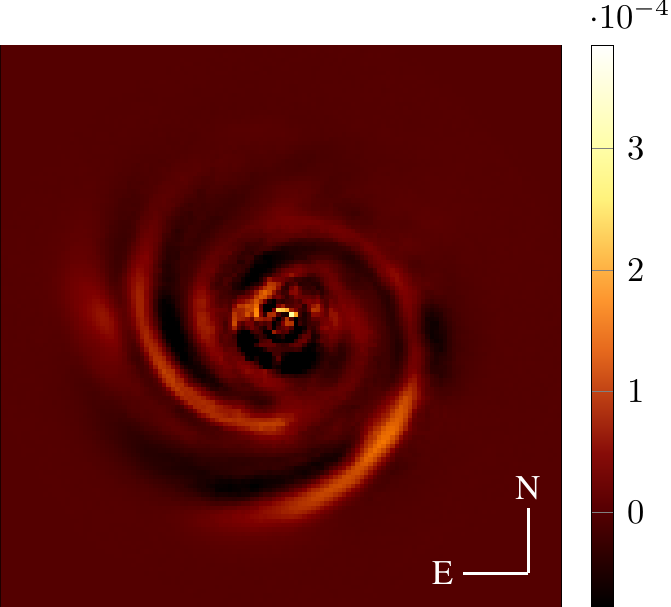}}
  {\includegraphics[width=0.31\textwidth]{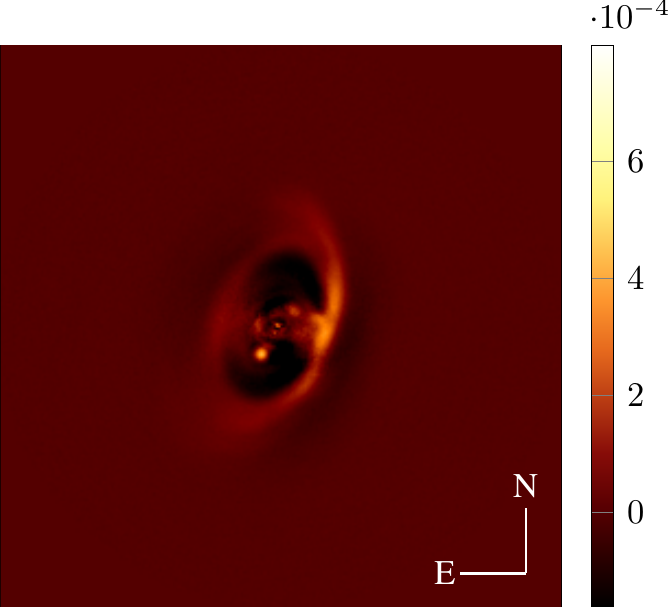}}  
\caption{Top: processed frames using PCA-SFS on the datasets of HR~4796A (left), SAO~206462 (center), and PDS~70 (right). HR~4796A and SAO~206462 are both surrounded by bright disks, however, their shapes are poorly rendered on the processed frames of PCA-SFS. In particular, we note the presence of non-physical negative valued regions (dark areas). Bottom: processed frames using NMF-SFS for the same datasets.} 
\label{fig:pca}
\end{figure*}

In addition, because PCA-SFS is based on ADI, the centro-symmetric part of the signal is seemingly not rotating and hence estimated as being part of the starlight residuals $\bs L$, since they are included in the first PCs. This effect prevents the detection of centered face-on disks and provokes typical distortion visible on Fig.~\ref{fig:pca} on the part of the disk that is closer to the center (\ie rotating less), whereas models predict that this should be the brightest part of the disk \citep[e.g.~][]{milli2012impact}.

Fig.~\ref{fig:pca} (top row) shows the processed frames obtained by applying a PCA-SFS on three targets hosting bright disks: the ellipsoidal disk surrounding HR~4796A~\citep{milli2017near}, the spiral disk surrounding SAO~206462~\citep{maire2017testing}, and the protoplanetary disk PDS~70~\citep{keppler2018discovery} showing a large cavity. These datasets were all taken with the VLT/SPHERE high-contrast instrument (see App.~\ref{app:data_set} for details on the datasets used). The recovered disks incorporate non-physical negative valued regions (dark areas in Fig.~\ref{fig:pca}, top row). In addition the morphology of the disks is substantially altered, as expected from the self-absorption effect \citep{milli2012impact}. In comparison to the expected morphology of the disks obtained from radiative transfer models \citep{milli2017near,maire2017testing,keppler2018discovery} and from polarized imaging~\citep{milli2019hr4796pol,stolker2016sao206,keppler2018discovery}, the flux distribution in the disk is not preserved.

\subsection{PCA as an SVD truncation} 
The PCA was introduced in the high-contrast imaging literature from a statistical point of view with the definition of the (temporal) Karhunen-Lo\`{e}ve basis~\citep{soummer2012detection}. We consider it here for its ability to provide us with a low-rank approximation of data~\citep{eckartapproximation,abdi2010principal}. Projecting a matrix $\bs X \in \bb R^{T \times n^2}$ with $T \leq n^2$ (as assumed for the considered ADI datasets) on the first $r$ vectors of its Karhunen-Lo\`{e}ve basis 
(\ie the eigenvectors of $\bs X\bs X^\top$ associated with the singular vectors $\bs X$) is equivalent to computing the best rank-$r$ approximation $\Hsvd{r}(\bs X)$ of $\bs X$ with respect to the Frobenius norm:
\begin{equation}
\ts \Hsvd{r}(\bs X) := \argmin_{\bs U} \frac{1}{2}\| \bs X  - \bs U \|^2_2\ \st\ \rank(\bs U) \leq r.
\label{eq:low-rank-frobenius}
\end{equation}
The solution of Eq.~\eqref{eq:low-rank-frobenius} is closed form and is found using the singular value decomposition (SVD) of $\bs X$. The SVD of a matrix $\bs X\in \bb R^{T\times n^2 }$ writes $\bs X = \bs U \bs \Sigma \bs V^\top$, with $\bs U \in \bb R^{T\times T }$, $\bs \Sigma \in \bb R^{T\times T }$, and $\bs V \in \bb R^{n^2 \times T}$. In this decomposition, the matrix $\bs U$ is orthogonal, the columns of $\bs V$ are orthonormal, and $\bs \Sigma$ is a diagonal matrix whose entries $\sigma_i = \Sigma_{ii}$ (with $\sigma_i  \geq \sigma_{i+1}$) are called the singular values of $\bs X$. The solution of the problem~\eqref{eq:low-rank-frobenius} with $r\leq T$ is then given by $\Hsvd{r}(\bs X) = \bs U_{[r]} \bs U_{[r]}^\top \bs X = \bs X \bs V_{[r]} \bs V_{[r]}^\top = \bs U_{[r]} \bs \Sigma_{rr} \bs V_{[r]}^\top$.

In the context of PCA-SFS applied to an image cube $\bs Y$, we can interpret (up to a reshaping) the matrix $\bs V^\top \in \bb R^{T \times n^2}$ obtained from the SVD of $\bs Y$ as a list of $T$ images $\bs v_i \in \bb R^{n \times n}$, $i \in [T]$, the \emph{singular images} of $\bs Y$. Moreover, the matrix $\bs W := \bs U \bs \Sigma \in \bb R^{T \times T}$, including the temporal evolution of $\bs Y$ (as encoded in $\bs U$), weights each singular image in the synthesis of the $T$ $n$-by-$n$ frames $\{\bs y_i\}_{i=1}^T$ of $\bs Y$, \ie $\bs y_i = \sum_j w_{ij} \bs v_j$ for $i \in [T]$. 

\begin{figure}
  \centering
  {\includegraphics[width=0.3\textwidth]{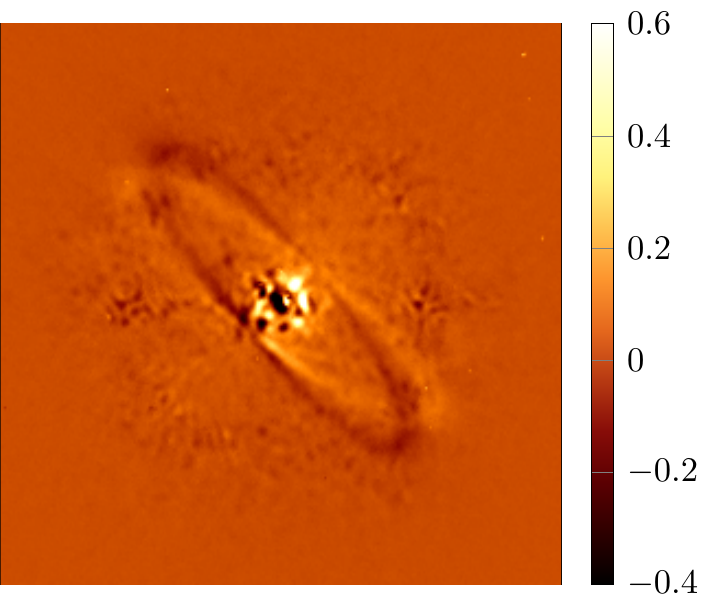}}
\caption{Sixth singular image, \ie $\bs v_6$, of HR~4796A produced by the SVD of $\bs Y$. We clearly see rotated positive (bright) and negative (dark) copies of the disks. Note that the absolute intensity of the singular images does not convey any astronomical significance. We thus do not use the common zero in the colorbar.}
\label{fig:V_init_vs_V_algo}
\vskip -3ex
\end{figure}

For instance, by inspecting the singular images of the HR~4796A image cube, we observe that they contain a significant fraction of the disk signal, appearing as negative and positive copies of the disk (Fig.~\ref{fig:V_init_vs_V_algo}). Now consider the third step (SFS-3) of speckle subtraction in the context of PCA-SFS. Using Eq.~\eqref{eq:low-rank-frobenius} and given a prescribed rank $r \in [T]$, we can rewrite this step as
\begin{equation}
  \label{eq:3rd-step-PCA-SFS}
  \bs S^{(r)} := \bs Y - \Hsvd{r}(\bs Y),  
\end{equation}
and this matrix is simply composed of a set of $T$ images $\bs s_{i} = \sum_{j=r+1}^T w_{ij} \bs v_{j}$; the first $r$ images of $\bs V$ are disregarded.

In addition we mention a PCA-SFS refinement, the nonnegative matrix factorization \citep[NMF, see][]{ren2018nmf, gonzalez2017vip}, which consists in  computing $\bs L$ as a factorization $\bs H_T \bs H_N^{\top}$, where $\bs H_T\in \mathbb R^{r\times T}, \bs H_N \in \mathbb R^{r\times N}$ are nonnegative. 
The parallel with the SVD truncation is then immediate: $\bs H_T$ ($\bs H_N$, respectively) can be thought as the nonnegative version of $\bs W_{[r]}$ ($\bs V_{[r]}$ respectively). Note that only $\bs H_T$ and $\bs H_N$ are nonnegative, the processed frames produced by NMF-SFS  is not nonnegative. 

Fig.~\ref{fig:pca} (bottom row) shows the same objects processed with the NMF-SFS approach \citep[as implemented in the VIP package,][]{gonzalez2017vip}. The NMF-SFS processed frames of both SAO~206462 and PDS~70 are identical to their PCA-SFS counterparts. For HR~4796A, the intensity of the disk is slightly larger in the NMF-SFS processed frame than in the PCA-SFS one. However, the shape of the disk is still heavily distorted. We obtained similar results using the NMF implementation used in~\citep{ren2018nmf}.

\section{Acquisition model and greedy algorithm}
\label{sec:acqu-model-greedy}

\subsection{Acquisition model}
\label{subsec:acquisition-model}

We start from an ADI sequence $\bs Y \in \bb R^{T\times n^2}$ that we model as the sum of two terms: the starlight $\bs Y_{\stlight}$ and the rotating circumstellar signal $\bs Y_{\cirsig}$. In addition, even after image reduction (bad pixel suppression, flat and dark subtraction, etc.), noise due to the acquisition process still remains, as well as photon noise inherent to the brightness of the starlight residuals. We model these imperfections by an additive noise term $\NoiseDetector \in \bb R^{T \times n^2}$. 

As reminded previously, most of the star signal is blocked by the coronagraph, but residual instrumental aberrations provoke leakage of starlight from the coronagraph, resulting in the presence of bright quasi-static speckles that are gathered in $\bs Y_{\stlight}$. These speckles can be modeled as the sum of two terms encoding their temporal behavior, a static term $\Lmodel$ and a non-static term $\NoiseSpeckles$: 
\begin{equation}
\textstyle \bs Y_{\stlight} = \Lmodel+ \NoiseSpeckles,
\end{equation}
where $\Lmodel$ is assumed to be a rank-$r$ matrix with $r \ll T$. 
The circumstellar signals (such as disks or exoplanets) have a low intensity and the effect of the residual instrumental aberrations are below the noise level. The intensity of the circumstellar signals can be assumed constant in time, and we model $\bs Y_{\cirsig}$ as a single rotating image $\xmodel \in \bb R^{n^2}$. In addition, because the light is diffracted when it enters the telescope and passes through the instrument, the circumstellar signal is convolved (that is, blurred) by the instrumental response. We thus model $\bs Y_{\cirsig}$ as 
\begin{equation}
  \label{eq:circumstellar-signal-model}
  \bs Y_{\cirsig} = \cl T ( \rotop[\bs 1_T \xmodel^\top ]),
\end{equation}
where $\rotop: \bb R^{T\times n^2} \rightarrow \bb R^{T\times n^2} $ is the linear operator that rotates each frame of the volume according to the parallactic angles, and $\cl T: \bb R^{T\times n^2} \rightarrow \bb R^{T\times n^2} $ is a 2-D convolutive operator applied separately on each image of $\rotop[\bs 1_T \xmodel^\top ]$ (see Sec.~\ref{priors-identification} for its exact definition). Note that since the images are represented on a pixel grid, $\rotop$ includes an interpolation, which implies some numerical technicalities discussed in Sec.~\ref{sec:numerical-aspects}.

The final acquisition model of the ADI sequence $\bs Y$ is then:
\begin{equation}
\bs Y = \Lmodel +\NoiseSpeckles +  \cl T( \rotop[\bs 1_T \bs \xmodel^\top]) + \NoiseDetector.
\label{eq:complete_forward_model}
\end{equation} 
This acquisition model will guide us to formulate estimation algorithms for both $\xmodel$ and $\Lmodel$. In the following, in a first step, we propose a fixed-point algorithm relying on a simplification of the direct model given at Eq.~\eqref{eq:complete_forward_model}. Later, we incorporate the complete acquisition model of Eq.~\eqref{eq:complete_forward_model} into our source separation algorithm (Sec.~\ref{sec:unmixing_algo}).  

\subsection{Fixed-point algorithm}
\label{sec:GreeDS}

In this section, we present a novel algorithm which is a straightforward improvement of PCA-SFS based on the model presented at Eq.~\eqref{eq:complete_forward_model}. First, we assume a low noise framework, meaning that $\NoiseSpeckles$ and $\NoiseDetector$ are small compared to $\Lmodel$ and $\xmodel$. Also, we neglect the diffracting effects of the telescope. Thus, Eq.~\eqref{eq:complete_forward_model} now becomes:
\begin{equation}
\bs Y =  \Lmodel +  \rotop[\bs 1_T \xmodel^\top].  
\end{equation}
Denoting by $\rotop^{-1}$ the inverse rotation operator of $\rotop$ (and assuming $\rotop^{-1} \circ \rotop [\bs X] = \bs X$ for any image cube $\bs X$ despite the implicit interpolation operation), we can then solve this equation for $\bs x$:
\begin{equation}
\ts \xmodel^\top = \frac{1}{T} \bs 1_T^\top \rotop^{-1} [ \bs Y - \Lmodel ],
\label{GreeDS_solve_d}
\end{equation}
and for $\Lmodel$:
 \begin{equation}
 \Lmodel = \Hsvd{r}(\bs Y - \rotop[\bs 1_T \bs \xmodel ]),
\label{GreeDS_solve_L}
 \end{equation}
 where we used the fact that, since $\Lmodel$ is assumed to be a rank $r$ matrix, we can write $\Lmodel = \Hsvd{r}(\Lmodel) =   \Hsvd{r}(\bs Y - \rotop[\bs 1_T \bs \xmodel ])$.

Injecting Eq.~\eqref{GreeDS_solve_L} into Eq.~\eqref{GreeDS_solve_d}, we get the fixed point equation for $\xmodel$:
\begin{equation}
\ts \xmodel^\top = \frac{1}{T} \bs 1_T^\top \rotop^{-1} \left[ \bs Y - \Hsvd{r}(\bs Y - \rotop[\bs 1_T \xmodel^\top ]) \right].
\label{GreeDS_fixed_point_equation}
\end{equation}
The form of Eq.~\eqref{GreeDS_fixed_point_equation} suggests that we can recover $\xmodel$ using the fixed-point algorithm
\begin{equation}
\ts \bs x_{k+1}^\top = \bs f(\bs x_k; \bs Y, \rho)^\top := \frac{1}{T} \bs 1_T^\top \rotop^{-1} \big[ \bs Y - \Hsvd{\rho}(\bs Y - \rotop[\bs 1_T \bs x_k^\top ])\big],
\label{GreeDS_fixed_point_iteration}
\end{equation}
where $\rho \in [r]$ is the number of principal components used to build the speckle field model and is set by the user.
We note that, starting with $\bs x_0 = 0$ and recalling (SFS-2) of PCA-SFS, $\bs S = \bs Y - \bs L$, the first iterate,
\begin{equation}
\ts  \bs x_1 = \frac{1}{T} \bs 1_T^\top \rotop^{-1} [\bs Y - \Hsvd{\rho}(\bs Y)] = \frac{1}{T} \bs 1_T^\top \rotop^{-1} \bs S,
\end{equation}
is nothing but the rank-$\rho$ PCA processed frame, resulting of the fourth step (SFS-4) of PCA-SFS.  

As we have seen in the previous section, PCA-SFS introduces strong distortions in the processed frame. These distortions are stronger when $\rho$ is larger. We thus expect the iterative procedure proposed in Eq.~\eqref{GreeDS_fixed_point_iteration} to perform better if we set $\rho=1$ and then progressively increase its value until $\rho = r$.
Therefore, we compute $l$ iterations of Eq.~\eqref{GreeDS_fixed_point_equation} with $\bs x_0 = 0$ and $\rho=1$, then we increase $\rho$ and compute again $l$ iterations following Eq.~\eqref{GreeDS_fixed_point_equation}, until $\rho =r$. Moreover, to shed the negative valued artifacts, we impose the positivity by computing $\ts \bs x_{k+1}^\top =\chi_+( \bs f(\bs x_k; \bs Y, \rho)^\top)$, where $\chi_+(x) = 1$ if $x>0$ and $0$ otherwise, and is applied pixel-wise on images.

We name the resulting algorithm Greedy Disk Subtraction (GreeDS) (see its summary in Alg.~\ref{alg:GreedyAlgoDisk}). The GreeDS algorithm is a variation of the algorithm presented in~\cite{pairet2018reference}. The convergence of GreeDS is not guaranteed theoretically. However, as reported below, extensive simulations showed that it is a reliable method for estimating $\Lmodel$ and $\bs N$, which, as discussed in Sec.~\ref{sec:unmixing_algo}, is needed for the MAYO pipeline.

\begin{figure*}
  \centering
  {\includegraphics[width=0.31\textwidth]{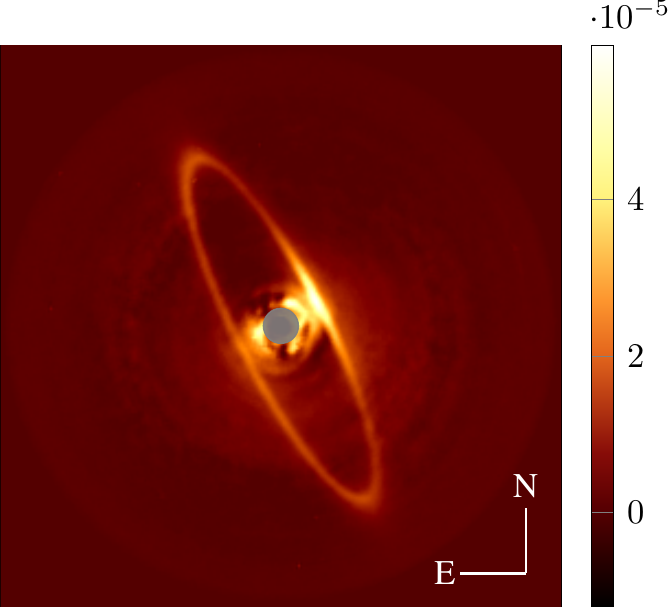}}
  {\includegraphics[width=0.31\textwidth]{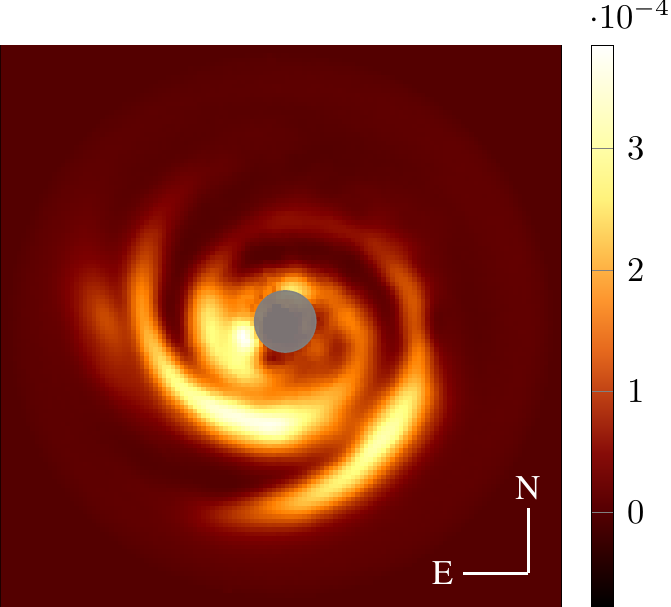}}
  {\includegraphics[width=0.31\textwidth]{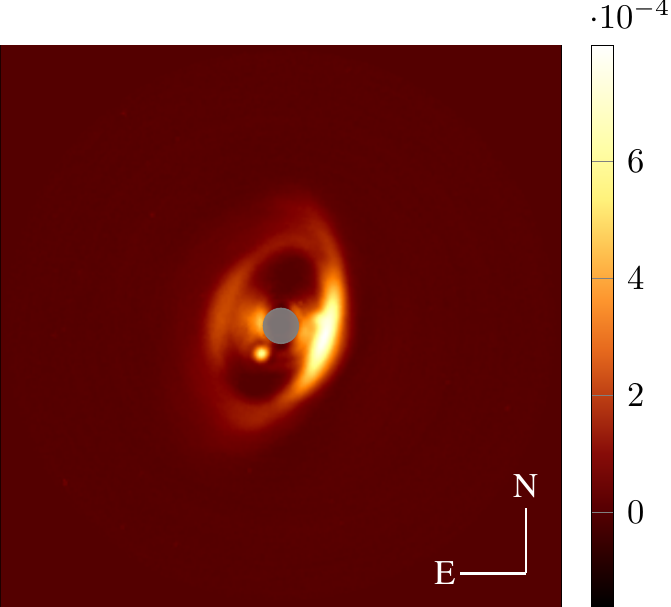}}
\caption{Processed frames produced by the GreeDS algorithm (Algorithm~\ref{alg:GreedyAlgoDisk}) for HR~4796A (left), SAO~206462 (center), and PDS~70 (right).}
\label{fig:illustration-greedy-algo}
\end{figure*}

Fig.~\ref{fig:illustration-greedy-algo} displays the processed frames obtained with GreeDS for both HR~4796A, SAO~206462, and PDS~70 with $\rho=l=10$. As expected, these processed frames are not plagued with the typical PCA-induced distortions mentioned at Sect.~\ref{sec:dataset} and shown in Fig.~\ref{fig:pca}, and by construction, they do not include negative values.

Although the GreeDS algorithm provides appealing images of disks, it comes with a few limitations. First, the noise is neglected, meaning that the GreeDS algorithm is not guaranteed to perform well for faint disks. Second, the GreeDS is unable to account for the diffractive effects of the telescope, \ie the operator $\cl T$ must be neglected. And finally, the GreeDS algorithm is not explicitly designed to capture circumstellar signals. In practice, the GreeDS algorithm tends to capture anything that can be considered rotating. In the noisy case, there are numerous artifacts in the processed frames produced by GreeDS. These artifacts mostly appear as circular-shaped noise (as it can be seen on the HR~4796A frame presented in Fig.~\ref{fig:illustration-greedy-algo}, left). Although the applicability of the algorithm is limited to bright exoplanets and disks, it will play an important role in the source separation algorithm presented in Sec.~\ref{sec:unmixing_algo}.

In the next section, we formulate the disk and exoplanet restoration as a source separation task. This approach takes the form of a general framework that leverages prior knowledge of the expected object structures to improve the quality of the restoration task. This knowledge includes the acquisition model of Eq.~\eqref{eq:complete_forward_model} and other priors discussed in Sec.~\ref{priors-identification}.

\begin{algorithm}[t]
\caption{GreeDS algorithm for ADI dataset}
\label{alg:GreedyAlgoDisk}
\begin{algorithmic}[1]
\Procedure{GreeDS}{$\bs Y, \rho, l$}\\
\algorithmicrequire{ Dataset $\bs Y \in \bb R^{T \times n^2}$, $\rho, l \in \bb N$}
\State $\bs x \leftarrow \bs 0_{n^2}$
      \For{$r=1,2,\dots, \rho$}
      \For{$i=1,2,\dots, l$}
  \State $\bs x^\top \leftarrow \chi_+\left(\frac{1}{T} \bs 1_T^\top \rotop^{-1} \left[ \bs Y - \Hsvd{r}(\bs Y - \rotop[\bs 1_T \bs x^\top ]) \right] \right),$
      \EndFor 
      \EndFor
      \State return $\bs x$
\EndProcedure
\end{algorithmic}
\end{algorithm}

\section{Source separation algorithm for circumstellar disk imaging}
\label{sec:unmixing_algo}

In this section, we present the disk and planet restoring algorithm, the Morphological Analysis Yielding separated Objects iN Near infrAred usIng Sources Estimation (MAYONNAISE), that we name MAYO for short. The core of MAYO is a source separation problem~\citep[see, \eg][]{bobin20085,donoho2009geometric}, where we intend to separate the stellar speckle field from the circumstellar signals, itself separated into two distinct components, the extended signals (circumstellar disks) and the point source signals (planets). The architecture of our approach allows considering or neglecting specific characteristics of the ADI dataset at hand. For instance, by design, our framework is robust to the typical noise present in ADI datasets. Besides, we consider the response of the instrument by performing a signal deconvolution, using the empirical PSF of the instrument (non-coronagraphic, unsaturated image of the target star), which could be potentially replaced by a more accurate linear response model, if available.

As a general rule, the source separation task is ill-posed as there exists an infinite number of solutions. However, the underlying physics of the acquisition setup and the expected morphology of the circumstellar signals help us identify prior information about each component. Constraining the outputs of our algorithm to respect this prior information then improves the well posedness of the source separation task.

The source separation problem relies on a few key parameters, most of which can be estimated from the output of GreeDS. For this reason, MAYO has the form of a pipeline, where we first run GreeDS, then estimate all the required parameters and finally, we solve the source separation problem. As the latter is the core of MAYO, Sections~\ref{sec:source-separ-algor} to~\ref{sec:numerical-aspects} are solely dedicated to it. Then in Sec.~\ref{sec:MAYO-pipeline}, we present the full MAYO pipeline.

\subsection{Source separation algorithm}
\label{sec:source-separ-algor}

We aim to formulate an algorithm yielding two terms, $\hat{\bs L}$ and $\hat{\bs x}$, estimating the static speckle field $\Lmodel$ and the circumstellar signals $\xmodel$, respectively. The quality of these estimates depends on the signal-to-noise ratio and the nature of the noise terms $\NoiseSpeckles$ and $\NoiseDetector$ in Eq.~\eqref{eq:complete_forward_model}.

From an estimation theory standpoint, given a general family $\cl F \subset \bb R^{T \times n^2} \times \bb R^{n^2}$ of admissible estimates (\eg composed of structural constraints, as detailed it in Sec.~\ref{priors-identification}), and assuming that the distribution of $\NoiseTotal = \NoiseSpeckles + \NoiseDetector$ is known, appropriate estimates $(\hat{\bs L}, \hat{\bs x}) \in \cl F$ must minimize a cost function (or fidelity term) $\cl E(\bs L, \bs x) := \cl L\big(\bs Y - \bs L  -  \cl T( \rotop[\bs 1_T \bs x^\top ] )\big)$, with $\cl L$ the negative log-likelihood of the noise density. We assume that $\cl L$ is \emph{separable}, that is, there exists a function $\cl L'$ such that $\cl L(\bs A) = \sum_{i,j} \cl L'(A_{i,j})$ for any $\bs A \in \bb R^{T \times n^2}$. For instance, assuming that the noise is white and Gaussian, our estimates should minimize $\frac{1}{2} \| \bs Y - \bs L  -  \cl T( \rotop[\bs 1_T \bs x^\top ]) \|_2^2$. In Sec.~\ref{priors-identification}, condition \ref{sec:statistics-noise}, we will derive a separable $\cl L$ from a more accurate noise model than the Gaussian distribution, more suited to the actual speckle statistics. 

In general, the source separation task reads as 
\begin{equation}
 \{\hat{\bs L},\hat{\bs x}\}\ =\ \argmin_{ (\bs L, \bs x) \in \cl F }  \cl L\big(\bs Y -\bs L  -  \cl T( \rotop[\bs 1_T \bs x^\top ] )\big).
\label{eq:opt_no_priors}
\end{equation}

Without any structural constraint on the estimate (if $\cl F = \bb R^{T \times n^2} \times \bb R^{n^2}$), the problem \eqref{eq:opt_no_priors} is ill-posed since an infinity of solutions exists. For instance, if $\{\hat{\bs L},\hat{\bs x}\}$ is a solution, then $\{\hat{\bs L} + \alpha \cl T( \rotop[\bs 1_T \hat{\bs x}^\top ]), (1-\alpha) \hat{\bs x}\}$ is also a solution for all $\alpha \in \bb R$. Furthermore, most of the solutions are not physical (\eg they could be locally negative or have infinite energy) and the produced outputs $\{\hat{\bs L}$, $\hat{\bs x}\}$ are not necessarily good estimates of the static speckle field and the circumstellar signal, respectively. To constrain the solutions, $\cl F$ must include realistic regularization on $\hat{\bs L}$ and $\hat{\bs x}$, coming from their expected properties. By enforcing these different regularizations on the disk and the exoplanetary signals, we are able to separate the image $\xmodel$ into $\Diskmodel$ and $\Planetmodel$, representing the disk and the exoplanetary signals respectively. In this context, the general constrained optimization~\eqref{eq:opt_no_priors} becomes
\begin{subequations}
\label{eq:opt_general_priors}
\begin{align}
\label{eq:opt_general_priors_objective}
\{\hat{\bs L},\hat{\bs x}_\inddisk,\hat{\bs x}_\indpla\}\ =\ \argmin_{\bs L,\bs x_\inddisk,\bs x_\indpla} \quad& \cl L(\bs Y - \bs L  - \cl T( \rotop[\bs 1_T (\bs x_\inddisk+\bs x_\indpla)^\top ]) )\\
 \st\ \quad&\bs L \in \cl C_{\rm sf},\, \bs x_\inddisk \in \cl C_\inddisk,\, \bs x_\indpla \in \cl C_\indpla,
\end{align}
\end{subequations}
where $\cl F$ is implicitly defined from the sets $ \cl C_{\rm sf} \subset \bb R^{T \times n^2}$, $ \cl C_\inddisk, \cl C_\indpla  \subset \bb R^{n^2}$ of physically plausible speckles fields, disk images, and exoplanet images, respectively. The precise meaning of these sets is given in the next section. 
Ideally, the form of $ \cl C_{\rm sf}$ and $ \cl C_\inddisk \cup \cl C_\indpla  $ is such that for $\{\hat{\bs L},\hat{\bs x}_\inddisk,\hat{\bs x}_\indpla\}$, a solution of~\eqref{eq:opt_general_priors}, the quantity $ \cl T( \rotop[\bs 1_T (\hat{\bs x}_\inddisk+\hat{\bs x}_\indpla)^\top ])$ does not belong to $\cl C_{\rm sf}$, which implies that any estimate $\{\hat{\bs L} + \alpha \cl T( \rotop[\bs 1_T (\hat{\bs x}_\inddisk+\hat{\bs x}_\indpla)^\top ]), (1-\alpha) (\hat{\bs x}_\inddisk+\hat{\bs x}_\indpla)\}$, displaying the same value for the cost~\eqref{eq:opt_general_priors_objective}, is not a solution of~\eqref{eq:opt_general_priors}. 

\subsection{Structures and Priors Identification}
\label{priors-identification}

In this section, we identify the crucial structures and priors respected by the different deterministic components of the model Eq.~\eqref{eq:complete_forward_model}, thus specifying the family $\cl F$ of valid estimates. This will help us to stabilize the formulation of our source separation problem, and thus the estimation of $\Lmodel$, $\Diskmodel$, and $\Planetmodel$.

\paragraph{Static part of the speckle field:}
\label{sec:static-part-speckle}
In the computer vision literature, the technique of background-foreground separation enforcing a low-rank background has been extensively used for its efficiency in separating static scenes from a moving foreground~\citep[see, \eg][]{zhou2011godec}. The results obtained by the GreeDS algorithm showed us that a low-rank representation is still appropriate for modeling the speckle field from an ADI dataset. The artifacts produced by PCA-SFS in the estimation $\Lmodel$ are not due to a flaw in this representation but are induced by an inaccurate integration of the influence of the rotating structures, such as the disk and the exoplanets (see Sec.~\ref{sec:dataset}). For this reason, we enforce the rank of any estimate $\bs L$ not to exceed a given $r \in \bb N$, \ie $\text{rank}(\bs L) \leq r$.

\paragraph{Spatial structure of the extended sources:}
\label{sec:spat-struct-disks}

Our source separation problem must be further stabilized by an appropriate regularization of the disk component. First, this problem relies on similar ingredients to the GreeDS algorithm that is prone to circular artifacts (see Sec.~\ref{sec:GreeDS}). Second, part of our estimation translates into a deconvolution operation, a procedure that is prohibitively sensitive to additive noise without regularization~~\citep[see, \eg][]{starck2002deconvolution}. A wavelets basis is a common choice to enforce specific, correlated structures within a signal. However, in spite of its success in many practical applications, this basis is not appropriate when dealing with multivariate data~\citep[see, \eg][]{donoho2001sparse}. Intuitively, while optimal in representing piecewise smooth 1-D signals, the wavelet transform of an image poorly captures edges and curved structures; these are in general not aligned with the vertical, horizontal and (bi-) diagonal directions probed by the wavelet basis, and require a sub-optimal number of coefficients to be accurately represented~\citep{jacques2011panorama}. Over the last 20 years, a large variety of  ``*-lets'' transforms (most of them overcomplete) have been devised to remove the directional limitations of the wavelet transform, such as the curvelets~\citep{candes2000curvelets}, the contourlets~\citep{do2002contourlets}, and many others~\citep{jacques2011panorama}. In our context, we chose to use the shearlets transform ~\citep{kutyniok2012shearlets}, for which the directionality limitation is solved by an efficient shearing operator~\citep{kutyniok2016shearlab}. 

Given the matrix representation of a shearlets transform $\bs \Psi \in \bb R^{d\times n^2}$, with $d \geq n^2$ as this transform is overcomplete, we consider that the disk component $\Diskmodel$ is sparsely represented in the shearlet domain. Mathematically, this comes down to assuming that $\| \bs \Psi \bs \Diskmodel\|_0$ is small, where $\| \bs a \|_0$ returns the sum of non-zero entries of $\bs a$. We will thus impose that an estimate $\bs x_\inddisk$ of $\Diskmodel$ respects the constraints $\| \bs \Psi \bs x_\inddisk\|_0 \leq s$ (or its convex relaxation, as described in the next paragraph) for a prescribed number $s$ of non-zero coefficients in the shearlet domain $\bs \Psi$.

\paragraph{Separating the point sources from the extended sources:}
\label{sec:incl-exopl-sign}

Although the main objective of the present paper is disk imaging, observations of stellar systems likely include both disks and exoplanets, as illustrated by the multi-planetary system PDS~70 (see Fig.~\ref{fig:illustration-greedy-algo}). Disks and exoplanets are morphologically distinct as the former are spatially extended while the latter are point-like sources. This morphological difference suggests we can leverage the principle of the Morphological Components Analysis~\citep[MCA,][]{starck2005morphological} to separate the two sources.

Consider an image $\bs x \in \bb R^{n^2}$ that is the sum of $k$ images $\bs x_i$, \ie $\bs x = \sum_{i=1}^k \bs x_i \in \bb R^{n^2}$, along with the following assumptions.
\begin{itemize}
\item[\emph{(i)}] For each $\bs x_i$, there exists a $n^2 \times d_i$ dictionary --- namely, an overcomplete system including orthonormal bases ($d_i = n^2$) and \emph{frames} ($d_i \geq n^2$) --- $\bs \Phi_{(i)}$ of $d_i \geq n^2$ columns (or atoms) such that $\bs x_i$ can be sparsely represented (or approximated) by a few columns of $\bs \Phi_{(i)}$.
\item[\emph{(ii)}] Each dictionary $\bs \Phi_{(i)}$ cannot sparsely represent the images of another source $\bs x_j$ for $j \neq i$.
\end{itemize}
The principle of MCA is to separate the $k$ images by enforcing the sparsity of each image in its respective dictionary. 
In our context, there are $k=2$ images corresponding to the exoplanet and the disk signals, each sparsely representable in a distinct dictionary. Indeed, the direct (or pixel) domain provides an optimally sparse representation for exoplanets due to their point-like appearance. Conversely, the shearlets are less optimal for representing point sources than extended structures, since the shearlet transform of a point source is spread over more coefficients than a few pixels.  We thus propose to minimize the cost~\eqref{eq:opt_general_priors_objective} under the condition that each component of the circumstellar signal is constrained independently to be sparse in its own representation, \ie $\| \bs \Psi^\top \bs x_\inddisk\|_0 \leq s_\inddisk$ and $ \| \bs x_\indpla\|_0 \leq s_\indpla$.

\paragraph{Response of the instrument:}
\label{sec:response-telescope}
As mentioned in Sec.~\ref{subsec:acquisition-model}, the telescope has a diffractive effect on $\Diskmodel$ and $\Planetmodel$ that we aim to take into account thanks to an integrated deconvolution procedure. We introduced the operator $\cl T$ in Eq.~\eqref{eq:circumstellar-signal-model} for this specific purpose. Without the coronagraph, the effect of $\cl T$ amounts to convolving the objects of interest with the point-spread function (PSF) of the instrument. However, in high-contrast imaging, the presence of the coronagraphic device breaks the linearity of the convolutive model: the coronagraph obstructs the central peak of the star, with a non-null inner working angle (IWA).

As a complete model of the coronagraph effect is complex \citep[\eg][]{Herscovici2017anal} and out of the scope of our study, we propose for now to simply discard all pixels at very small angular separation (that is, below the coronagraph IWA) by applying a circular binary mask to our images: below the IWA of the coronagraph we do not expect to find a realistic exploitable signal, and above the IWA of the coronagraph, the classic convolutive model holds. We define the mask as an operator $\maskop: \bb R^{T \times n^2} \to \bb R^{T \times n^2}$ associated with a radius $\omega > 0$ (in pixel unit) relatively to the center $(c_1,c_2) \in [n] \times [n]$ of the images. Representing $\maskop$ in 3-D for convenience (so that a matrix $\bs A \in \bb R^{T \times n^2}$ is unfolded in $\bb R^{T \times n \times n}$), we have
\begin{equation}
  \label{eq:mask-def}
  \maskop(\bs A)_{t,i,j} =
  \begin{cases}
    A_{t,i,j}&\text{if $\|(i,j) - (c_1,c_2)\| \geq \omega$}\\
    0&\text{otherwise}.
  \end{cases}
\end{equation}

We then adjust $\omega$ so that, in the region unmasked by $\maskop$, $\cl T$ is well approximated by a convolution with the empirical off-axis unsaturated PSF $\psf$, that is
$$
[\cl M \circ \cl T]( \bs A) \approx \cl M[\psf \convOp \bs A],
$$
for any matrix $\bs A \in \bb R^{T \times n^2}$, where $\convOp$ denotes the 2-D convolution operation, applied to each of the $T$ images composing $\bs A$. In practice, we set the value of $\omega$ to the IWA of the coronagraph, which can be measured empirically with the instrument or theoretically determined via simulations with an accurate model of coronagraph \cite{Soummer2007mft}. 
This observation allows us to adapt the cost function in Eq.~\eqref{eq:opt_general_priors_objective} as 
\begin{equation}
  \label{eq:new-cost-with-mask}
  \cl L\big(\maskop( \bs Y - \bs L  - \psf \convOp \rotop[\bs 1_T \bs  (\bs x_\inddisk + \bs x_\indpla)^\top ]  )\big).  
\end{equation}

Note that, while minimizing the cost \eqref{eq:new-cost-with-mask}, the masking operator $\maskop$ does not set to zero the spatial pixels with small angular separation of the estimates $\hat{\bs L}$, $\hat{\bs x}_\inddisk$, and $\hat{\bs x}_\indpla$. From the separability of $\cl L$ (see Sec.~\ref{sec:source-separ-algor}), this region is simply not considered in fidelity cost $\cl L$. Its corresponding pixels can thus take any value compatible with our sparsity constraints; the estimates realize a form of intensity interpolation in this small angle area, that is a sparsity-regularized image \emph{inpainting}~\citep{fadili2009inpainting}.

\paragraph{Positivity of the images:}
\label{sec:positivity-images} 

As a last constraint, since the signals of interest represent intensity values, their pixel values must also be positive. Therefore, we enforce the estimates to respect $\bs L, \bs x \geq 0$ (entrywise and componentwise, respectively). 

\subsection{Derivation of the cost function from the noise statistics}
\label{sec:statistics-noise}

In the acquisition model~\eqref{eq:complete_forward_model}, the noise $\NoiseTotal = \NoiseSpeckles+\NoiseDetector$ is regarded as the sum of the (non-static) speckle noise~$\NoiseSpeckles$ and the residual detector noise~$\NoiseDetector$. We briefly discuss these noise terms and provide an adequate fidelity term suited to their statistics. 

The speckles originate mainly from slowly evolving optical aberrations in the instrument and marginally from short-lived atmospheric residuals post-adaptive optics correction. In a long-exposure image, the speckles intensity distribution is accurately modeled by a Modified Rician (MR) distribution $\MR(I_c,I_s)$ with density
\begin{equation}
\ts p_{\text{MR}}(I , I_c,I_s) = \frac{1}{I_s}\exp \left(-\frac{I+I_c}{I_s}\right) \cl{I}_0 \left(\frac{2\sqrt{II_c}}{I_s}\right),
\label{MR_pdf}
\end{equation}
where $\cl{I}_0$ is the modified Bessel function of the first kind~\citep[][and references therein]{fitzgerald2006speckle, soummer2007speckle,marois2008confidence}. \cite{pairet2019stim} showed that the MR distribution is sub-exponential~\citep{vershynin2010introduction}, meaning that its tail decays as a Laplace distribution. Mathematically, we have, for a random variable  $X\sim \MR(I_c,I_s)$,  
\begin{equation}
\bb P[|X| > \epsilon] \leq C \exp(-c \epsilon)
\label{eq:subexp_decay}
\end{equation}
where $c>0$. From a maximum likelihood estimation (MLE) standpoint, and in absence of any other noise source, this would suggest setting an $\ell_1$-norm for the data fidelity term $\cl L$ for high intensity observations.

The detector noise encompasses disturbances of different nature occurring at the detector level. As $\bs Y$ goes through a data reduction pipeline (including flat, dark and background corrections), a thorough description of the distribution of the detector noise is delicate and beyond the scope of the present paper. We argue that most of the disturbances introduced by the detector emanate from the addition of many independent random variables. From the central limit theorem, their combined action thus follows an additive Gaussian noise; the fidelity term should therefore be set to an $\ell_2$-norm, the negative log-likelihood of the Gaussian distribution. Besides this Gaussian noise contribution, there remain a few high-intensity defective pixels (or hot pixels) in $\bs Y$ after the data reduction.  As these hot pixels have high values and are sparsely distributed, an $\ell_1$-norm is suitable to capture their impact in the fidelity term.

From these considerations, it is appropriate to regroup the noise sources according to their suitable fidelity terms. We thus rewrite $\NoiseTotal = \bs N_{\ell_1} + \bs N_{\ell_2}$, where $\bs N_{\ell_1} $ encompasses both $\NoiseSpeckles$ and the hot pixel impact, whereas $\bs N_{\ell_2}$ denotes all the other Gaussian-like detector noise sources. Accordingly, an appropriate fidelity term should combine the $\ell_1$-norm and $\ell_2$-norm. Note that, since the temporally correlated noise is absorbed by $\bar{\bs L}$, we assume that the total noise has zero expectation ($\bb E\,\NoiseTotal = \bs 0$). However, the mean and variance of the speckle noise depend on the radial distance from the star~\citep[see, \eg][]{soummer2007speckle}. Therefore, the optimal norm to minimize the fidelity term $\cl L$ is radius-dependent, which was empirically highlighted in the recent work of~\cite{dahlqvist2020regime}.   

Rather than establishing the precise theoretical form of our fidelity term -- which would require a possibly unreachable statistical analysis of all noise sources -- we propose to model $\NoiseTotal$ with a parametric \emph{Huber} density \cite[p. 86]{huber1981robust}, \cite[p. 279]{dennisessential}. This specific choice leverages the larger probability to find high pixel intensities in $\bs N_{\ell_1}$ than in $\bs N_{\ell_2}$.

Below a given intensity threshold $\delta \geq 0$, the Huber PDF is indeed Gaussian, while above this threshold, it follows a Laplace distribution; the Huber density is a mixture of a truncated Gaussian and a truncated Laplace distribution. Mathematically, a zero mean and unit variance Huber PDF is defined~by:
\begin{align}
\cl H_{\delta} \sim Z(\delta)^{-1} \exp(- c \, |x|_\delta),
\label{eq:PDFassumed}
\end{align}
where $c>0$ is a universal constant, and $Z$ ensures the normalization of $\cl H_{\delta}$. The function ${|\cdot|_\delta}$ is (up to a constant shift) the negative log-likelihood of $\cl H_{\delta}$, or Huber-loss, defined by
\begin{align}
|x |_\delta := 
\begin{cases} 
\inv{2} x^2 &\text{if } |x| \leq \delta,\\
\delta (| x | - \inv{2}\delta) &\text{if } |x| > \delta. 
\label{eq:Huber-loss}
\end{cases}
\end{align}
The Huber-loss \eqref{eq:Huber-loss} is convex, and both continuous and differentiable everywhere for any value of $\delta \geq 0$.

Our main hypothesis for setting the fidelity cost $\cl L$ in the optimization \eqref{eq:opt_general_priors} consists in assuming that each voxel $N_{ij}$ of the noise $\NoiseTotal$, when properly normalized by its standard deviation $\xi_{ij}>0$ (which could vary with the radial distance), is identically and independently distributed ($\iid$) as 
\begin{equation}
  \label{eq:huber-iid-assumption}
  N^{\rm n}_{ij} := \xi^{-1}_{ij}\,N_{ij} \sim_{\iid} \cl H_{\delta}.
\end{equation}
The normalized noise $\NoiseTotal^{\rm n}$ is then a Huber noise with threshold $\delta$ and unit variance (on each voxel), a direct generalization of both the additive Gaussian ($\delta \to +\infty$) or Laplace ($\delta \ll 1$) noise models. 
Under this assumption, the PDF of $\NoiseTotal^{\rm n}$ is the product of the $Tn^2$ normalized voxel PDFs, and its negative log-likelihood corresponds to the normalized Huber norm (up to a constant shift):     
  \begin{equation}
    \label{eq:normalized-huber-norm}
    \ts \cl L(\bs N) = \Hnorm{\bs N} := \sum_{ij} \xi_{ij} | \xi_{ij}^{-1} N_{ij} |_\delta,   
  \end{equation}
with $\bs \Xi$ the matrix whose entries are $\xi_{ij}$. This endows our source separation with the convex fidelity cost 
\begin{equation}
\cl E^{\rm H}(\bs L, \bs x_\inddisk, \bs x_\indpla) := \Hnorm{\maskop(\bs Y - \bs L  - \psf \convOp \rotop[\bs 1_T (\bs x_\inddisk + \bs x_\indpla)^\top ] )}.
\label{eq:huberfidelity}
\end{equation}

The fidelity \eqref{eq:huberfidelity} assumes that the Huber density parameters $\bs \Xi$ and $\delta$ are given. While their exact values are unknown, we described in App.~\ref{app:normalized-huber-loss} an automatic procedure to estimate them from the GreeDS residual computed from $\bs Y$, a proxy of $\NoiseTotal$.  We show in this appendix that this residual, once properly normalized by the entries of $\bs \Xi$, is well fitted by a single Huber density \eqref{eq:PDFassumed} with a unique threshold $\delta$, thus validating the hypothesis \eqref{eq:huber-iid-assumption}. We also compare in Sec.~\ref{sec:numercial-exp} the results obtained by the cost $\cl E^{\rm H}$ to those reached by setting $\cl L$ to either a square $\ell_2$-norm or an $\ell_1$-norm, and show that \eqref{eq:huberfidelity} yields a more accurate signal separation.

\subsection{Final formulation of the source separation problem}
\label{sec:final-form-optim}

From the previous sections, our source separation task is tantamount to solving the following ideal constrained optimization problem:
\begin{subequations}
\label{eq:final-objective}
\begin{align}
\argmin_{\bs L, \bs x_\inddisk, \bs x_\indpla} \quad &\ts \Hnorm{\maskop(\bs Y - \bs L  - \psf \convOp \rotop[\bs 1_T (\bs x_\inddisk + \bs x_\indpla)^\top ] )},  \\
\st\quad&\text{rank}(\bs L) \leq r, \label{eq:final-objective-rank}\\
&\| \bs \Psi^\top \bs x_\inddisk\|_0 \leq s_\inddisk, \label{eq:final-objective-l0-Psi}\\
&\| \bs x_\indpla\|_0 \leq s_\indpla, \label{eq:final-objective-l0-Id}\\
&\bs L,  \bs x_\inddisk, \bs x_\indpla \geq 0.
\end{align}
\end{subequations}
This optimization scheme is unfortunately non-convex and NP-hard in general \citep{natarajan1995sparse}. To tackle this problem, we could develop a greedy algorithm to approximate the solution of Eq.~\eqref{eq:final-objective}, such as a variant of iterative hard thresholding \citep{blumensath2008iterative}, but such a procedure is often highly sensitive to initialization. Another strategy is to relax Eq.~\eqref{eq:final-objective} into a convex optimization by replacing $\rank(\cdot)$ in~\eqref{eq:final-objective-rank} by the nuclear norm $\|\cdot\|_*$ \citep[summing the singular values of the tested matrix;][]{candes2011robust}, and $\|\cdot \|_0$ in Eq.~\eqref{eq:final-objective-l0-Psi} and Eq.~\eqref{eq:final-objective-l0-Id} by the $\ell_1$-norm, \ie adding up the absolute components of the tested vector. This would lead to the new constraints
\begin{subequations}
\label{eq:convex_relaxed_objective}
\begin{numcases}{}
\ts \|\bs L\|_* \leq \tau_L\label{eq:nuclear-low-rank-constraint}\\
\ts \| \bs \Psi^\top \bs x_\inddisk\|_1 \leq \tau_\inddisk,\ \|\bs x_\indpla\|_1 \leq \tau_\indpla,\ \bs L, \bs x_\inddisk, \bs x_\indpla \geq 0.\label{eq:all-but-nuclear-low-rank-constraint}
\end{numcases}
\end{subequations}

The resulting convex optimization, which is common in the signal and image processing literature~\citep{bobin20085}, is appealing but highly sensitive to the value of the parameters $\tau_L, \tau_\inddisk, \tau_\indpla > 0$; the constraints \eqref{eq:nuclear-low-rank-constraint} and \eqref{eq:all-but-nuclear-low-rank-constraint} crucially depend on the intensity of their input. This effect is also amplified by the large contrast existing between the starlight and the circumstellar signals. This new optimization is therefore unfit for estimating the circumstellar signals $\bs x_\inddisk$ and $\bs x_\indpla$; a small variation in our estimation of $\tau_L$ quickly yields unfaithful estimates for $\bs x_\inddisk$ and $\bs x_\indpla$. We illustrate this sensititvity in App.~\ref{app:dummy-convex-problem} in a simple noiseless \emph{low-rank plus sparse} decomposition that both non-convex and convex approach solve easily when the components have similar intensities. We then consider the case of a low-rank component with significantly larger intensity than the sparse component. While the non-convex approach reaches similar performance in this case, the convex relaxation fails to faithfully reproduce the sparse component when $\tau_L$ slightly deviates from the optimal value (see Fig.~\ref{fig:dummy_convex_low_rank_plus_sparse} in this appendix). 

To alleviate this sensitivity issue, we consider another convex relaxation of the rank constraint \eqref{eq:final-objective-rank} based on the weighted nuclear norm proposed in \cite{eftekhari2018weighted}. In a nutshell, assuming we can estimate the subspace spanned by the columns of the target low-rank matrix $\Lmodel$, its column space, this method amounts to constraining any estimate of this matrix to live in the same space.

 Mathematically, assuming $\Lmodel$ has rank $r$ and using the SVD decomposition $\Lmodel$ = $\bar{\bs U} \bar{\bs \Sigma} \bar{\bs V}^\top$, the column space of $\Lmodel$ is spanned by the $r$ first (orthonormal) columns of $\bar{\bs U}$ since $\bar{\bs \Sigma}$ is made of $r$ non-zero values on its diagonal. Therefore, $\Lmodel$ respects the constraint
\begin{equation}
  \label{eq:other-convex-constraint-for-low-rank-L}
  \Lmodel \in \cl P_{r} := \{\bs A \in \bb R^{T \times n^2}\!: \bs A = \bar{\bs P}^{(r)} \bs A\},
\end{equation}
with $\bar{\bs P}^{(r)} := \bar{\bs U}_{[r]} \bar{\bs U}^\top_{[r]}$. Geometrically, $\cl P_{r}$ is the column space of $\Lmodel$, a subspace composed of rank-$r$ matrices. The matrix $\bar{\bs P}^{(r)}$ is also the projection (by left multiplication) of any matrix $\bs A$ on $\cl P_{r}$.

Interestingly, conversely to~\eqref{eq:nuclear-low-rank-constraint}, the constraint \eqref{eq:other-convex-constraint-for-low-rank-L} does not require any additional parameter and is invariant under rescaling: if $\bs A \in \cl P^{(r)}$, then $\lambda \bs A \in \cl P^{(r)}$ for any $\lambda \in \bb R$. Therefore, assuming we have access to a reliable estimate $\hat{\bs P}^{(r)}$ of $\bar{\bs P}^{(r)}$,  we can replace the low-rank constraint~\eqref{eq:nuclear-low-rank-constraint} by
\begin{equation}
  \label{eq:other-convex-constraint-for-low-rank-L-approx}
  \bs L \in \hat{\cl P}^{(r)} := \{\bs A \in \bb R^{T \times n^2}: \bs A = \hat{\bs P}^{(r)} \bs A\}. 
\end{equation}

To estimate $\bar{\bs P}^{(r)}$, we propose to leverage the capacity of GreeDS in extracting the rotating signals from $\bs Y$. If $\gdsest{\bs x}$ is the fixed point (or an approximation) reached by this algorithm, we compute $\gdsest{\bs L}$ from Eq.~\eqref{GreeDS_solve_L} (replacing $\bar{\bs x}$ by $\gdsest{\bs x}$), and consider the SVD decomposition $\gdsest{\bs L} = \gdsest{\bs U} \gdsest{\bs \Sigma} \gdsest{\bs V}^{\top}$  of that matrix. As we expect $ \gdsest{\bs U}_{[r]}$ to reliably approximate $ \bar{\bs U}_{[r]}$, we set our estimated projector to $\hat{\bs P}^{(r)} := \gdsest{\bs U}_{[r]} (\gdsest{\bs U}_{[r]})^\top$. We assess numerically the quality of the estimate $\hat{\bs P}^{(r)}$ of $\bar{\bs P}^{(r)}$ in Sec.~\ref{sec:TestprojOnU}.

The final formulation of the optimization problem, the one solved in the MAYONNAISE (or MAYO) pipeline, is then
\begin{subequations}
\label{eq:convex_relaxed_objective-final_formulation}
\begin{align}
\label{eq:objective-final_formulation}
\Delta(\bs Y) := \argmin_{\bs L, \bs x_\inddisk, \bs x_\indpla} \quad&\ts \Hnorm{\maskop(\bs Y - \bs L  - \psf \convOp \rotop[\bs 1_T (\bs x_\inddisk + \bs x_\indpla)^\top ] )},  \\
\label{eq:final-objective-rank-constraint}
\st\quad&\ts  \bs L \in \hat{\cl P}^{(r)},\\
\label{eq:final-objective-disk-constraint}
&\ts \| \bs \Psi^\top \bs x_\inddisk\|_1 \leq \tau_\inddisk,\\
\label{eq:final-objective-planet-constraint}
&\ts \|\bs x_\indpla\|_1 \leq \tau_\indpla \,,\\
\label{eq:final-objective-positivity}
& \ts \bs L,  \bs x_\inddisk, \bs x_\indpla \geq 0.
\end{align}
\end{subequations}
The constraint~\eqref{eq:final-objective-rank-constraint} does not depend on the intensity of its input, hence solving the sensitivity issue explained previously.

\subsection{Numerical implementation of the minimization program}
\label{sec:numerical-aspects}

The MAYO pipeline solves the convex problem~\eqref{eq:convex_relaxed_objective-final_formulation}. It uses a primal-dual algorithm called Primal-Dual Three-Operator splitting~\citep[PD3O][]{yan2018new}. This is an iterative algorithm, closely related to the projected gradient descent algorithm, that is tailored to minimize non-smooth convex problems of the form of~\eqref{eq:convex_relaxed_objective-final_formulation}. The PD3O algorithm requires to compute the gradient of $\cl E^{\rm H}$ in Eq.~\eqref{eq:huberfidelity} at each iteration. 

At first sight, the computation of the gradient of the cost $\cl E^{\rm H}$ with respect to $\bs x_\inddisk$ or $\bs x_\indpla$ is rather complex. Indeed, up to a convenient reshaping, the evaluation of any function $f: \bb R^{T \times n^2} \to \bb R$ (such as the restriction of $\cl E^{\rm H}$ to $\bs x_\inddisk$ or $\bs x_\indpla$) on the volume obtained from the rotation of a static configuration $\bs x \in \bb R^{n^2}$ can always be rewritten as $f(\rotop[\bs 1_T \bs x^T]) = \bar f(\rotopmat \bs x)$, for some appropriate function $\bar f: \bs z \in \bb R^{T\!n^2} \to \bar f(\bs z) \in \bb R$ and matrix $\rotopmat \in \bb R^{T\!n^2 \times n^2}$. In this context,
\[
  \bs \nabla_{\bs x} f(\rotop[\bs 1_T \bs x^T]) = \bs \nabla_{\bs x} \bar f( \rotopmat \bs x) =  \rotopmat^\top [\bs \nabla_{\bs z} \bar f]( \rotopmat \bs x). 
  \]
This shows that the explicit computation of $\rotopmat^\top$ is required, which is slow and inefficient since $\rotop$ includes an interpolation to comply with the pixel grid rotation. 

In this work, we follow another strategy provided by recent achievement in machine learning and automatic differentiation~(AD). AD is a technique that repeatedly applies the chain rules to compute derivatives of numerical functions. The resulting derivatives are exact (up to numerical errors) and computed with the same complexity as the cost function (up to a multiplicative factor), see~\eg~\citep{baydin2017automatic}. 

We have used the \textsc{PyTorch} toolbox to compute the gradient of $\cl E^{\rm H}$ with respect to $\bs x_\inddisk$ and $\bs x_\indpla$, thanks to its {\tt autograd} capability~\citep{paszke2017automatic}. More specifically, we use \textsc{kornia}, a computer vision toolbox compatible with the {\tt autograd} functionality of \textsc{PyTorch}~\citep{riba2020kornia}. 
Regarding the shearlets transform, we rely on the \textsc{Python} package \textsc{pyShearlab}, the python version of the \textsc{ShearLab 3D} toolbox~\citep{kutyniok2016shearlab}. Our numerical developments heavily rely on the \textsc{python} scientific computing libraries of \textsc{numpy} \citep{2020numpy}, and \textsc{scipy}~\citep{2020SciPy-NMeth}. In addition, the \textsc{VIP} toolbox is extensively used~\citep{2016ascl.soft03003G} for all the numerical experiments.

\subsection{Presentation of the MAYO pipeline}
\label{sec:MAYO-pipeline}

As we have seen, the output of GreeDS is used to estimate the parameters of the Huber-distribution, $\delta$ and $\bs \Xi$, and the set in constraint~\eqref{eq:final-objective-rank-constraint}. The complete MAYO pipeline is thus as follows: we run GreeDS, compute the SVD of $\gdsest{\bs L}$ to get $\hat{\bs P}^{(r)}$, run the procedure \textsc{HuberFit} (see App.~\ref{app:normalized-huber-loss}) to estimate $\delta$ and $\bs \Xi$, and finally we solve the problem~\ref{eq:objective-final_formulation}, that is, we compute $\Delta(\bs Y)$. The complete pipeline is described in Alg.~\ref{algo:mayo-pipeline}. The code of MAYO is available at~\url{https://github.com/bpairet/mayo_hci}.

\begin{algorithm}
\caption{MAYO pipeline}
\label{algo:mayo-pipeline}
\begin{algorithmic}[1]
\Procedure{MAYO}{$\bs Y$, $r$, $\tau_d$, $\tau_p$, $\omega$,  $\rho$, $l$}\\
\algorithmicrequire{ $\bs Y$, $r$, $\tau_d$, $\tau_p$, $\omega$,  $\rho$, $l$}
\State $\gdsest{\bs L}, \gdsest{\bs x} \leftarrow \textsc{GreeDS}(\bs Y, \rho, l)$
\State $\gdsest{\bs U} \gdsest{\bs \Sigma} \gdsest{\bs V}^{\top}  \leftarrow \textsc{SVD}( \gdsest{\bs L} )$
\State $\hat{\bs P}^{(r)} \leftarrow \gdsest{\bs U}_{[r]} \gdsest{\bs U}_{[r]}^\top$
\State $\delta, \bs \Xi \leftarrow \textsc{HuberFit}(\bs Y - \gdsest{\bs L} - \rotop[\bs 1_T (\gdsest{\bs x})^\top])$
\State $\hat{\bs L}, \hat{\bs x}_\inddisk, \hat{\bs x}_\indpla \leftarrow \Delta(\bs Y ; r,\tau_d, \tau_p,\omega,\delta, \bs \Xi,\hat{\bs P}^{(r)}) $
\State return $\hat{\bs x}_\inddisk$, $\hat{\bs x}_\indpla$ 
\EndProcedure
\end{algorithmic}
\end{algorithm}

Although the MAYO pipeline is mostly automatic, there remain a few user parameters to set: the mask radius $\omega$, the rank~$r$, and the sparsity constraints $\tau_\inddisk$ and $\tau_\indpla$. The radius $\omega$ depends on the type of coronagraph used for the observation and its transmission profile and can be set as the IWA of the coronagraph. For the rank parameter $r$, it is the same parameter used in a classic PCA-SFS and thus, selecting an adequate value for $r$ is done routinely in the literature. Regarding the sparsity constraints for the disks $\tau_\inddisk$ and the planets $\tau_\indpla$, we provide in App.~\ref{app:set-taus} a heuristic method to help the practitioner to choose correct values, as their setting is not straightforward.

\section{Numerical validation and performance estimation}
\label{sec:numercial-exp}

In this section, we assess the performance of our algorithm on an empty data cube from the VLT/SPHERE-IRDIS instrument \citep[courtesy of the SHINE guaranteed time survey,][]{chauvin2017shine}, in which we injected synthetic disk and planet signals. This injection was performed with the opposite parallactic angles to smear the presence of potential real signals in the considered dataset while preserving the spatiotemporal statistics of the starlight residuals. We injected the signals using the \textsc{VIP} toolbox \citep{2016ascl.soft03003G} in which models of disks can be produced using a light version of the GRaTeR \citep[GRenoble RAdiative TransfER,][]{augereau1999grater} tool. We consider a VLT/SPHERE-IRDIS dataset that is free from circumstellar signal, denoted Empty in Table~\ref{table:list_data_set}. We denote this volume by $\Lmodel$. We then inject a synthetic disk signal $\bs D$ as $\bs D = \psf \convOp  \rotop[\bs 1_T^\top \bs  x_\inddisk]$ to $\Lmodel$, forming $\bs Y = \Lmodel  + \bs D$. Except when otherwise stated, synthetic disks are only characterized by their inclination and contrast.

We also apply our method to the VLT/SPHERE-IRDIS data of three emblematic disks showing different structures: the debris disk HR~4796~A, the transition disk SAO~206462 and the protoplanetary disk PDS~70. We compare the results obtained from our method with PCA-SFS, and show that our approach has deeper detection capabilities, recovers the intensity distribution of the disk and accurately separates the planet from the disk signal.

\subsection{Assessment of the convex low-rank constraint}
\label{sec:TestprojOnU}

Since this fact is essential to our source separation method, we demonstrate here that the convex low-rank constraint~\eqref{eq:other-convex-constraint-for-low-rank-L-approx} is an appropriate surrogate for the constraint~\eqref{eq:other-convex-constraint-for-low-rank-L}. This amounts to showing that the projector $\hat{\bs P}^{(r)}$ (and the column space $\hat{\cl P}^{(r)}$) estimated from GreeDS is a reliable estimate of $\bar{\bs P}^{(r)}$ (resp. $\cl P^{(r)}$).

We consider $\gdsest{\bs L}$ from Eq.~\eqref{GreeDS_solve_L} recovered from the output of the GreeDS algorithm. As the number of frames $T$ in our dataset is only 48, we limit the value of $\rho$ in \eqref{GreeDS_fixed_point_iteration} to 10.  
As explained in Sec.~\ref{sec:final-form-optim}, we then set the constraint~\eqref{eq:other-convex-constraint-for-low-rank-L-approx} from the computation of $\hat{\bs P}^{(r)} := \gdsest{\bs U}_{[r]} (\gdsest{\bs U}_{[r]})^\top$, where $\gdsest{\bs U}$ is deduced from the SVD decomposition $\gdsest{\bs L}= \gdsest{\bs U} \gdsest{\bs \Sigma} (\gdsest{\bs V})^{\top}$.

As a first proximity measure between the column spaces $\hat{\cl P}^{(r)}$ and $\cl P^{(r)}$ (and their associated projectors), we show that they are well aligned. By construction, we know that these spaces are generated by the $r$-first columns of $\hat{\bs U}$ and $\bar{\bs U}$, respectively. However, if these spaces are identical, these columns may differ by an unknown rotation. Therefore, to verify the announced alignment, we simply test if the $T-r$ last columns of $\hat{\bs U}$ are orthogonal to the $r$ first ones of $\bar{\bs U}$, and conversely. Mathematically, we study if
\begin{equation}
  \bar{\bs u}_i \cdot \gdsest{\bs u}_j  \approx 0,\quad \bar{\bs u}_j \cdot \gdsest{\bs u}_i \approx 0,\quad  \forall i \in [r], j \in [T]\setminus [r],
   \label{eq:dot_product_r_span_ok}
\end{equation}
with $\bar{\bs u}_k$ and $\hat{\bs u}_k$ the columns of $\bar{\bs U}$ and $\hat{\bs U}$, respectively.

\begin{figure}
  \centering
  {\includegraphics[width=0.23\textwidth]{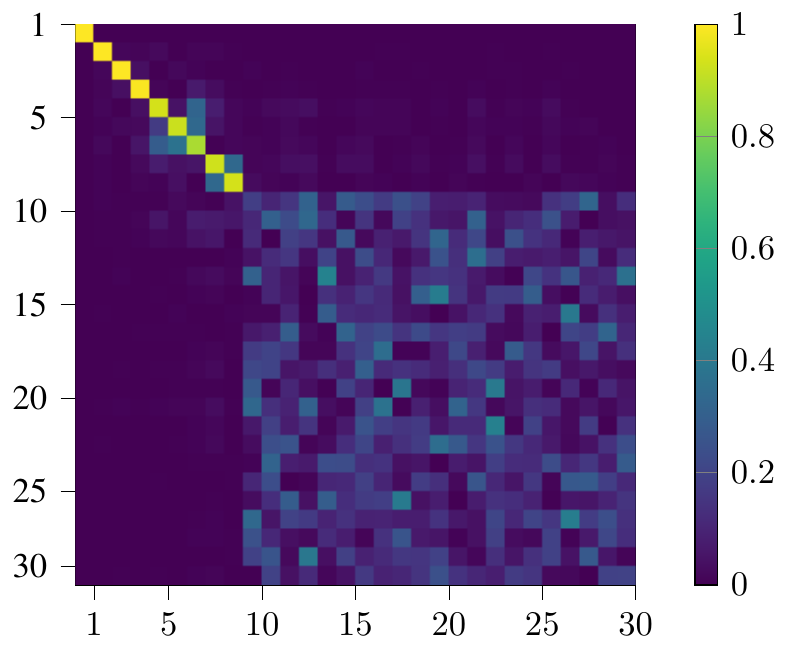}}
  {\includegraphics[width=0.23\textwidth]{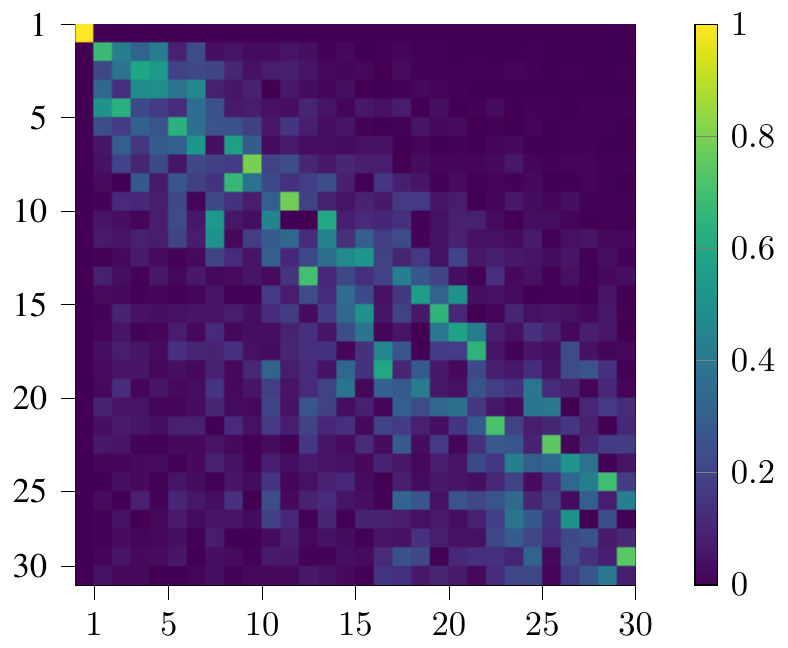}}
\caption{Dot product of $\bar{\bs U}$ and $\gdsest{\bs U}$ (left) compared to the dot product of $\bar{\bs U}$ and $\bs U_{\bs Y}$ (right). Only the dot products of the 30 first columns of each matrix are displayed for clarity. We can see that the spaces spanned by $\bar{\bs U}$ and $\gdsest{\bs U}$ are almost identical for the first $\rho-1$ columns, while for $\bs U_{\bs Y}$, only its first column spans a space similar to that of $\bar{\bs U}$.}
\label{fig:proj_onU_disk_no_disk_xl}
\end{figure}

On Fig.~\ref{fig:proj_onU_disk_no_disk_xl} (left), we display (a zoom on) the $T \times T$ matrix obtained from the dot products between the $T$ columns $\bar{\bs U}$ and the $T$ columns $\gdsest{\bs U}$. The block structure of this figure shows that~\eqref{eq:dot_product_r_span_ok} is approximately met for $r \leq \rho -1 = 9$. Note that for $i,j \leq \rho-1$, this matrix is close to the identity: the vectors $\bar{\bs u}_i$ and $\gdsest{\bs u}_j$ are thus almost identical, which is stronger than~\eqref{eq:dot_product_r_span_ok}. As a comparison, we also display on Fig.~\ref{fig:proj_onU_disk_no_disk_xl} (right) the dot products between the columns of $\bar{\bs U}$ and the ones of $\bs U_{\bs Y}$ (obtained from the SVD of $\bs Y$) to highlight that, as expected, the space spanned by the two matrices are not similar, apart from the first column. On Fig.~\ref{fig:proj_onU_disk_no_disk_xl}, we only display the result for a disk of inclination 60 and contrast $5.3\times 10^{-5}$. However, we have observed a similar trend for a wide variety of disks, as outlined by the average score of our second proximity evaluation.

\begin{figure}
  \centering
  {\includegraphics[width=0.4\textwidth]{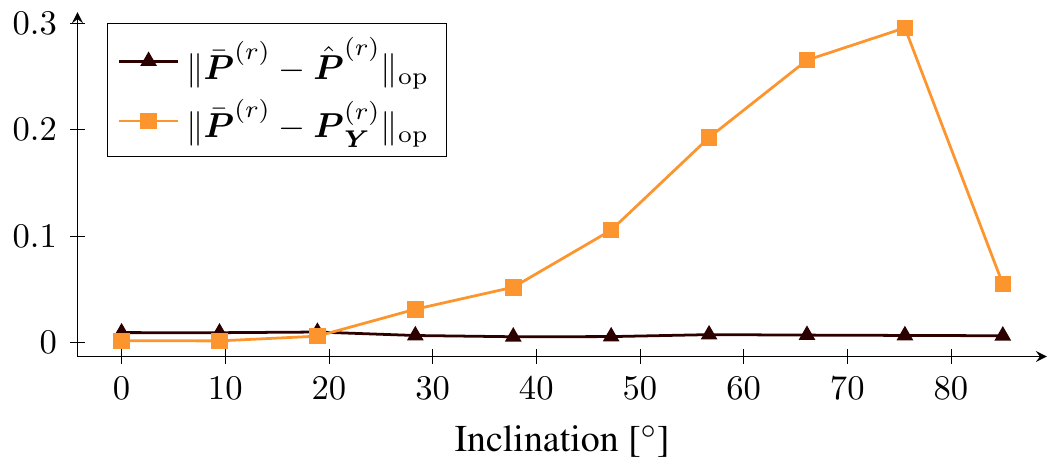}}
\caption{Evolution of $\|\bar{\bs P}^{(r)}-\hat{\bs P}^{(r)}\|_{\rm op}$ and $\|\bar{\bs P}^{(r)} - \bs P^{(r)}_{\bs Y}\|_{\rm op}$ with respect to the disk inclination and for disk contrast ranging from $3.5\times 10^{-6}$ to $7.0\times 10^{-5}$. For each inclination and contrast value, ten disks were injected with different orientations in the image.  
The value of $\|\bar{\bs P}^{(r)}-\hat{\bs P}^{(r)}\|_{\rm op}$ is always low, meaning that the projector $\hat{\bs P}^{(r)}$ estimated with the GreeDS algorithm is a good approximation of the groundtruth projector $\bar{\bs P}^{(r)}$.}
\label{fig:distance_P_y_vs_P_greeds_and_P_Y}
\end{figure}

Our second evaluation of the proximity between $\hat{\cl P}^{(r)}$ and $\cl P^{(r)}$ aims at showing that the largest relative distance between any $\bs L \in \hat{\cl P}^{(r)}$ and the true column space $\cl P^{(r)}$ is small. Since $\bar{\bs P}^{(r)} \bs L \in \cl P^{(r)}$ is the closest matrix of $\cl P^{(r)}$ to $\bs L$ and $\|\bs B \bs C\|_2 \leq \|\bs B\|_{\rm op} \|\bs C\|_2$ for any $\bs B \in \bb R^{T \times T}$ and $\bs C \in \bb R^{T \times n^2}$, this largest distance is bounded by
  \begin{equation}
    \label{eq:column-space-estimate-quality}
    \ts \sup_{\bs L \in \hat{\cl P}^{(r)}} \big(\|\bs L\|^{-1}_2\, \|\bs L - \bar{\bs P}^{(r)} \bs L\|_2\big)\ \leq\ \|\hat{\bs P}^{(r)} - \bar{\bs P}^{(r)}\|_{\rm op}.
  \end{equation}
We have analyzed this bound empirically by comparing $\|\bar{\bs P}^{(r)}-\hat{\bs P}^{(r)}\|_{\rm op}$ and $\|\bar{\bs P}^{(r)} - \bs P^{(r)}_{\bs Y}\|_{\rm op}$ over a large variety of synthetic disk configurations. We applied the GreeDS algorithm on a total of 400 synthetic disks, with inclination ranging from 0$^{\circ}$ (face-on) to 85$^{\circ}$ (edge-on) and contrast ranging from $3.5\times 10^{-6}$ to $ 7.0\times 10^{-5}$. We display the resulting average values as a function of the inclination in Fig.~\ref{fig:distance_P_y_vs_P_greeds_and_P_Y}. 

We observe that $\|\bar{\bs P}^{(r)}-\hat{\bs P}^{(r)}\|_{\rm op}$ is always significantly smaller than 1, with an average value of $0.007$, a tiny fraction of the upper bound $\|\bar{\bs P}^{(r)}-\hat{\bs P}^{(r)}\|_{\rm op} \leq \|\bar{\bs P}^{(r)}\|_{\rm op} +\|\hat{\bs P}^{(r)}\|_{\rm op} = 2$ (since the operator norm of a projector is one). This illustrates the suitability of estimating $\cl P_r$ with $\hat{\cl P}_r$. We note that for small inclination, the $\|\bar{\bs P}^{(r)} - \bs P^{(r)}_{\bs Y}\|_{\rm op}$ is slightly above $\|\bar{\bs P}^{(r)}-\hat{\bs P}^{(r)}\|_{\rm op}$. This is not surprising as the little rotational diversity of small inclination disks implies that they are almost entirely included in the first PC of $\bs Y$. As we can see on Fig.~\ref{fig:proj_onU_disk_no_disk_xl}, the first column of $\bs U_Y$ and $\bar{\bs U}$ are almost identical. Hence, for these disks, $\bar{\bs P}^{(r)} \approx \bs P^{(r)}_{\bs Y} $ and thus, even though $\hat{\bs P}^{(r)}$ is a good estimate, it does not bring an improvement over $ \bs P^{(r)}_{\bs Y} $ in this case.

Our two measures of proximity show that $\hat{\bs P}^{(r)}$ is a reliable estimate of $\bar{\bs P}^{(r)}$ in all large variety of disk configurations. We will thus consider that imposing the rank constraint using condition~\eqref{eq:final-objective-rank-constraint} is valid for any $r\leq \rho-1$ in the other datasets processed in this work.

\subsection{Huber-loss comparison with $\ell_2$ and $\ell_1$-norms}
\label{sec:huber-vs-l2-or-l1}

We challenge here the suitability of the Huber-loss developed in Sec.~\ref{sec:statistics-noise} as the minimized cost of our source separation algorithm. We compare this choice to the case of a cost function $\cl L$ set to the square $\ell_2$-norm (suited to a Gaussian noise) and the $\ell_1$-norm (suited to a Laplace noise). As detailed in Alg.~\ref{algo:mayo-pipeline}, the parameters of the Huber-loss are fixed using the \textsc{HuberFit} procedure (see App.~\ref{app:normalized-huber-loss}).
 
\begin{figure*}
  \centering
   \vspace{0.3cm}
    {\includegraphics[width=0.31\textwidth]{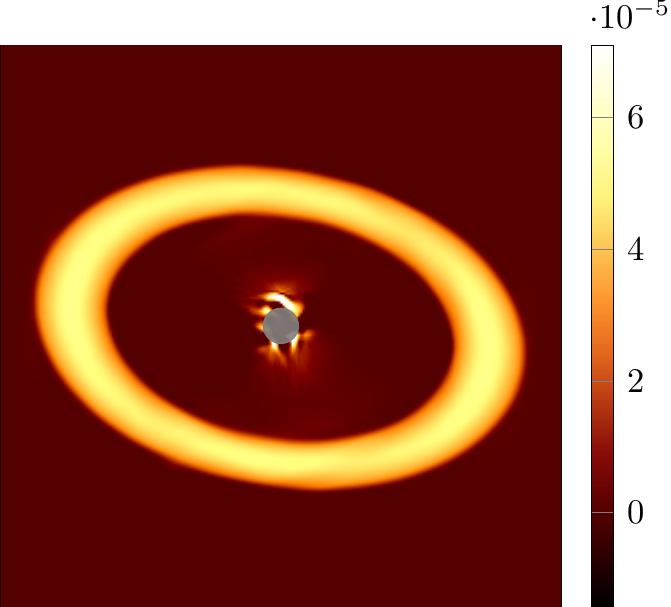}}\hspace{5mm}
    {\includegraphics[width=0.31\textwidth]{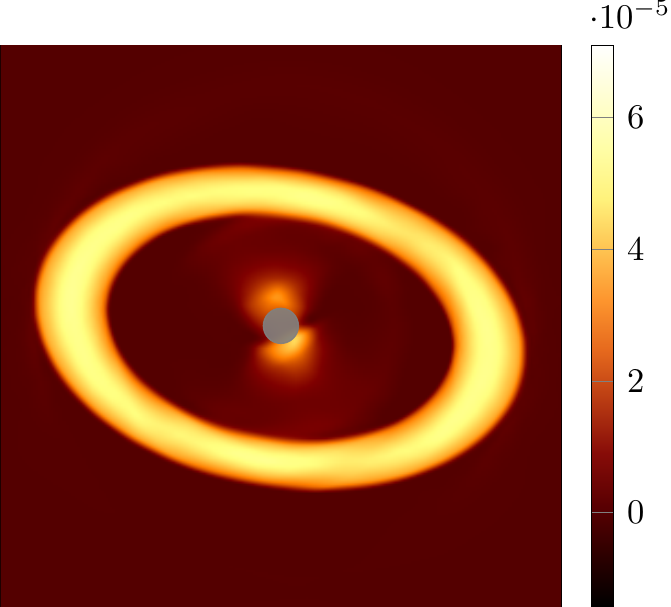}}\hspace{5mm}
    {\includegraphics[width=0.31\textwidth]{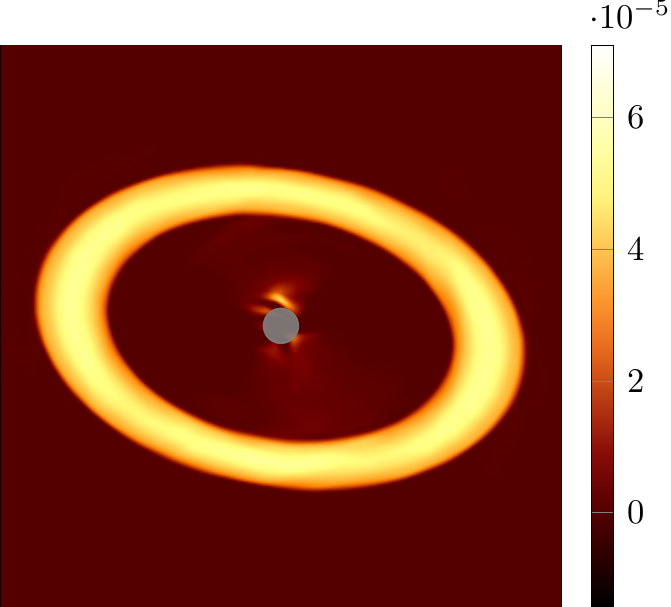}}
  \caption{Comparison of the estimated images, $\hat{\bs x}_\inddisk$, obtained with different fidelity terms. From left to right: $\ell_2$-norm, $\ell_1$-norm, and Huber-loss.
  The difference is mostly noticeable for the residual speckles in the center. Unsurprisingly from a MLE perspective, the $\ell_2$-norm frame is plagued with excessive speckles while the $\ell_1$-norm is capable of eliminating them. Although the superiority of the Huber-loss compared to the $\ell_1$-norm is difficult to assess visually, the LSE scores of Table~\ref{table:huber_l2} show that using the Huber-loss yields better results than either an $\ell_2$-norm or $\ell_1$-norm. }
   \label{fig:comparison_huber_l2_proximal_unmixing}
\end{figure*}

To make the comparison, we injected a synthetic disk of inclination 60 degrees and of contrast $5.3 \times 10^{-5}$ and attempted to recover it with the Huber-loss, an $\ell_2$-norm ($\delta \rightarrow +\infty$) and an $\ell_1$-norm ($\delta \ll 1$). In Fig.~\ref{fig:comparison_huber_l2_proximal_unmixing} we show the corresponding estimated images $\hat{\bs x}_\inddisk$. The comparison is quantified with two different scores: (1) using the relative error between the processed frame and the ground-truth on the full image, \ie $\| \hat{\bs x}_\inddisk - \xmodel_\inddisk \|_2 / \| \xmodel_\inddisk \|_2$, and (2) using the relative error restricted to the locations of the disk to emphasize the quality of the reconstruction of the disk, regardless of the residual speckles that are mostly located at small angular separation. Given the operator 
\[
(\cl D_{\bs v}(\bs u))_i = \cl D_{\bs v}(\hat u_i) = 
\begin{cases} 
 \hat u_i  &\text{ if }  v_i \geq 0 \\
 0 &\text{ otherwise,} 
 \end{cases}
\]
the second score is defined as $\| \cl D_{\xmodel_\inddisk}(\hat{\bs x}_\inddisk) - \xmodel_\inddisk \|_2 / \| \xmodel_\inddisk \|_2$. 
These two scores, corresponding to the images of Fig.~\ref{fig:comparison_huber_l2_proximal_unmixing}, are shown in Table~\ref{table:huber_l2}. 
The Huber-loss improves the quality of $\hat{\bs x}_\inddisk$ compared to both the $\ell_2$ and $\ell_1$-norms, regardless of the score considered.

\begin{table}
\centering
	\begin{tabular}{l c c c } 
\hline	
Score & $\ell_2$-norm & $\ell_1$-norm & Huber-loss  \\
\hline
(1): $\| \hat{\bs x}_\inddisk - \xmodel_\inddisk \|_2 / \| \xmodel_\inddisk \|_2$ & 0.247  & 0.235 &  0.180 \\
(2): $\| \cl D_{\xmodel_\inddisk}(\hat{\bs x}_\inddisk) - \xmodel_\inddisk \|_2 / \| \xmodel_\inddisk \|_2$ & 0.149  & 0.141 & 0.132  \\
        \hline
	\end{tabular}
\caption{LSE between the ground-truth (injected disk) and our reconstruction (estimated disk), when using three different norms to run the algorithm (the $\ell_2$-norm, the $\ell_1$-norm and the huber-loss). The score (1) is for the whole image (top row), whereas score (2) is restricted to the disk signal only (bottom row).}
	\label{table:huber_l2}
\end{table}

\subsection{Performance analysis on synthetic data and comparison to PCA-SFS}
\label{sec:comparison-with-pca}

To assess the performance and limits of our algorithm, we applied it on VLT/SPHERE-IRDIS data in which we injected synthetic disk signals beforehand and we compared the results to those obtained from the classical PCA-SFS algorithm. We injected four different synthetic disk signals presented in Fig.~\ref{fig:results_syntdatasets} (left column): (a) one disk with inclination 50$^\circ$ and peak contrast of $5.3\times 10^{-5}$ (top), (b) one disk with inclination 70$^\circ$ and peak contrast of $5.3\times 10^{-5}$ (middle-top), (c) one faint disk with inclination 50$^\circ$ and peak contrast of $3.5\times 10^{-6}$ (middle-bottom), and (d) a face-on disk slightly de-centered (of $60~\mathrm{mas}$ in x- and y-direction), with peak contrast of $5.3\times 10^{-5}$. The injected disks (ground truth) shown in Fig.~\ref{fig:results_syntdatasets} (left) are before the convolution with the off-axis instrumental PSF.

The results obtained with the classical PCA-SFS algorithm are shown in Fig.~\ref{fig:results_syntdatasets} middle panel, and the results from MAYO are shown in the right panel. To ease the qualitative comparison between PCA-SFS and MAYO, Fig.~\ref{fig:results_syntdatasets} (right) displays the profiles of each image along one horizontal line (white dotted line in the images of Fig.~\ref{fig:results_syntdatasets}). In each of these four cases, the two least-square scores are displayed on Table~\ref{table:comparison_PCA_MAYO}.

\newcommand{\scaleExAll}{0.6}
\begin{figure*}
  \centering
  \scalebox{0.93}{
\begin{tabular}{llll}
   \vspace{0.3cm}
    {\includegraphics[scale=\scaleExAll]{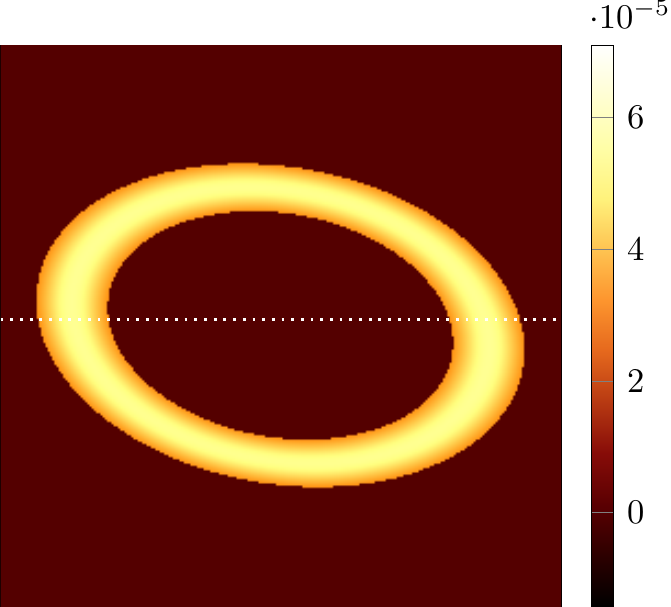}}&
    {\includegraphics[scale=\scaleExAll]{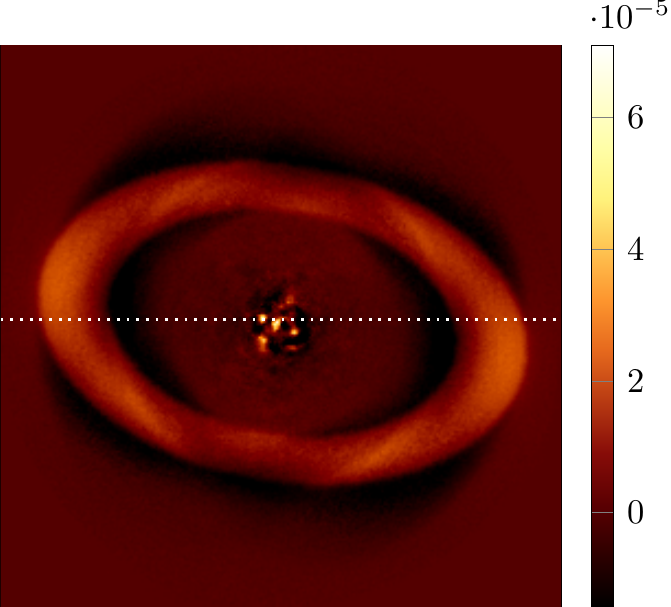}}&
    {\includegraphics[scale=\scaleExAll]{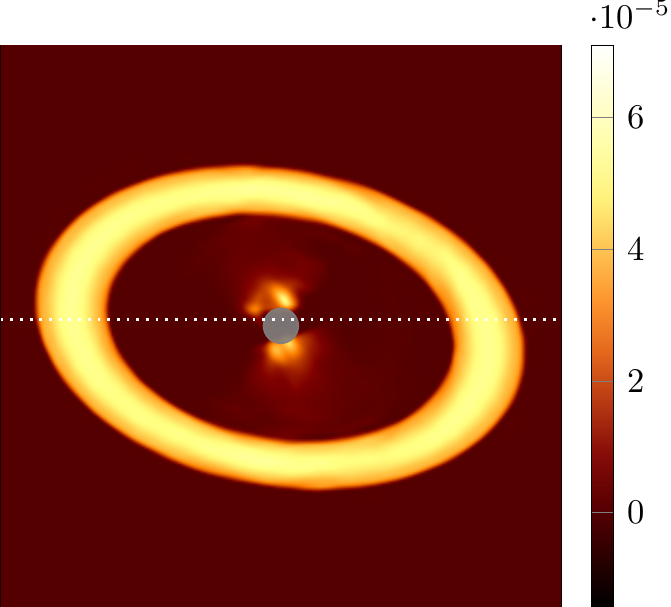}}&
   \hspace{-0.5cm}
    {\includegraphics[scale=\scaleExAll]{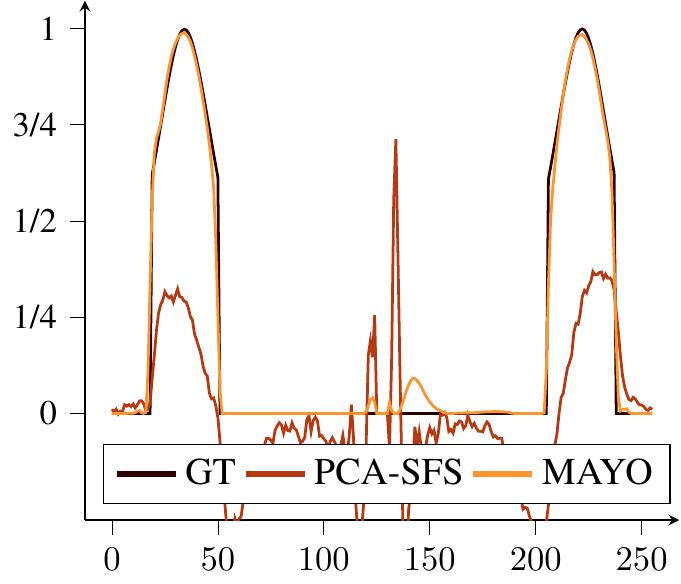}}\\
   \vspace{0.3cm}
    {\includegraphics[scale=\scaleExAll]{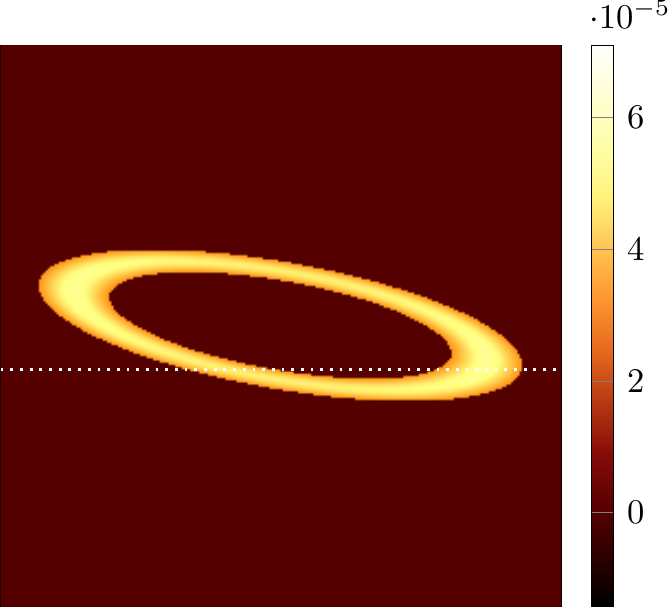}}&
    {\includegraphics[scale=\scaleExAll]{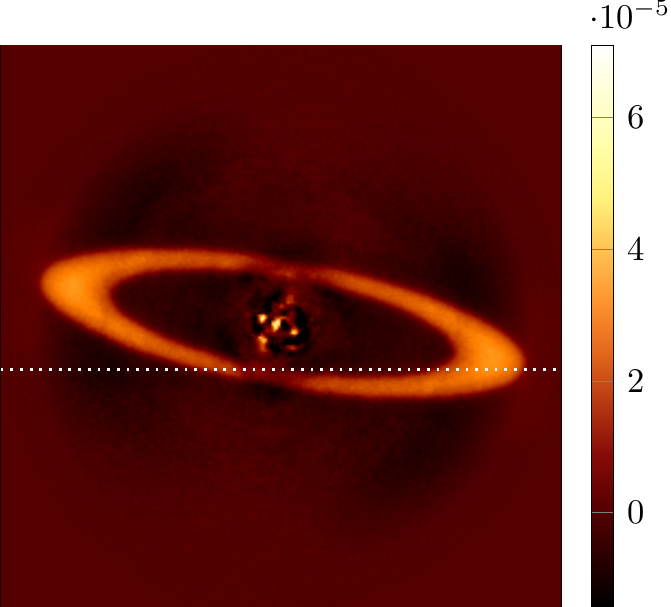}}&
    {\includegraphics[scale=\scaleExAll]{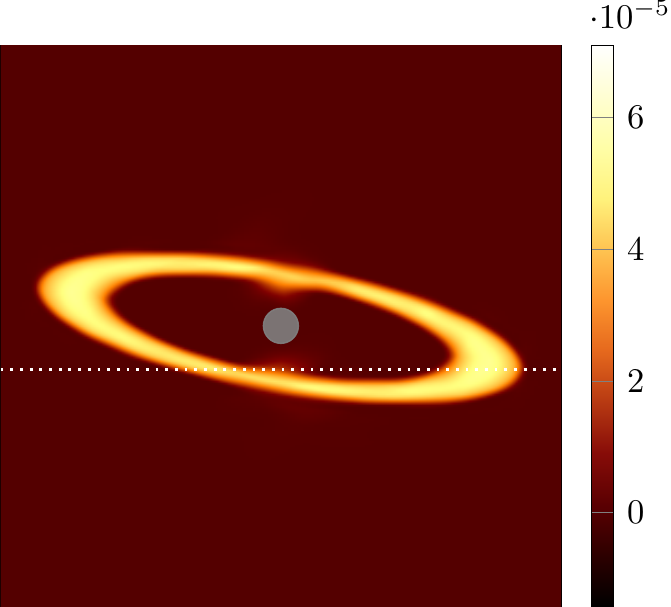}}&
   \hspace{-0.5cm}
    {\includegraphics[scale=\scaleExAll]{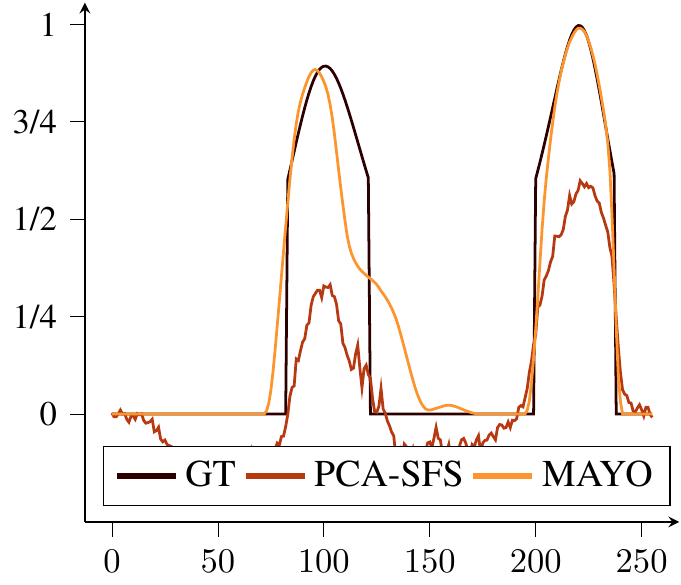}}\\
\vspace{0.3cm}
    {\includegraphics[scale=\scaleExAll]{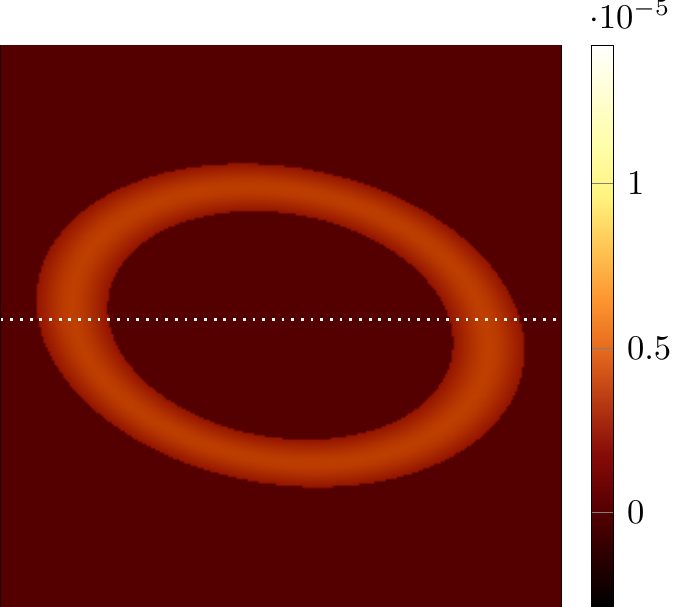}}&
    {\includegraphics[scale=\scaleExAll]{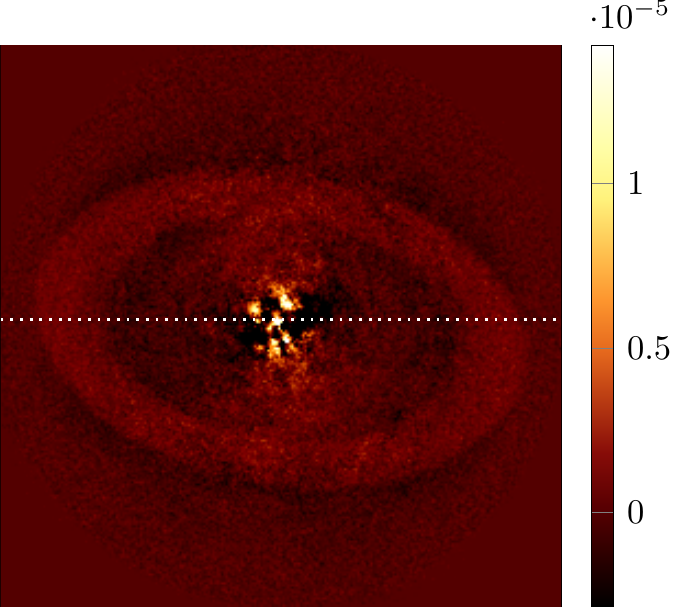}}&
    {\includegraphics[scale=\scaleExAll]{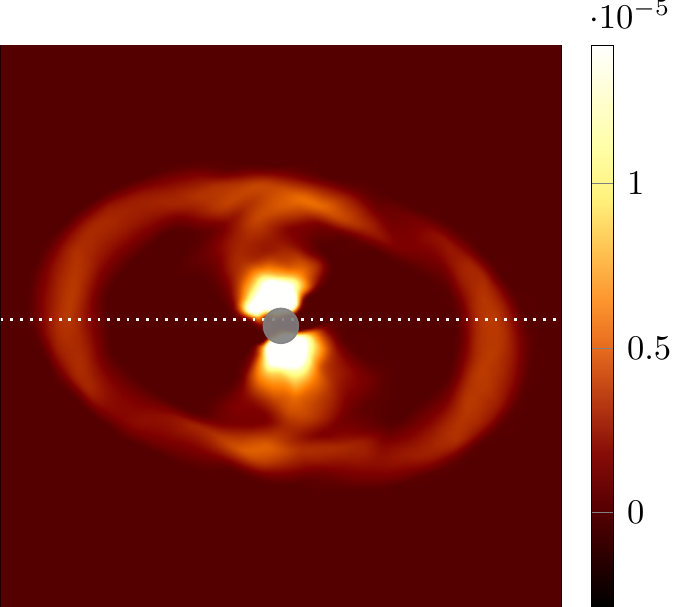}}&
   \hspace{-0.5cm}
    {\includegraphics[scale=\scaleExAll]{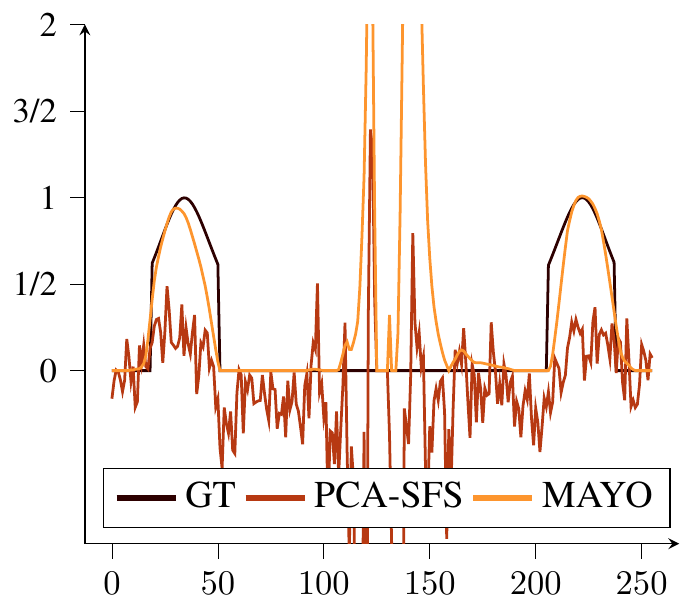}}\\
\vspace{0.3cm}
    {\includegraphics[scale=\scaleExAll]{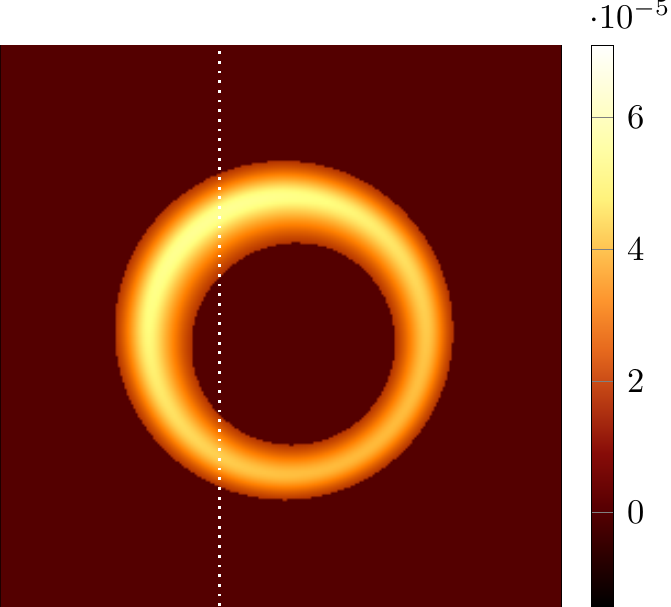}}&
    {\includegraphics[scale=\scaleExAll]{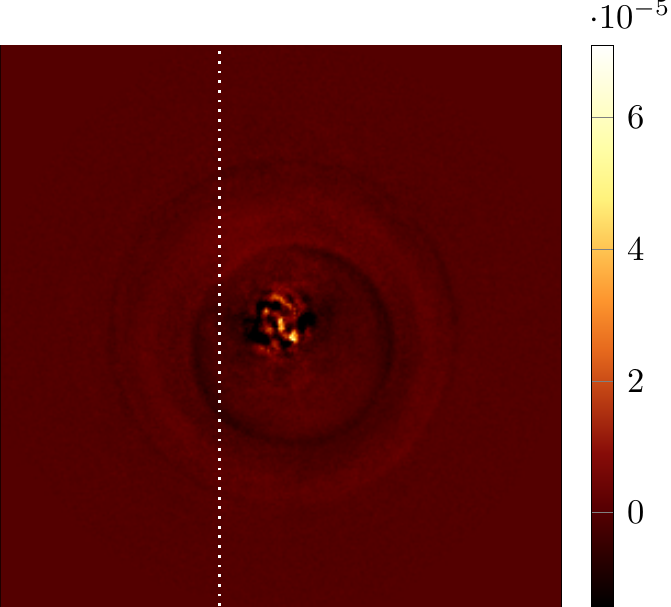}}&
    {\includegraphics[scale=\scaleExAll]{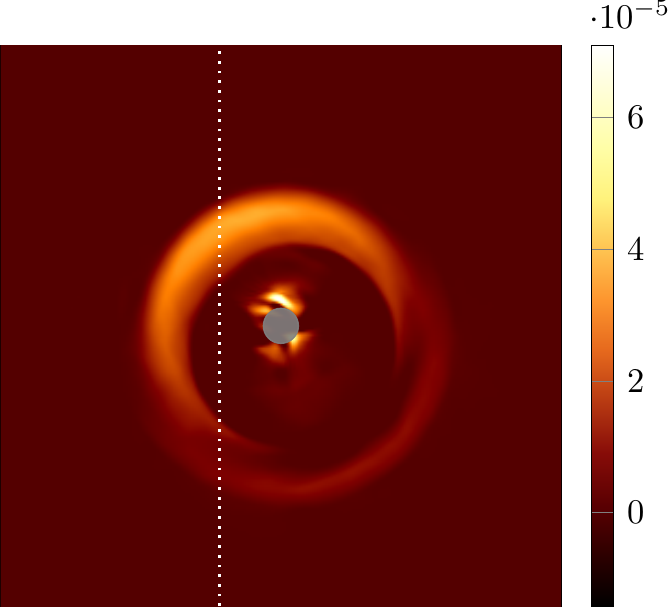}}&
   \hspace{-0.5cm}
    {\includegraphics[scale=\scaleExAll]{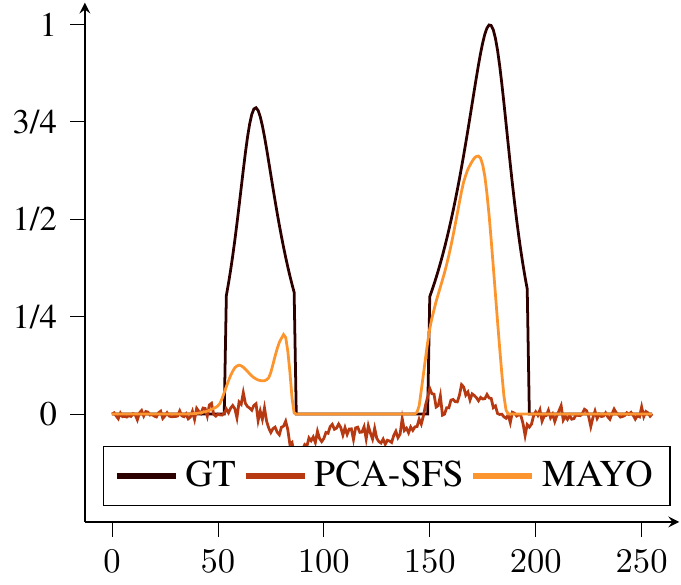}}
 \end{tabular}}
  \caption{Results of our proposed source separation algorithm and comparison with PCA-SFS in the case of four disks injected in an empty VLT/SPHERE-IRDIS data cube. 
  From top to bottom: (a) inclination of $50^{\circ}$ and contrast of $5.3\times 10^{-5}$, (b) inclination of $75^{\circ}$ and contrast of $5.3\times 10^{-5}$, (c) inclination of $50^{\circ}$ and contrast of $3.5\times 10^{-6}$ and, (d) inclination of $0^{\circ}$ and contrast of $5.3\times 10^{-5}$. 
  Left: injected disk signal (ground-truth, GT), before the convolution by the instrumental PSF. 
  Middle-left: PCA-SFS processed frame. 
  Middle-right: MAYO processed frame. 
  Right: Intensity profile along the white dotted line for the ground-truth (dark solid line), the PCA-SFS estimation (red solid line) and the MAYO estimation (orange solid line).
  In the images, the central gray area indicates the mask $\maskop$ used to run the MAYO algorithm, as defined in Eq.~(\ref{eq:mask-def}), which corresponds to the inner working angle of the Lyot coronagraph used during the observations.
  }
 \label{fig:results_syntdatasets}
\end{figure*}

\begin{table*}
\centering
	\begin{tabular}{|l|c|c|c|c|} 
		\hline
 & \multicolumn{2}{|c|}{Score (1): $\| \hat{\bs x}_\inddisk - \xmodel_\inddisk \|_2 / \| \xmodel_\inddisk \|_2$} &  \multicolumn{2}{|c|}{Score (2): $\| \cl D_{\xmodel_\inddisk}(\hat{\bs x}_\inddisk) - \xmodel_\inddisk \|_2 / \| \xmodel_\inddisk \|_2$} \\	
	\hline
	Injected disk	 & PCA-SFS  & MAYO  & PCA-SFS & MAYO  \\
		\hline
(a) $i=50^\circ$, $C=5.3\times10^{-5}$  & 0.85 & 0.24 & 0.81 & 0.12 \\
(b) $i=75^\circ$, $C=5.3\times10^{-5}$  & 0.68 & 0.28 & 0.61 & 0.21 \\
(c) $i=50^\circ$, $C=3.5\times10^{-6}$  & 1.18 & 2.72 & 0.88 & 0.36 \\
(d) $i=0^\circ$,   $\,\,C=5.3\times10^{-5}$  & 1.00 & 0.76 & 1.00 & 0.75 \\ 
        \hline
	\end{tabular}
\caption{Least-square error between the ground truth (injected disk) and the reconstruction (estimated disk) from PCA-SFS and MAYO, for the four test-cases described in Sec.~\ref{sec:comparison-with-pca} and shown in Fig.~\ref{fig:results_syntdatasets}. The score (1) is for the whole image (left columns), whereas score (2) is restricted to the disk signal only (right columns).}
	\label{table:comparison_PCA_MAYO}
\end{table*}

For the bright disk with a $50^\circ$ inclination (Fig.~\ref{fig:results_syntdatasets}, top row), the disk signal is fully recovered with our method, while the PCA-SFS processing distorts the disk intensity profile with a flux about three times lower than the ground truth. 
For the bright disk with a $75^\circ$ inclination (Fig.~\ref{fig:results_syntdatasets}, middle-top row),
our method preserves the intensity profile, while PCA-SFS again strongly distorts the intensity profile. Moreover, about a third of the flux is lost (to highlight this, the 1D plot in Fig.~\ref{fig:results_syntdatasets}, right, focuses on the location where PCA-SFS shows the strongest distortion). 

In the case of the fainter disk (Fig.~\ref{fig:results_syntdatasets}, middle-bottom row), our method again fully preserves the signal but the intensity profile is slightly distorted in some regions, while the PCA-SFS signal is close to zero with a noisy intensity profile. Note that the LSE on the full image (score 1) for this disk is very large for MAYO. This is because, near the center, $\| \hat{\bs x}_\inddisk - \xmodel_\inddisk \|_2 = \| \hat{\bs x}_\inddisk \|_2$ is as large as for the brighter disk. The quantity $\| \xmodel_\inddisk \|_2$ is, however, much smaller since this disk is 15 times fainter. It follows that the $\| \hat{\bs x}_\inddisk - \xmodel_\inddisk \|_2 /\| \xmodel_\inddisk \|_2$ is large. The second score, that only consider the disk shows that MAYO clearly outperforms PCA-SFS in terms of disk restoration; at the cost of excessive speckle residuals near the center.

At last, in the challenging case of the face-on disk (Fig.~\ref{fig:results_syntdatasets}, bottom row), our approach detects the disk signal with up to three-quarter of its intensity but the overall profile intensity is highly distorted. However, as expected, PCA-SFS fails to detect the disk signal. In this specific case, a forward modeling approach, as done in the literature when using classical techniques such as PCA-SFS \citep{milli2012impact}, is still required to evaluate the distortion induced by MAYO.

Thanks to these four typical test cases, we show that MAYO always globally or partially recovers the disk signal, and preserves the intensity profile within the disk, if we exclude the case of face-on disks suffering from a lack of angular diversity in ADI sequences. In addition, by construction, MAYO does not show any negative values in the reconstructed image compared to PCA-SFS.

\subsection{Selectivity of the source separation method}
\label{sec:numerical_MCA}
Our algorithm consists in separating point source signals from extended signals by using a MCA approach, that is to say by estimating the sparse point source planetary signals in the pixel domain and the extended disk signals in the shearlet domain. A challenging separation occurs when the planetary signal is blended with the disk signal, such as for the companion PDS~70~c. This object is fully embedded in the protoplanetary disk surrounding PDS~70 \citep{Mesa2019pds70} and a different technique than ADI was required to confirm it \citep{Haffert2019pds70}.

\begin{figure*}
  \centering
  {\includegraphics[width=0.31\textwidth]{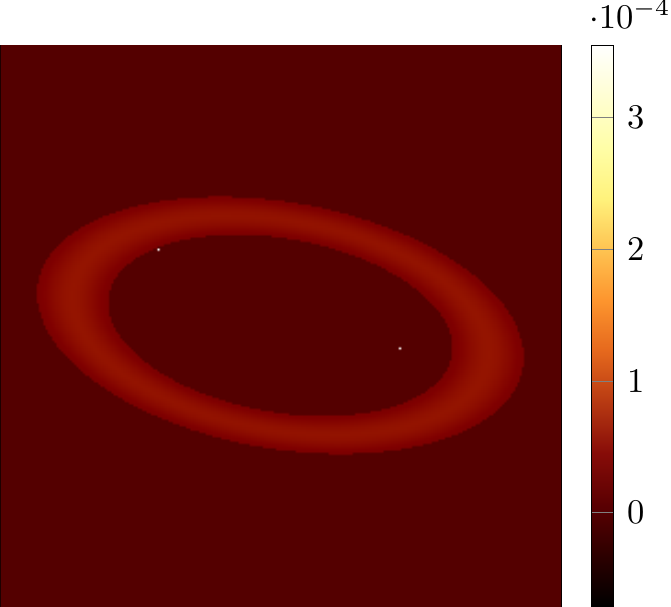}}
  {\includegraphics[width=0.31\textwidth]{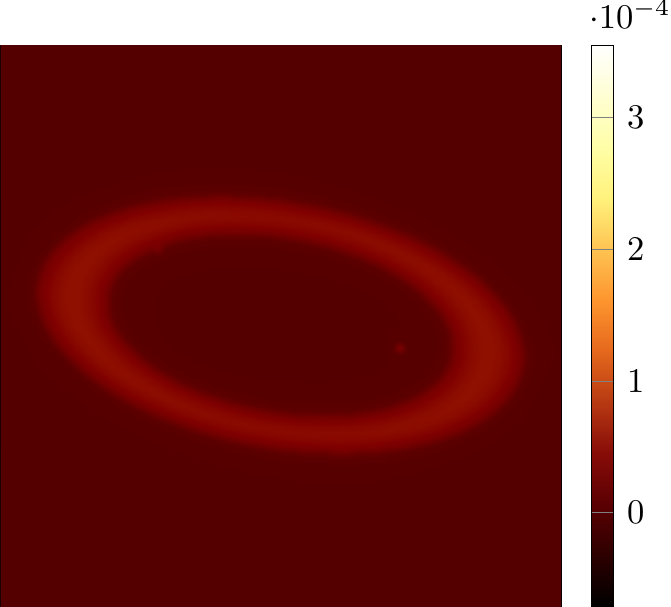}}
  {\includegraphics[width=0.31\textwidth]{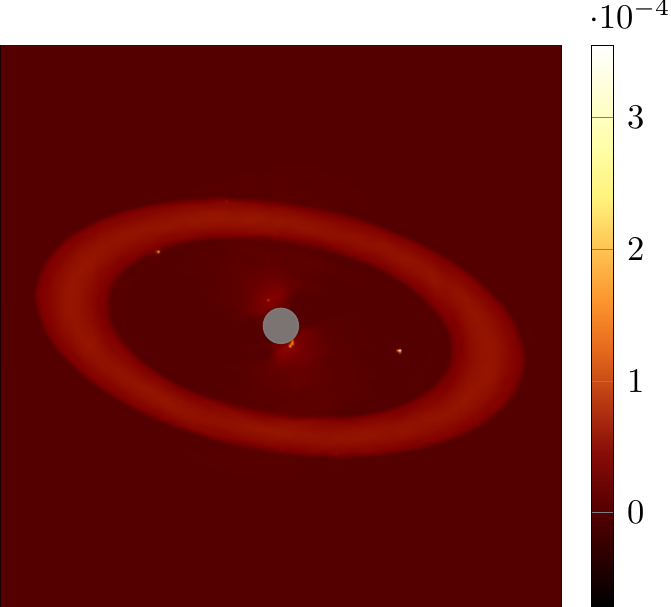}}
\caption{Capability of MAYO to extract a point source within an extended source. 
Left: disk and exoplanet signals injected in the data, before convolution with the instrumental PSF. 
Center: disk and exoplanet signals injected in the data, after convolution with the instrumental PSF. 
Right: result from the MAYO algorithm, showing that the disk signal is preserved and the planetary signal recovered.}
\label{fig:disk_planet}
\end{figure*}

In this section, we apply MAYO to a synthetic dataset that contains a disk and two planets, as shown in Fig.~\ref{fig:disk_planet} (left). The disk is of inclination $60^\circ$ and contrast $5.3\times 10^{-5}$ and the first planet (P1) has a contrast of $7\times 10^{-5}$. The second exoplanet (P2), of contrast $3.5\times 10^{-5}$, is injected very close to the disk, to check whether MAYO is capable of separating a planetary signal embedded in a disk signal. Note that the contrast is given after the convolution with the instrumental PSF. As seen in Fig.~\ref{fig:disk_planet} (middle), the convolution has the effect to drastically decrease the intensity of exoplanetary signals compared to the disk signal. The estimated image by MAYO, $\hat{\bs x} = \hat{\bs x}_\inddisk + \hat{\bs x}_\indpla$, is shown in  Fig.~\ref{fig:disk_planet} (right). 

As shown by the previous experiment in Section~\ref{sec:comparison-with-pca}, the disk is fully recovered and its intensity profile is preserved. In addition, the two injected companions are clearly detected in Figure~\ref{fig:disk_planet} (right). For P1, 90$\%$ of its intensity is recovered in $\hat{\bs x}_\indpla$. For P2, because the injected circumstellar signal is blurred, the signal of the exoplanet next to the disk is mixed with the signal of the disk, see Fig.~\ref{fig:disk_planet} (middle). There is thus a morphological ambiguity between the disk and the exoplanet. However, even in this challenging situation, MAYO is still capable of recovering 60$\%$ of P2 in $\hat{\bs x}_\indpla$.

\subsection{Residual artifacts at small angular separation}

The images estimated by MAYO can display high-intensity residuals at close angular separation. This can be seen most notably in Figs.~\ref{fig:results_syntdatasets} and~\ref{fig:results_MAYO_real_data}.
We can understand this limitation by observing that at small angular separation, ADI datasets exhibit a strong ambiguity between temporally static and rotating signals: speckles can be efficiently represented either as part of $\bs L$ or $\bs x$. We thus reach the conflicting objectives of simultaneously minimizing self-absorption of circumstellar signals while rejecting the presence of speckles in $\bs x$ near the image center.

One way to mitigate this effect could be to extend our approach to RDI in addition to ADI. As RDI does not rely on the rotation of the circumstellar signal, it does not suffer from an increased ambiguity near the center of the field of view. Another solution would consist in statistically quantifying the level of certainty of any features displayed in the MAYO outputs~\cite{pereyra2017maximum, repetti2019scalable}. This would allow us to either discriminate between residual speckles and actual circumstellar signals, or to conclude that a detected feature is not reliable. Such an extension is, however, beyond the scope of the present paper and we left it as a future work. 

Currently, the intensity near the center of the frame of either $\hat{\bs x}_\inddisk$ or $\hat{\bs x}_\indpla$ must thus be considered with cautious. The presence of any feature reconstructed by MAYO must be confirmed through multiple ADI acquisition campaigns.

\subsection{Results on emblematic disks}
\label{sec:illustr-results-real}

\begin{figure*}
  \centering
  {\includegraphics[width=0.31\textwidth]{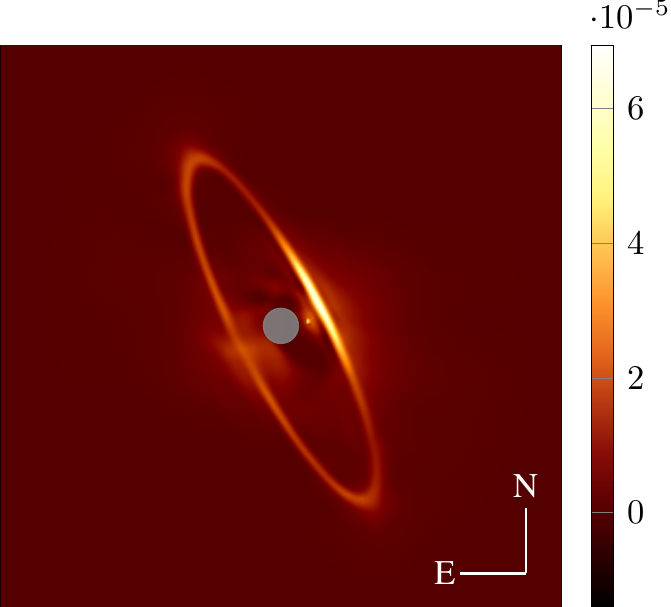}}
  {\includegraphics[width=0.31\textwidth]{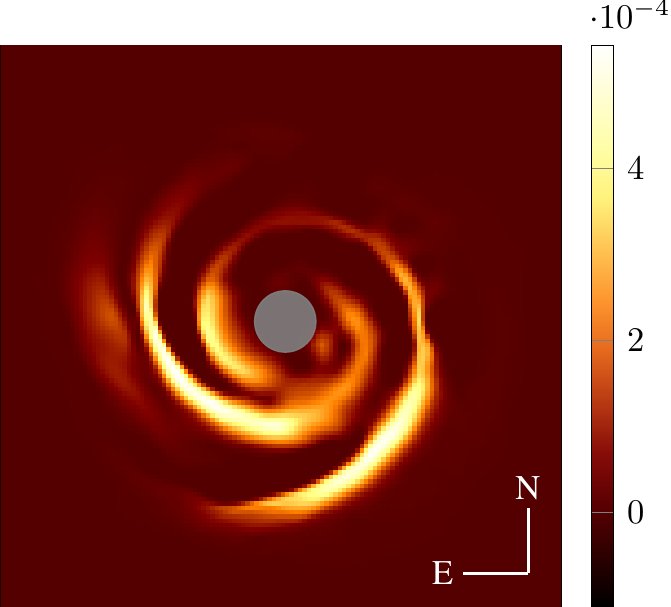}}
  {\includegraphics[width=0.31\textwidth]{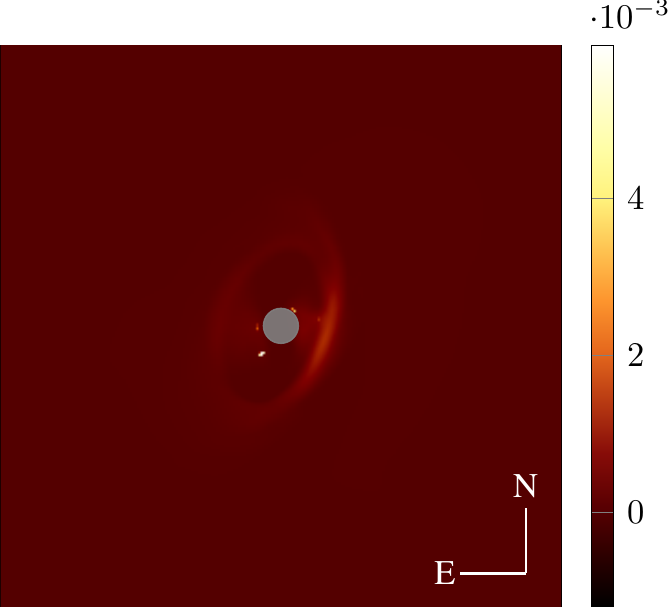}}\\
  {\includegraphics[width=0.31\textwidth]{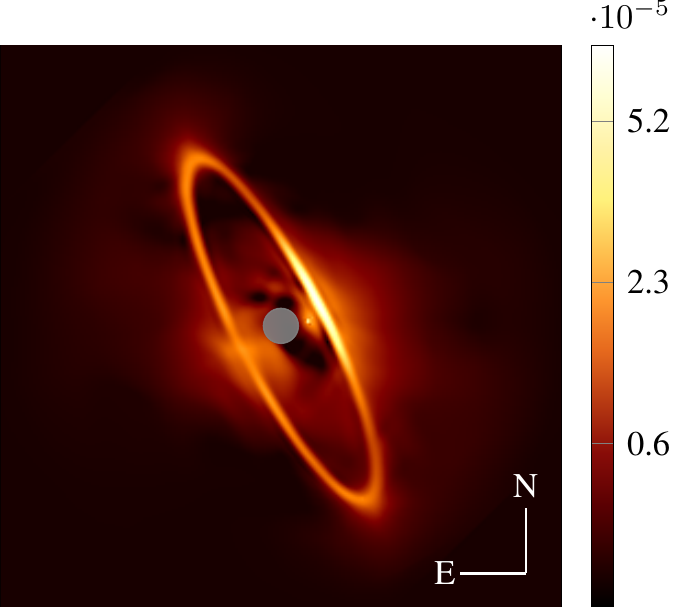}}
  {\includegraphics[width=0.31\textwidth]{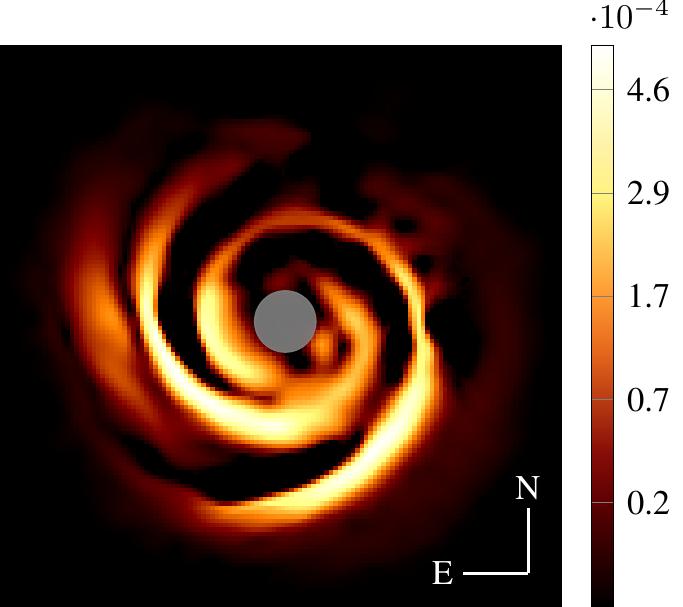}}
  {\includegraphics[width=0.31\textwidth]{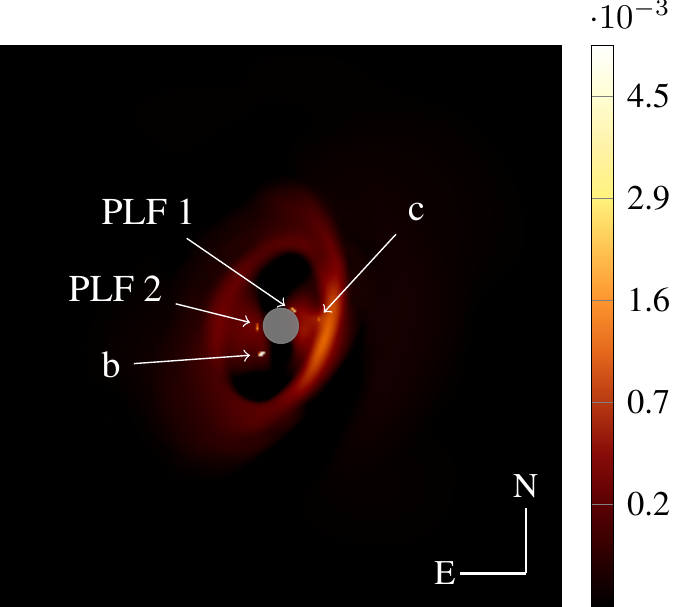}}
\caption{Results the MAYO algorithm applied to the SPHERE data of HR~4796A (left), SAO~206462 (center) and PDS~70 (right), in linear scale (top) and square root scale (bottom), for which we do not use the common zero in the colorbar for clarity. For the three cases, the central gray area indicates the mask $\maskop$ used to run the MAYO algorithm, as defined in Eq.~(\ref{eq:mask-def}), which corresponds to the inner working angle of the Lyot coronagraph under use during the observations.}
\label{fig:results_MAYO_real_data}
\end{figure*}

We here apply MAYO to SPHERE data containing real disk and planet signals to further highlight the capability of MAYO to disentangle point sources from the extended source. We process the three emblematic targets already mentioned in Sec.~\ref{sec:dataset}. These were observed with the high-contrast instrument SPHERE (see App.~\ref{app:data_set} for details about the three datasets) and correspond to the bright debris disk surrounding HR~4796A \citep{milli2019hr4796pol}, the spiral transition disk SAO~206462 \citep{maire2017testing}, and the protoplanetary disk PDS~70 \citep{keppler2018discovery}. The signals reconstructed by MAYO ($\hat{\bs x}_\inddisk+ \hat{\bs x}_\indpla$) are shown in Fig.~\ref{fig:results_MAYO_real_data}. 

For the case of the debris disk surrounding HR~4796A \citep{Schneiderhr4796}, the image reconstructed by MAYO (Fig.~\ref{fig:results_MAYO_real_data}, left) directly highlights three major points: \emph{(i)} the forward scattering side of the disk (the brightest part of the ring) is clearly the northern side as demonstrated in \cite{Milli2017hr4796}, \emph{(ii)} the ansae (extremities of the disk) show the typical scattering structure expected by the radiative transfer models and demonstrated in \citep{Lagrange2012hr4796}, and \emph{(iii)}, the brightest part shows a slight warp in intensity, also expected from the radiative transfer models \citep{Milli2017hr4796}. Those results are straightforward using our reconstructed image with MAYO whereas, when using classical post-processing techniques, they arose from analyses back and forth with scattered light models of the disk. The analysis of the surface brightness distribution of the disk and the extraction of the scattering phase function from the MAYO reconstructed images will be published in a future paper (Milli et al., in prep.).

Regarding the transition disk SAO~206462 \citep{grady2009sao206462}, the image reconstructed with MAYO (Fig.~\ref{fig:results_MAYO_real_data}, middle) shows the two spiral arms with unprecedented details. This high spatial resolution of the disk allows: \emph{(i)} constraining the scattering properties of the grains constituting the spirals, \emph{(ii)} constraining the potential presence of a planet launching the spirals through hydro-dynamical simulations, and \emph{(iii)}, constraining the exact origin of the spirals through follow-up of their trace with time, as their motion will be different if caused by gravitational instability, or an embedded companion on either a circular or eccentric orbit~\citep[\eg][]{Ren2020mwc758,Calcino2020}.

At last, for the case of the multiplanetary system around PDS~70, the reconstructed image with MAYO (Fig.~\ref{fig:results_MAYO_real_data}, right) does show very clearly the companion PDS~70~b unveiled from the same SPHERE images with classical methods \citep{Muller2018pds70}. We note that, despite the deconvolution procedure integrated in MAYO, the signal of this companion is not located on a single pixel but on a clump of pixels. It is not clear whether this is due to the circumplanetary disk surrounding PDS~70~b, unveiled with high-resolution spectroscopy~\citep{Christiaens2019cpd} and possibly related to the sub-mm continuum signal detected with ALMA~\citep{Isella2019cpd}, or due to a smearing effect to which MAYO is potentially sensitive. 
In addition, the planet PDS~70~c, identified thanks to high-spectral resolution technique \citep{Haffert2019pds70}, which is embedded in the disk signal and therefore difficult to recover with classical methods \citep{Mesa2019pds70}, is clearly recovered by MAYO. The \emph{point-like feature}~(PLF 1, in Fig.~\ref{fig:results_MAYO_real_data}) recently pointed out in \citep{Mesa2019pds70} from a thorough exploration of all of the SPHERE near-infrared images is also detected, as well as another structure very close to the star~(PLF 2, in Fig.~\ref{fig:results_MAYO_real_data}), that was also seen, at a lower signal-to-noise ratio in \cite{Mesa2019pds70} using different speckle subtraction algorithms and inverse problems approaches. Note that these features could also be the results of optical interactions between the inner-disk~\citep{hashimoto2012pds70,keppler2018discovery} and the coronagraph. Moreover, as see in Fig.~\ref{fig:disk_planet}, MAYO is prone to false positives close to the star, and these features should be investigated in other epochs.  

As for the circumstellar disk~\citep{Riaud2006pds70}, the large gap within the circumstellar disk is an obvious result of planet-disk interactions, but the MAYO image highlights more structures within the outer disk, pointing towards the highly dynamic nature of the system: \emph{(i)} a possible spiral arm structure located in the north and \emph{(ii)} a flux asymmetry along the outer disk. To check whether the spiral is real and if the asymmetry is due to a phase function effect or a shadow cast by the inner disk, the MAYO images will be further investigated, along with additional data (Desgrange et al., in prep.).

These three examples highlight the scientific interest of this new post-processing method for disk imaging: we do not need iterations between a radiative transfer model and the final speckle subtracted image to access the intensity distribution of the disk, and the convolution process is directly taken into account in MAYO. Therefore, images provided by MAYO are directly ready for astrophysical interpretation. 
In addition, for these three examples, the shape of the reconstructed disk intensity image obtained by MAYO is highly similar to the shape of the disk polarimetry image obtained from polarimetric differential imaging \citep[PDI,][]{kuhn2001pdi} that is free of stellar residuals, as shown in \cite{milli2019hr4796pol,perrin2015hr4796pol}, \cite{stolker2016sao206,takami2014polar}, and \cite{keppler2018discovery,takami2014polar} for HR~4796A, SAO~206462 and PDS~70 respectively.

\section{Conclusion}
\label{sec:conclusion}

In this paper, we introduced MAYO, an innovative image processing pipeline designed to restore both exoplanets and circumstellar disk signals from high-contrast images taken in pupil-tracking mode (ADI dataset) with ground-based instruments. We grounded our approach on a specific source separation task that leverages the morphological diversity between disks and exoplanets to distinguish them. MAYO also includes a deconvolution, which is, to the best of our knowledge, a first for the processing of such data. This allows MAYO to faithfully preserve the shape and the flux distribution of extended sources and accurately distinguish point source signals in the images. Moreover, the deconvolution allows distinguishing abrupt and smooth transitions in the disk matter distribution. 

As a secondary contribution, we introduce the mathematical model of the ADI acquisition process upon which our main algorithm is based. From this model, we derived the GreeDS algorithm. Although it is used as a preliminary step in our pipeline, GreeDS can also be used as a standalone algorithm.

To validate the capability of MAYO to reconstruct circumstellar disks, we applied it to real high-contrast images (from the VLT/SPHERE-IRDIS instrument), in which we injected synthetic disks signals in different configurations (in terms of contrast and inclination). For a contrast of $5.3\times10^{-5}$ and inclinations 50 or 75 degrees, MAYO restores a faithful image of the disks. MAYO is able to restore disks with contrast as high as $3.5\times10^{-6}$, although in this case, the image is plagued with residual speckles and there are some distortions in the recovered shape of the disk. Finally, we show that MAYO is also able to recover some signal from a face-on disk, which is the main geometrical restriction of current post-processing techniques. MAYO is therefore more sensitive to disk signals than any other method and using it could help increase the number of detected circumstellar disks.
To validate the effectiveness of MAYO in separating point sources from extended structures, we injected two exoplanets in addition to a synthetic disk. From this experiment, we show that MAYO succeeds in extracting the exoplanetary signals, even when the exoplanet lies in close vicinity from the disk.

At last, we applied MAYO to data containing real disk and exoplanet signals, highlighting the gain of MAYO compared to the current state-of-the-art post-processing methods based on ADI: thanks to MAYO, there is no need for forward modeling, which consists in creating a disk model, reducing it with ADI and iterate over the model parameters until the residuals are minimized at the disk location. 
In particular, for debris disk MAYO makes it possible for the first time to extract the scattering phase function without requiring any parametrization, as in \cite{Olofsson2020}, applied so far only in polarimetry due to the inherent biases of ADI.
In conclusion, MAYO stands as a complementary tool, along with polarimetry images and modeling tools, to understand the nature of the micro-sized grain population in circumstellar disks.

\textbf{Future work:} The MAYO pipeline is built upon a general framework providing a versatile scheme to include or disregard important physical properties of ADI datasets. In future works, a more accurate model of the noise distribution could be integrated if available, as well as a more realistic model of the telescope transfer function (including, \eg diffractive effects, coronagraph perturbation). This last case would impose us a more complex, non-convolutive linear operator $\cl T$ in the forward imaging model, accounting for the spatial dependence of the telescope response, especially at small angular separation. As noted, the deconvolution makes MAYO potentially subject to smearing effects, due to long integration times. If this is confirmed in a subsequent study, MAYO can be extended to include a circular motion in the convolution, thus canceling the smearing effects. 

By design, the MAYO algorithm can be easily extended to reference star differential imaging; in this context, we would have to replace the set $\hat{\cl P}^{(r)}$, which accounts for the low-rank quasi-static noise in ADI, with a set constructed from a reference star catalog \citep{xuan2018rdi,ruane2019rdi,bohn2020rdi}. With this extension, we expect to achieve significant gains with face-on disks. 

The choice of the shearlets transform is also arbitrary and future work could consider curvelets~\citep{candes2000curvelets}, contourlets~\citep{do2002contourlets} or even starlets~\citep{starck2011starlet}. Another promising option is to learn a suitable, low-complexity representation of the disk and planet signals, either from a dictionary learning strategy~\citep[see, \eg][]{mairal2009supervised,mairal2009online,tosic2011dictionary}, or using deep generative (adversarial) networks \citep{goodfellow2016nips,ulyanov2018deep,yip2019pushing}. These two types of learned representations reach appealing performance in many imaging inverse problems (\eg in computer vision, biomedical and astronomical imaging). For ADI or RDI processing, the associated learning task would, however, require a library of physically sound disk-like images.

\section*{Acknowledgements}

We would like to thank the anonymous reviewers for their insights and comments that greatly improved the presentation of the manuscript.

We thank our colleagues, including O.~Absil, V.~Christiaens, C.~Desgrange, M.~Keppler, A.-L.~Maire, and J.~Milli for the many fruitful discussions that greatly contributed to the research.

We would like to express our gratitude to J.~Milli, A.-L.~Maire and the SPHERE-DC team for kindly providing the reduced SPHERE datasets used in this study. We also thank J.~Milli, V.~Christiaens and M.~Keppler for valuable inputs about the astrophysical interpretation of our reconstructed images. 

\section*{Data availability}

The data underlying this article were provided by J.~Milli, A.-L.~Maire and the SPHERE-DC team. Data will be shared on request to the corresponding author with permission of J.~Milli, A.-L.~Maire and the SPHERE-DC team.

\newpage
\bibliographystyle{mnras}
\interlinepenalty=1000
\bibliography{biblio}

\begin{thebibliography}{}
\makeatletter
\relax
\def\mn@urlcharsother{\let\do\@makeother \do\$\do\&\do\#\do\^\do\_\do\%\do\~}
\def\mn@doi{\begingroup\mn@urlcharsother \@ifnextchar [ {\mn@doi@}
  {\mn@doi@[]}}
\def\mn@doi@[#1]#2{\def\@tempa{#1}\ifx\@tempa\@empty \href
  {http://dx.doi.org/#2} {doi:#2}\else \href {http://dx.doi.org/#2} {#1}\fi
  \endgroup}
\def\mn@eprint#1#2{\mn@eprint@#1:#2::\@nil}
\def\mn@eprint@arXiv#1{\href {http://arxiv.org/abs/#1} {{\tt arXiv:#1}}}
\def\mn@eprint@dblp#1{\href {http://dblp.uni-trier.de/rec/bibtex/#1.xml}
  {dblp:#1}}
\def\mn@eprint@#1:#2:#3:#4\@nil{\def\@tempa {#1}\def\@tempb {#2}\def\@tempc
  {#3}\ifx \@tempc \@empty \let \@tempc \@tempb \let \@tempb \@tempa \fi \ifx
  \@tempb \@empty \def\@tempb {arXiv}\fi \@ifundefined
  {mn@eprint@\@tempb}{\@tempb:\@tempc}{\expandafter \expandafter \csname
  mn@eprint@\@tempb\endcsname \expandafter{\@tempc}}}

\bibitem[\protect\citeauthoryear{Abdi \& Williams}{Abdi \&
  Williams}{2010}]{abdi2010principal}
Abdi H.,  Williams L.~J.,  2010, Wiley interdisciplinary reviews: computational
  statistics, 2, 433

\bibitem[\protect\citeauthoryear{Almeida \& Figueiredo}{Almeida \&
  Figueiredo}{2013}]{almeida2013parameter}
Almeida M.~S.,  Figueiredo M.~A.,  2013, IEEE Transactions on Image Processing,
  22, 2751

\bibitem[\protect\citeauthoryear{Amara \& Quanz}{Amara \&
  Quanz}{2012}]{amara2012pynpoint}
Amara A.,  Quanz S.~P.,  2012, Monthly Notices of the Royal Astronomical
  Society, 427, 948

\bibitem[\protect\citeauthoryear{{Andrews}}{{Andrews}}{2020}]{Andrews2020rev}
{Andrews} S.~M.,  2020, \mn@doi [\araa] {10.1146/annurev-astro-031220-010302},
  \href {https://ui.adsabs.harvard.edu/abs/2020ARA&A..58..483A} {58, 483}

\bibitem[\protect\citeauthoryear{{Augereau}, {Lagrange}, {Mouillet},
  {Papaloizou}  \& {Grorod}}{{Augereau} et~al.}{1999}]{augereau1999grater}
{Augereau} J.~C.,  {Lagrange} A.~M.,  {Mouillet} D.,  {Papaloizou} J.~C.~B.,
  {Grorod} P.~A.,  1999, \aap, \href
  {https://ui.adsabs.harvard.edu/abs/1999A&A...348..557A} {348, 557}

\bibitem[\protect\citeauthoryear{Bae \& Zhu}{Bae \& Zhu}{2018}]{bae2018planet}
Bae J.,  Zhu Z.,  2018, The Astrophysical Journal, 859, 119

\bibitem[\protect\citeauthoryear{Baydin, Pearlmutter, Radul  \& Siskind}{Baydin
  et~al.}{2017}]{baydin2017automatic}
Baydin A.~G.,  Pearlmutter B.~A.,  Radul A.~A.,   Siskind J.~M.,  2017, The
  Journal of Machine Learning Research, 18, 5595

\bibitem[\protect\citeauthoryear{{Beuzit} et~al.,}{{Beuzit}
  et~al.}{2019}]{Beuzit2019}
{Beuzit} J.~L.,  et~al., 2019, \mn@doi [\aap] {10.1051/0004-6361/201935251},
  \href {https://ui.adsabs.harvard.edu/abs/2019A&A...631A.155B} {631, A155}

\bibitem[\protect\citeauthoryear{Blumensath \& Davies}{Blumensath \&
  Davies}{2008}]{blumensath2008iterative}
Blumensath T.,  Davies M.~E.,  2008, Journal of Fourier analysis and
  Applications, 14, 629

\bibitem[\protect\citeauthoryear{Bobin, Starck, Moudden  \& Fadili}{Bobin
  et~al.}{2008}]{bobin20085}
Bobin J.,  Starck J.-L.,  Moudden Y.,   Fadili M.~J.,  2008, Advances in
  Imaging and Electron Physics, 152, 221

\bibitem[\protect\citeauthoryear{{Bohn} et~al.,}{{Bohn}
  et~al.}{2019}]{bohn2020rdi}
{Bohn} A.~J.,  et~al., 2019, \mn@doi [\aap] {10.1051/0004-6361/201834523},
  \href {https://ui.adsabs.harvard.edu/abs/2019A&A...624A..87B} {624, A87}

\bibitem[\protect\citeauthoryear{Boos \& Leonard}{Boos \&
  Leonard}{2013}]{dennisessential}
Boos D.~B.,  Leonard L.~S.,  2013, Essential Statistical Inference: Theory and
  Methods.
Springer Texts in Statistics

\bibitem[\protect\citeauthoryear{{Calcino}, {Christiaens}, {Price}, {Pinte},
  {Davis}, {van der Marel}  \& {Cuello}}{{Calcino} et~al.}{2020}]{Calcino2020}
{Calcino} J.,  {Christiaens} V.,  {Price} D.~J.,  {Pinte} C.,  {Davis} T.~M.,
  {van der Marel} N.,   {Cuello} N.,  2020, \mn@doi [\mnras]
  {10.1093/mnras/staa2468}, \href
  {https://ui.adsabs.harvard.edu/abs/2020MNRAS.498..639C} {498, 639}

\bibitem[\protect\citeauthoryear{Candes \& Donoho}{Candes \&
  Donoho}{2000}]{candes2000curvelets}
Candes E.~J.,  Donoho D.~L.,  2000, Technical report, Curvelets: A surprisingly
  effective nonadaptive representation for objects with edges.
Stanford Univ Ca Dept of Statistics

\bibitem[\protect\citeauthoryear{Cand{\`e}s, Li, Ma  \& Wright}{Cand{\`e}s
  et~al.}{2011}]{candes2011robust}
Cand{\`e}s E.~J.,  Li X.,  Ma Y.,   Wright J.,  2011, Journal of the ACM
  (JACM), 58, 1

\bibitem[\protect\citeauthoryear{{Cantalloube}, {Dohlen}, {Milli}, {Brandner}
  \& {Vigan}}{{Cantalloube} et~al.}{2019}]{cantalloube2019msgr}
{Cantalloube} F.,  {Dohlen} K.,  {Milli} J.,  {Brandner} W.,   {Vigan} A.,
  2019, \mn@doi [The Messenger] {10.18727/0722-6691/5138}, \href
  {https://ui.adsabs.harvard.edu/abs/2019Msngr.176...25C} {176, 25}

\bibitem[\protect\citeauthoryear{Carbillet et~al.,}{Carbillet
  et~al.}{2011}]{carbillet2011aplc}
Carbillet M.,  et~al., 2011, Experimental Astronomy, 30, 39

\bibitem[\protect\citeauthoryear{{Chauvin} et~al.,}{{Chauvin}
  et~al.}{2017}]{chauvin2017shine}
{Chauvin} G.,  et~al., 2017, in {Reyl{\'e}} C.,  {Di Matteo} P.,  {Herpin} F.,
  {Lagadec} E.,  {Lan{\c c}on} A.,  {Meliani} Z.,   {Royer} F.,  eds,
  SF2A-2017: Proceedings of the Annual meeting of the French Society of
  Astronomy and Astrophysics. pp 331--335

\bibitem[\protect\citeauthoryear{{Christiaens}, {Cantalloube}, {Casassus},
  {Price}, {Absil}, {Pinte}, {Girard}  \& {Montesinos}}{{Christiaens}
  et~al.}{2019}]{Christiaens2019cpd}
{Christiaens} V.,  {Cantalloube} F.,  {Casassus} S.,  {Price} D.~J.,  {Absil}
  O.,  {Pinte} C.,  {Girard} J.,   {Montesinos} M.,  2019, \mn@doi [\apjl]
  {10.3847/2041-8213/ab212b}, \href
  {https://ui.adsabs.harvard.edu/abs/2019ApJ...877L..33C} {877, L33}

\bibitem[\protect\citeauthoryear{Dahlqvist, Cantalloube  \& Absil}{Dahlqvist
  et~al.}{2020}]{dahlqvist2020regime}
Dahlqvist C.-H.,  Cantalloube F.,   Absil O.,  2020, Astronomy \& Astrophysics,
  633, A95

\bibitem[\protect\citeauthoryear{Defrere et~al.,}{Defrere
  et~al.}{2014}]{defrere2014lbthci}
Defrere D.,  et~al., 2014, in Adaptive Optics Systems IV. p. 91483X

\bibitem[\protect\citeauthoryear{Do \& Vetterli}{Do \&
  Vetterli}{2002}]{do2002contourlets}
Do M.~N.,  Vetterli M.,  2002, in Proceedings. International Conference on
  Image Processing. pp~I--I

\bibitem[\protect\citeauthoryear{Donoho}{Donoho}{2001}]{donoho2001sparse}
Donoho D.~L.,  2001, Constructive Approximation, 17, 353

\bibitem[\protect\citeauthoryear{Donoho \& Kutyniok}{Donoho \&
  Kutyniok}{2009}]{donoho2009geometric}
Donoho D.,  Kutyniok G.,  2009, in SAMPTA09, Marseille, France.

\bibitem[\protect\citeauthoryear{Eckart \& Young}{Eckart \&
  Young}{1936}]{eckartapproximation}
Eckart Y.,  Young G.,  1936, Psychometrika. v1, pp 211--218

\bibitem[\protect\citeauthoryear{Eftekhari, Yang  \& Wakin}{Eftekhari
  et~al.}{2018}]{eftekhari2018weighted}
Eftekhari A.,  Yang D.,   Wakin M.~B.,  2018, IEEE Transactions on Information
  Theory, 64, 4044

\bibitem[\protect\citeauthoryear{Esposito, Fitzgerald, Graham  \&
  Kalas}{Esposito et~al.}{2013}]{esposito2013fm}
Esposito T.~M.,  Fitzgerald M.~P.,  Graham J.~R.,   Kalas P.,  2013, The
  Astrophysical Journal, 780, 25

\bibitem[\protect\citeauthoryear{Fadili, Starck  \& Murtagh}{Fadili
  et~al.}{2009}]{fadili2009inpainting}
Fadili M.-J.,  Starck J.-L.,   Murtagh F.,  2009, The Computer Journal, 52, 64

\bibitem[\protect\citeauthoryear{Fitzgerald \& Graham}{Fitzgerald \&
  Graham}{2006}]{fitzgerald2006speckle}
Fitzgerald M.~P.,  Graham J.~R.,  2006, The Astrophysical Journal, 637, 541

\bibitem[\protect\citeauthoryear{{Gomez Gonzalez}, {Wertz}, {Christiaens},
  {Absil}  \& {Mawet}}{{Gomez Gonzalez} et~al.}{2016}]{2016ascl.soft03003G}
{Gomez Gonzalez} C.~A.,  {Wertz} O.,  {Christiaens} V.,  {Absil} O.,   {Mawet}
  D.,  2016, {VIP: Vortex Image Processing pipeline for high-contrast direct
  imaging of exoplanets}, Astrophysics Source Code Library (\mn@eprint {ascl}
  {1603.003})

\bibitem[\protect\citeauthoryear{Gonzalez et~al.,}{Gonzalez
  et~al.}{2017}]{gonzalez2017vip}
Gonzalez C. A.~G.,  et~al., 2017, The Astronomical Journal, 154, 7

\bibitem[\protect\citeauthoryear{Goodfellow}{Goodfellow}{2016}]{goodfellow2016nips}
Goodfellow I.,  2016, arXiv preprint arXiv:1701.00160

\bibitem[\protect\citeauthoryear{Grady et~al.,}{Grady
  et~al.}{2009}]{grady2009sao206462}
Grady C.,  et~al., 2009, The Astrophysical Journal, 699, 1822

\bibitem[\protect\citeauthoryear{{Haffert}, {Bohn}, {de Boer}, {Snellen},
  {Brinchmann}, {Girard}, {Keller}  \& {Bacon}}{{Haffert}
  et~al.}{2019}]{Haffert2019pds70}
{Haffert} S.~Y.,  {Bohn} A.~J.,  {de Boer} J.,  {Snellen} I.~A.~G.,
  {Brinchmann} J.,  {Girard} J.~H.,  {Keller} C.~U.,   {Bacon} R.,  2019,
  \mn@doi [Nature Astronomy] {10.1038/s41550-019-0780-5}, \href
  {https://ui.adsabs.harvard.edu/abs/2019NatAs...3..749H} {3, 749}

\bibitem[\protect\citeauthoryear{Hashimoto et~al.,}{Hashimoto
  et~al.}{2012}]{hashimoto2012pds70}
Hashimoto J.,  et~al., 2012, The Astrophysical Journal Letters, 758, L19

\bibitem[\protect\citeauthoryear{{Herscovici-Schiller}, {Mugnier}  \&
  {Sauvage}}{{Herscovici-Schiller} et~al.}{2017}]{Herscovici2017anal}
{Herscovici-Schiller} O.,  {Mugnier} L.~M.,   {Sauvage} J.-F.,  2017, \mn@doi
  [\mnras] {10.1093/mnrasl/slx009}, \href
  {https://ui.adsabs.harvard.edu/abs/2017MNRAS.467L.105H} {467, L105}

\bibitem[\protect\citeauthoryear{Huber}{Huber}{1981}]{huber1981robust}
Huber P.~J.,  1981, Robust statistics.
John Wiley \& Sons

\bibitem[\protect\citeauthoryear{Isella, Benisty, Teague, Bae, Keppler,
  Facchini  \& P{\'e}rez}{Isella et~al.}{2019}]{Isella2019cpd}
Isella A.,  Benisty M.,  Teague R.,  Bae J.,  Keppler M.,  Facchini S.,
  P{\'e}rez L.,  2019, The Astrophysical Journal Letters, 879, L25

\bibitem[\protect\citeauthoryear{Jacques, Duval, Chaux  \& Peyr{\'e}}{Jacques
  et~al.}{2011}]{jacques2011panorama}
Jacques L.,  Duval L.,  Chaux C.,   Peyr{\'e} G.,  2011, Signal Processing, 91,
  2699

\bibitem[\protect\citeauthoryear{{Jovanovic} et~al.,}{{Jovanovic}
  et~al.}{2015}]{Jovanovic2015}
{Jovanovic} N.,  et~al., 2015, \mn@doi [\pasp] {10.1086/682989}, \href
  {http://adsabs.harvard.edu/abs/2015PASP..127..890J} {127, 890}

\bibitem[\protect\citeauthoryear{Kenworthy, Hinz, Codona, Wilson, Skrutskie  \&
  Solheid}{Kenworthy et~al.}{2010}]{kenworthy2010lbthci}
Kenworthy M.~A.,  Hinz P.~M.,  Codona J.~L.,  Wilson J.~C.,  Skrutskie M.~F.,
  Solheid E.,  2010, in Optical and Infrared Interferometry II. p. 77342P

\bibitem[\protect\citeauthoryear{Keppler et~al.,}{Keppler
  et~al.}{2018}]{keppler2018discovery}
Keppler M.,  et~al., 2018, Astronomy \& Astrophysics, 617, A44

\bibitem[\protect\citeauthoryear{Kuhn, Potter  \& Parise}{Kuhn
  et~al.}{2001}]{kuhn2001pdi}
Kuhn J.,  Potter D.,   Parise B.,  2001, The Astrophysical Journal Letters,
  553, L189

\bibitem[\protect\citeauthoryear{Kutyniok \& Labate}{Kutyniok \&
  Labate}{2012}]{kutyniok2012shearlets}
Kutyniok G.,  Labate D.,  2012, Shearlets: Multiscale analysis for multivariate
  data.
Springer Science \& Business Media

\bibitem[\protect\citeauthoryear{Kutyniok, Lim  \& Reisenhofer}{Kutyniok
  et~al.}{2016}]{kutyniok2016shearlab}
Kutyniok G.,  Lim W.-Q.,   Reisenhofer R.,  2016, ACM Transactions on
  Mathematical Software (TOMS), 42, 1

\bibitem[\protect\citeauthoryear{{Lagrange} et~al.,}{{Lagrange}
  et~al.}{2012}]{Lagrange2012hr4796}
{Lagrange} A.~M.,  et~al., 2012, \mn@doi [\aap] {10.1051/0004-6361/201219187},
  \href {https://ui.adsabs.harvard.edu/abs/2012A&A...546A..38L} {546, A38}

\bibitem[\protect\citeauthoryear{Li, Banerjee, Pudritz, J{\o}rgensen, Shang,
  Krasnopolsky, Maury  \& Beuther}{Li et~al.}{2014}]{li2014protostars}
Li Z.,  Banerjee R.,  Pudritz R.,  J{\o}rgensen J.,  Shang H.,  Krasnopolsky
  R.,  Maury A.,   Beuther H.,  2014, Protostars and Planets VI

\bibitem[\protect\citeauthoryear{{Macintosh} et~al.,}{{Macintosh}
  et~al.}{2008}]{Macintosh2008}
{Macintosh} B.~A.,  et~al., 2008, in Adaptive Optics Systems. p. 701518,
  \mn@doi{10.1117/12.788083}

\bibitem[\protect\citeauthoryear{Mairal, Bach, Ponce  \& Sapiro}{Mairal
  et~al.}{2009a}]{mairal2009online}
Mairal J.,  Bach F.,  Ponce J.,   Sapiro G.,  2009a, in Proceedings of the 26th
  annual international conference on machine learning. pp 689--696

\bibitem[\protect\citeauthoryear{Mairal, Ponce, Sapiro, Zisserman  \&
  Bach}{Mairal et~al.}{2009b}]{mairal2009supervised}
Mairal J.,  Ponce J.,  Sapiro G.,  Zisserman A.,   Bach F.~R.,  2009b, in
  Advances in neural information processing systems. pp 1033--1040

\bibitem[\protect\citeauthoryear{Maire et~al.,}{Maire
  et~al.}{2017}]{maire2017testing}
Maire A.-L.,  et~al., 2017, Astronomy \& Astrophysics, 601, A134

\bibitem[\protect\citeauthoryear{Males et~al.,}{Males
  et~al.}{2018}]{males2018magao}
Males J.~R.,  et~al., 2018, in Adaptive Optics Systems VI. p. 1070309

\bibitem[\protect\citeauthoryear{Marois, Lafreniere, Doyon, Macintosh  \&
  Nadeau}{Marois et~al.}{2006}]{marois2006angular}
Marois C.,  Lafreniere D.,  Doyon R.,  Macintosh B.,   Nadeau D.,  2006, The
  Astrophysical Journal, 641, 556

\bibitem[\protect\citeauthoryear{Marois, Lafreniere, Macintosh  \&
  Doyon}{Marois et~al.}{2008}]{marois2008confidence}
Marois C.,  Lafreniere D.,  Macintosh B.,   Doyon R.,  2008, The Astrophysical
  Journal, 673, 647

\bibitem[\protect\citeauthoryear{Martinez, Dorrer, Carpentier, Kasper,
  Boccaletti, Dohlen  \& Yaitskova}{Martinez et~al.}{2009}]{martinez2009aplc}
Martinez P.,  Dorrer C.,  Carpentier E.~A.,  Kasper M.,  Boccaletti A.,  Dohlen
  K.,   Yaitskova N.,  2009, Astronomy \& Astrophysics, 495, 363

\bibitem[\protect\citeauthoryear{Mawet et~al.,}{Mawet
  et~al.}{2016}]{mawet2016kpic}
Mawet D.,  et~al., 2016, in Adaptive Optics Systems V. p. 99090D

\bibitem[\protect\citeauthoryear{{Mesa} et~al.,}{{Mesa}
  et~al.}{2019}]{Mesa2019pds70}
{Mesa} D.,  et~al., 2019, \mn@doi [\aap] {10.1051/0004-6361/201936764}, \href
  {https://ui.adsabs.harvard.edu/abs/2019A&A...632A..25M} {632, A25}

\bibitem[\protect\citeauthoryear{Milli, Mouillet, Lagrange, Boccaletti, Mawet,
  Chauvin  \& Bonnefoy}{Milli et~al.}{2012}]{milli2012impact}
Milli J.,  Mouillet D.,  Lagrange A.-M.,  Boccaletti A.,  Mawet D.,  Chauvin
  G.,   Bonnefoy M.,  2012, Astronomy \& Astrophysics, 545, A111

\bibitem[\protect\citeauthoryear{{Milli} et~al.,}{{Milli}
  et~al.}{2017b}]{Milli2017hr4796}
{Milli} J.,  et~al., 2017b, \mn@doi [\aap] {10.1051/0004-6361/201527838}, \href
  {https://ui.adsabs.harvard.edu/abs/2017A&A...599A.108M} {599, A108}

\bibitem[\protect\citeauthoryear{Milli et~al.,}{Milli
  et~al.}{2017a}]{milli2017near}
Milli J.,  et~al., 2017a, Astronomy \& Astrophysics, 599, A108

\bibitem[\protect\citeauthoryear{Milli et~al.,}{Milli
  et~al.}{2019}]{milli2019hr4796pol}
Milli J.,  et~al., 2019, Astronomy \& Astrophysics, 626, A54

\bibitem[\protect\citeauthoryear{{M{\"u}ller} et~al.,}{{M{\"u}ller}
  et~al.}{2018}]{Muller2018pds70}
{M{\"u}ller} A.,  et~al., 2018, \mn@doi [\aap] {10.1051/0004-6361/201833584},
  \href {https://ui.adsabs.harvard.edu/abs/2018A&A...617L...2M} {617, L2}

\bibitem[\protect\citeauthoryear{Natarajan}{Natarajan}{1995}]{natarajan1995sparse}
Natarajan B.~K.,  1995, SIAM journal on computing, 24, 227

\bibitem[\protect\citeauthoryear{Oliphant}{Oliphant}{2006}]{2020numpy}
Oliphant T.,  2006, {NumPy}: A guide to {NumPy}, USA: Trelgol Publishing, \url
  {http://www.numpy.org/}

\bibitem[\protect\citeauthoryear{{Olofsson}, {Milli}, {Bayo}, {Henning}  \&
  {Engler}}{{Olofsson} et~al.}{2020}]{Olofsson2020}
{Olofsson} J.,  {Milli} J.,  {Bayo} A.,  {Henning} T.,   {Engler} N.,  2020,
  \mn@doi [\aap] {10.1051/0004-6361/202038237}, \href
  {https://ui.adsabs.harvard.edu/abs/2020A&A...640A..12O} {640, A12}

\bibitem[\protect\citeauthoryear{Otazo, Candes  \& Sodickson}{Otazo
  et~al.}{2015}]{otazo2015low}
Otazo R.,  Candes E.,   Sodickson D.~K.,  2015, Magnetic resonance in medicine,
  73, 1125

\bibitem[\protect\citeauthoryear{Pairet, Cantalloube  \& Jacques}{Pairet
  et~al.}{2018}]{pairet2018reference}
Pairet B.,  Cantalloube F.,   Jacques L.,  2018, arXiv preprint
  arXiv:1812.01333

\bibitem[\protect\citeauthoryear{Pairet, Cantalloube, Gomez~Gonzalez, Absil  \&
  Jacques}{Pairet et~al.}{2019}]{pairet2019stim}
Pairet B.,  Cantalloube F.,  Gomez~Gonzalez C.~A.,  Absil O.,   Jacques L.,
  2019, Monthly Notices of the Royal Astronomical Society, 487, 2262

\bibitem[\protect\citeauthoryear{Paszke et~al.,}{Paszke
  et~al.}{2017}]{paszke2017automatic}
Paszke A.,  et~al., 2017, in NIPS 2017 2017.

\bibitem[\protect\citeauthoryear{Pereyra}{Pereyra}{2017}]{pereyra2017maximum}
Pereyra M.,  2017, SIAM Journal on Imaging Sciences, 10, 285

\bibitem[\protect\citeauthoryear{Perrin et~al.,}{Perrin
  et~al.}{2015}]{perrin2015hr4796pol}
Perrin M.~D.,  et~al., 2015, The Astrophysical Journal, 799, 182

\bibitem[\protect\citeauthoryear{Pueyo et~al.,}{Pueyo
  et~al.}{2012}]{pueyo2012dloci}
Pueyo L.,  et~al., 2012, The Astrophysical Journal Supplement Series, 199, 6

\bibitem[\protect\citeauthoryear{Ren, Pueyo, Zhu, Debes  \& Duch{\^e}ne}{Ren
  et~al.}{2018}]{ren2018nmf}
Ren B.,  Pueyo L.,  Zhu G.~B.,  Debes J.,   Duch{\^e}ne G.,  2018, The
  Astrophysical Journal, 852, 104

\bibitem[\protect\citeauthoryear{Ren, Pueyo, Chen, Choquet, Debes, Duch{\^e}ne,
  M{\'e}nard  \& Perrin}{Ren et~al.}{2020a}]{ren2020nmfmask}
Ren B.,  Pueyo L.,  Chen C.,  Choquet {\'E}.,  Debes J.~H.,  Duch{\^e}ne G.,
  M{\'e}nard F.,   Perrin M.~D.,  2020a, arXiv preprint arXiv:2001.00563

\bibitem[\protect\citeauthoryear{{Ren} et~al.,}{{Ren}
  et~al.}{2020b}]{Ren2020mwc758}
{Ren} B.,  et~al., 2020b, arXiv e-prints, \href
  {https://ui.adsabs.harvard.edu/abs/2020arXiv200704980R} {p. arXiv:2007.04980}

\bibitem[\protect\citeauthoryear{Repetti, Pereyra  \& Wiaux}{Repetti
  et~al.}{2019}]{repetti2019scalable}
Repetti A.,  Pereyra M.,   Wiaux Y.,  2019, SIAM Journal on Imaging Sciences,
  12, 87

\bibitem[\protect\citeauthoryear{{Riaud}, {Mawet}, {Absil}, {Boccaletti},
  {Baudoz}, {Herwats}  \& {Surdej}}{{Riaud} et~al.}{2006}]{Riaud2006pds70}
{Riaud} P.,  {Mawet} D.,  {Absil} O.,  {Boccaletti} A.,  {Baudoz} P.,
  {Herwats} E.,   {Surdej} J.,  2006, \mn@doi [\aap]
  {10.1051/0004-6361:20065232}, \href
  {https://ui.adsabs.harvard.edu/abs/2006A&A...458..317R} {458, 317}

\bibitem[\protect\citeauthoryear{Riba, Mishkin, Ponsa, Rublee  \& Bradski}{Riba
  et~al.}{2020}]{riba2020kornia}
Riba E.,  Mishkin D.,  Ponsa D.,  Rublee E.,   Bradski G.,  2020, in
  Proceedings of the IEEE/CVF Winter Conference on Applications of Computer
  Vision. pp 3674--3683

\bibitem[\protect\citeauthoryear{Ruane et~al.,}{Ruane
  et~al.}{2019}]{ruane2019rdi}
Ruane G.,  et~al., 2019, The Astronomical Journal, 157, 118

\bibitem[\protect\citeauthoryear{Schneider et~al.,}{Schneider
  et~al.}{1999}]{Schneiderhr4796}
Schneider G.,  et~al., 1999, \mn@doi [The Astrophysical Journal]
  {10.1086/311921}, 513, L127

\bibitem[\protect\citeauthoryear{Skrutskie et~al.,}{Skrutskie
  et~al.}{2010}]{skrutskie2010lmircam}
Skrutskie M.,  et~al., 2010, in Ground-based and Airborne Instrumentation for
  Astronomy III. p. 77353H

\bibitem[\protect\citeauthoryear{{Soummer}}{{Soummer}}{2005}]{soummer2005aplc}
{Soummer} R.,  2005, \mn@doi [\apj] {10.1086/427923}, \href
  {https://ui.adsabs.harvard.edu/#abs/2005ApJ...618L.161S} {618, L161}

\bibitem[\protect\citeauthoryear{{Soummer}, {Pueyo}, {Sivaramakrishnan}  \&
  {Vand erbei}}{{Soummer} et~al.}{2007a}]{Soummer2007mft}
{Soummer} R.,  {Pueyo} L.,  {Sivaramakrishnan} A.,   {Vand erbei} R.~J.,
  2007a, \mn@doi [Optics Express] {10.1364/OE.15.015935}, \href
  {https://ui.adsabs.harvard.edu/abs/2007OExpr..1515935S} {15, 15935}

\bibitem[\protect\citeauthoryear{Soummer, Ferrari, Aime  \& Jolissaint}{Soummer
  et~al.}{2007b}]{soummer2007speckle}
Soummer R.,  Ferrari A.,  Aime C.,   Jolissaint L.,  2007b, The Astrophysical
  Journal, 669, 642

\bibitem[\protect\citeauthoryear{Soummer, Pueyo  \& Larkin}{Soummer
  et~al.}{2012}]{soummer2012detection}
Soummer R.,  Pueyo L.,   Larkin J.,  2012, The Astrophysical Journal Letters,
  755, L28

\bibitem[\protect\citeauthoryear{Starck, Pantin  \& Murtagh}{Starck
  et~al.}{2002}]{starck2002deconvolution}
Starck J.-L.,  Pantin E.,   Murtagh F.,  2002, Publications of the Astronomical
  Society of the Pacific, 114, 1051

\bibitem[\protect\citeauthoryear{Starck, Moudden, Bobin, Elad  \&
  Donoho}{Starck et~al.}{2005}]{starck2005morphological}
Starck J.-L.,  Moudden Y.,  Bobin J.,  Elad M.,   Donoho D.,  2005, in Wavelets
  XI. p. 59140Q

\bibitem[\protect\citeauthoryear{Starck, Murtagh  \& Bertero}{Starck
  et~al.}{2011}]{starck2011starlet}
Starck J.,  Murtagh F.,   Bertero M.,  2011, The starlet transform in
  astronomical data processing: Application to source detection and image
  deconvolution

\bibitem[\protect\citeauthoryear{Stolker et~al.,}{Stolker
  et~al.}{2016}]{stolker2016sao206}
Stolker T.,  et~al., 2016, Astronomy \& Astrophysics, 595, A113

\bibitem[\protect\citeauthoryear{Takami et~al.,}{Takami
  et~al.}{2014}]{takami2014polar}
Takami M.,  et~al., 2014, The Astrophysical Journal, 795, 71

\bibitem[\protect\citeauthoryear{Tosic \& Frossard}{Tosic \&
  Frossard}{2011}]{tosic2011dictionary}
Tosic I.,  Frossard P.,  2011, IEEE Signal Processing Magazine, 28, 27

\bibitem[\protect\citeauthoryear{Ulyanov, Vedaldi  \& Lempitsky}{Ulyanov
  et~al.}{2018}]{ulyanov2018deep}
Ulyanov D.,  Vedaldi A.,   Lempitsky V.,  2018, in Proceedings of the IEEE
  Conference on Computer Vision and Pattern Recognition. pp 9446--9454

\bibitem[\protect\citeauthoryear{Vershynin}{Vershynin}{2010}]{vershynin2010introduction}
Vershynin R.,  2010, arXiv preprint arXiv:1011.3027

\bibitem[\protect\citeauthoryear{Vigan et~al.,}{Vigan et~al.}{2014}]{Vigan2014}
Vigan A.,  et~al., 2014, SPHERE/IRDIS: final performance assessment of the
  dual-band imaging and long slit spectroscopy modes,
  \mn@doi{10.1117/12.2055374}, \url {http://dx.doi.org/10.1117/12.2055374}

\bibitem[\protect\citeauthoryear{{Virtanen} et~al.,}{{Virtanen}
  et~al.}{2020}]{2020SciPy-NMeth}
{Virtanen} P.,  et~al., 2020, \mn@doi [Nature Methods]
  {https://doi.org/10.1038/s41592-019-0686-2}, \href {https://rdcu.be/b08Wh} {}

\bibitem[\protect\citeauthoryear{Xuan et~al.,}{Xuan et~al.}{2018}]{xuan2018rdi}
Xuan W.~J.,  et~al., 2018, The Astronomical Journal, 156, 156

\bibitem[\protect\citeauthoryear{Yan}{Yan}{2018}]{yan2018new}
Yan M.,  2018, Journal of Scientific Computing, 76, 1698

\bibitem[\protect\citeauthoryear{Yip et~al.,}{Yip
  et~al.}{2019}]{yip2019pushing}
Yip K.~H.,  et~al., 2019, in Joint European Conference on Machine Learning and
  Knowledge Discovery in Databases. pp 322--338

\bibitem[\protect\citeauthoryear{Zhang, Bergin, Schwarz, Krijt  \&
  Ciesla}{Zhang et~al.}{2019}]{zhang2019systematic}
Zhang K.,  Bergin E.~A.,  Schwarz K.,  Krijt S.,   Ciesla F.,  2019, The
  Astrophysical Journal, 883, 98

\bibitem[\protect\citeauthoryear{Zhou \& Tao}{Zhou \&
  Tao}{2011}]{zhou2011godec}
Zhou T.,  Tao D.,  2011, in Proceedings of the 28th International Conference on
  Machine Learning, ICML 2011.

\makeatother
\end{thebibliography}

\newpage
\appendix

\section{Presentation of the dataset used in this paper}
\label{app:data_set}
To demonstrate the retrieval capabilities of  MAYO, we applied it to three representative targets containing a disk (HR~4796, SAO~206462 and PDS~70) and one empty data set in which we injected synthetic disks and planets (empty dataset). These targets were observed with the VLT/SPHERE-IRDIS instrument \citep{Beuzit2019,Vigan2014}, using the apodized Lyot coronagraph \citep[APLC][]{soummer2005aplc,martinez2009aplc,carbillet2011aplc}. The properties of the data can be found in Tab.~\ref{table:list_data_set}. 

\begin{table*}
\centering
\caption{Description of the three VLT/SPHERE-IRDIS coronagraphic datasets used in this paper to test our MAYO approach. The total number of images constituting the data cube is noted $N_{\text{images}}$. The average seeing conditions and turbulence coherence time during the observation are noted $seeing$ and $\tau_0$ respectively.}
	\label{table:list_data_set}
	\begin{tabular}{l c c c c c c c c} 
		\hline
		Name & Observation date & Filter & $N_{\text{images}}$  & $N_{DIT} \times DIT$ [s] & Total field rotation [deg] & $seeing$ ["] & $\tau_0$ [ms] \\
		\hline
		HR~4796A    & 2015-02-02 & H2 ($1.59\;\mu$m)  & 110& $8\times 32$ & 48.6 & 0.60 & 11 \\ 
        SAO~206462 	& 2015-05-15 & K1 ($2.11\;\mu$m)  & 63 & $4\times 64$ & 63.6 & 0.58 & 10 \\
        PDS~70 	    & 2018-02-25 & K1 ($2.11\;\mu$m)  & 90 & $3\times 96$ & 95.7 & 0.87 & 4.3 \\
        Empty     	& -          & H2 ($1.59\;\mu$m)  & 48 & $4\times 64$ & 72.5 & 0.84 & 9.2 \\
        \hline
	\end{tabular}
\end{table*}

\section{Normalized Huber-loss and estimation of the parameters}
\label{app:normalized-huber-loss}

\begin{figure}
  \centering
  {\includegraphics[width=0.23\textwidth]{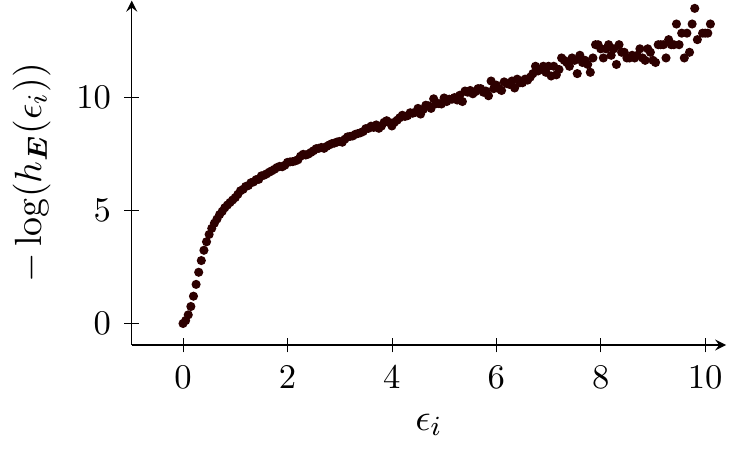}}
  {\includegraphics[width=0.23\textwidth]{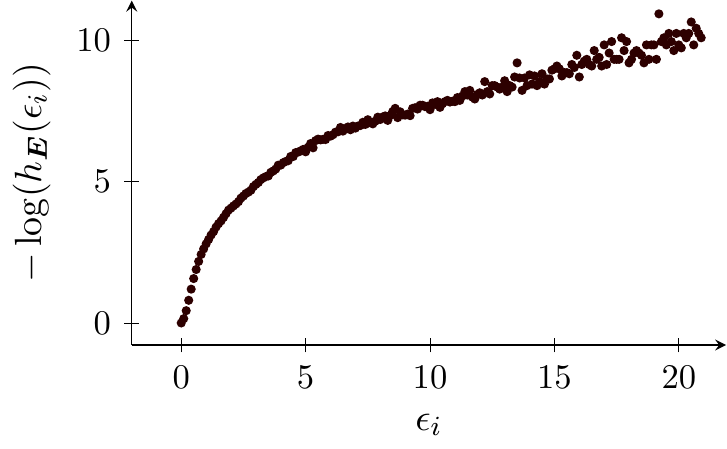}}\\
\caption{Negative logarithm of the computed PDF of the GreeDS residuals of PDS~70 (left) and  SAO~206462 (right).}
\label{fig:NegativeLogHistograms}
\end{figure}

From our model, the Huber-loss is the log-likelihood of the noise term $\NoiseTotal$. Thus, selecting the parameters of the Huber-loss requires to fit the negative logarithm of the PDF of the noise term $\NoiseTotal$. As $\NoiseTotal$ is unknown, we propose to estimate its PDF by computing the normalized histograms of the GreeDS residual, $\bs E = \bs Y - \gdsest{\bs L} -  \rotop[\bs 1_T (\gdsest{\bs x})^\top]$. We call the procedure described in this Appendix \textsc{HuberFit}. Given a (linear) regular partition of the intensities of $\bs E$ in $B$ bins $[\epsilon_i, \epsilon_{i+1})$ with $1\leq i \leq B$, and $\epsilon_1$ and $\epsilon_{B+1}$ the minimum and the maximum of the entries of $\bs E$, respectively, we denote $h_{\bs E}$ the normalized $B$-bin histogram of $\bs E$ defined by
\begin{equation}
\ts h_{\bs E}(\epsilon_i) = \frac{1}{T n^2} | \{ k,l : \epsilon_i \leq E_{k,l} < \epsilon_{i+1} \} |, \quad 1 \leq i \leq B+1.
\label{eq:computed-histograms}
\end{equation}
We computed in Fig.~\ref{fig:NegativeLogHistograms} the histogram of $\bs E$, setting $B=200$ to reach a good approximation of the PDF. As seen in this figure, since $h_{\bs E}$ is not convex, the PDF of $\NoiseTotal$ cannot follow a Huber density \eqref{eq:PDFassumed} whatever the value of $\delta$ and $\xi$. However, without reporting this experiment here, restricting $\bs E$ to specific annuli (with a width set to a multiple of the full width at half maximum (FWHM) of the instrumental PSF) leads to convex histograms compatible with a Huber density trend.

\begin{figure}
  \centering
  {\includegraphics[width=0.23\textwidth]{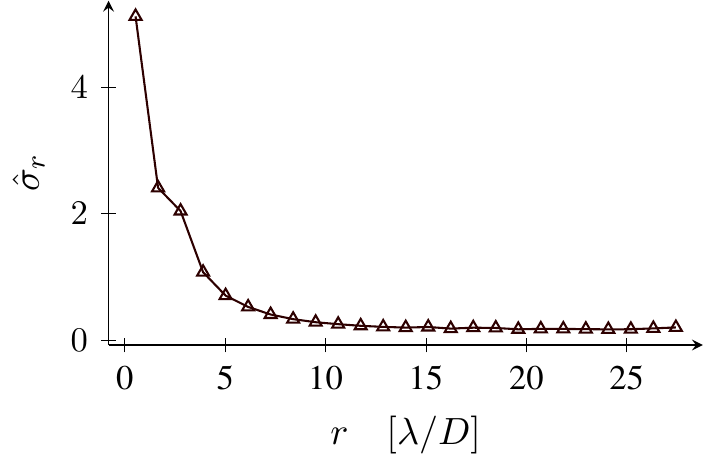}}
  {\includegraphics[width=0.23\textwidth]{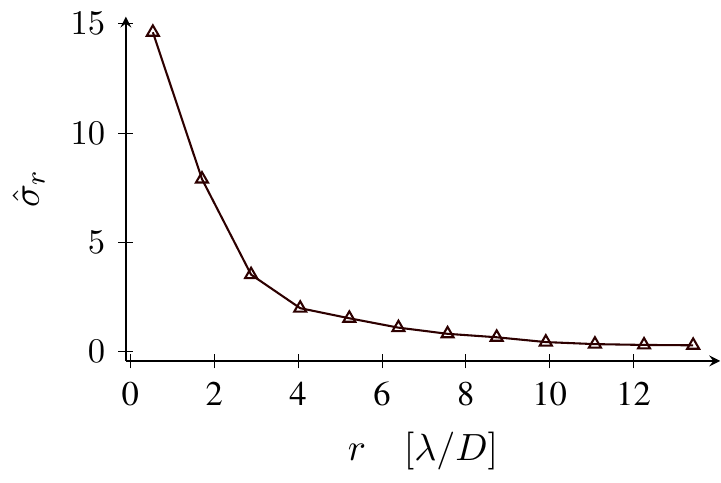}}
\caption{Radial profile of the computed standard deviation of the residuals for PDS~70 (left) and  SAO~206462 (right). Notice that $n$ is different for these datasets.}
\label{fig:sigma_by_annulus}
\end{figure}

We argue that the apparent misfit between the PDF of $\bs E$ and the Huber density is due to the heteroscedasticity of $\bs E$; it is well documented that the per voxel variance of ADI datasets has a strong radial dependency~\citep[see, \eg][]{soummer2007speckle}. This is confirmed by Fig.~\ref{fig:sigma_by_annulus} where we computed the empirical standard deviation $\hat{\sigma}_r$ of the voxels of $\bs E$ restricted to a given annulus of radius $r > 0$ and width $w$ set to the FWHM of the PSF. The displayed curves on the datasets PDS 70 and SAO 206462 are not constant and decay as the radius $r$ increases.

\begin{figure}
  \centering
  {\includegraphics[width=0.23\textwidth]{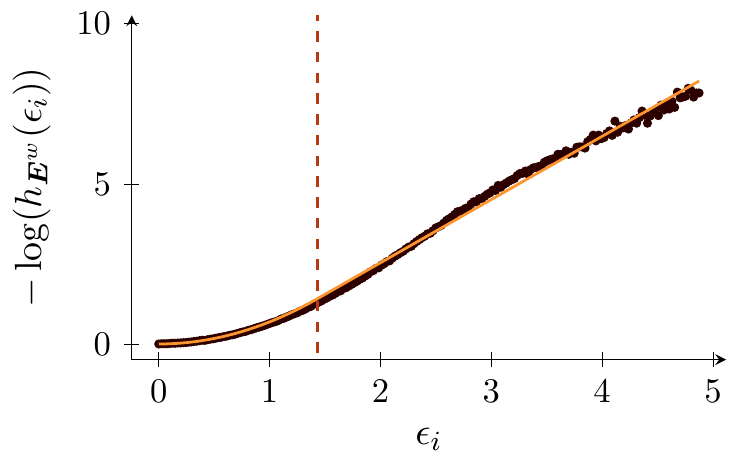}}
  {\includegraphics[width=0.23\textwidth]{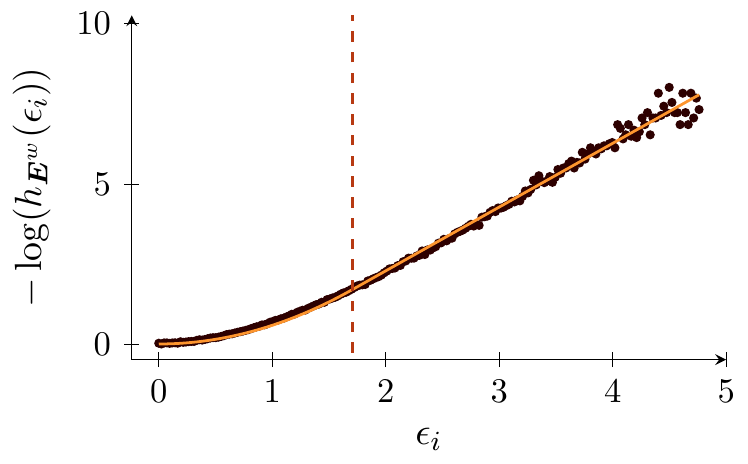}}
\caption{Negative logarithm of the computed PDF of the normalized GreeDS residuals of PDS~70 (left) and  SAO~206462 (right). The fit of the Huber-loss is overlaid and the vertical dashed line indicates the estimated value for $\delta$.}
\label{fig:normalized-NegativeLogHistograms}
\end{figure}

Interestingly, if we estimate the standard deviation of the noise $\NoiseTotal$ per voxel from $\hat{\sigma}_r$, setting
  $\hat{\xi}_{ij} = \hat{\sigma}_r$ if the voxel $(i,j)$ falls in the annulus of radius $r$ and width $w$, the histogram of the normalized residual $\bs E^{\rm w}$ such that $E^{\rm w}_{ij} = E_{ij} / \hat{\xi}_{ij}$ can be fit with a Huber density.  The negative logarithm of that histogram, \ie $-\log(h_{\bs E^{\rm w}} (\epsilon_i))$, is shown in Fig.~\ref{fig:normalized-NegativeLogHistograms} along with the least-square fit of the Huber-loss, with the corresponding threshold $\delta$ as a dashed vertical line. Two datasets, PDS~70 and SAO~206462, have been tested but similar results have been observed for all considered datasets (they are not reported here for conciseness). Compared to the histograms of the unweighted residuals (Fig.~\ref{fig:NegativeLogHistograms}), we can clearly fit the normalized histograms with a Huber-loss, despite a slightly increasing fitting error towards higher intensities for PDS 70. Before and after the estimated threshold, which is thus common to all voxels after normalization, we do see a quadratic and linear trend in the empiric estimation of the PDF of $\NoiseTotal$.

From this observation, we propose to set the normalizing parameters $\xi_{ij}$ of the Huber norm in Eq.~\eqref{eq:normalized-huber-norm} to the value of the estimates $\hat{\xi}_{ij}$ computed above. This provides an adequate candidate for the fidelity term, as demonstrated in Sec.~\ref{sec:huber-vs-l2-or-l1}.

\section{Convex low-rank plus sparse}
\label{app:dummy-convex-problem}

The ``low-rank plus sparse'' separation task finds applications in various fields, such as biomedical imaging \citep{otazo2015low} and background subtraction in static video \citep{zhou2011godec}. However, these applications most often consider low-rank and sparse components with similar intensities (\eg each with pixel value between 0 and 255). 

In high-contrast imaging, however, the star is significantly brighter than any circumstellar signal. Even with the coronagraph, the residual starlight (the speckle field) is typically at least $10^3$ times brighter than a companion or a disk. In this appendix, we show that this high-contrast between the low-rank component (the speckle field) and the sparse component (the circumstellar signal) is problematic when applying standard convexification of the low-rank plus sparse separation task. We show this in a simple noiseless separation task, exempt from rotation and convolution operations.

We consider the noiseless observation model $\bs Y =  \Lmodel + \Smodel$ made of a rank-one matrix $\Lmodel \in \bb R^{64\times 1024}$ ($r=1$) and a sparse matrix $\Smodel \in \bb R^{64\times 1024}$ with $\| \Smodel \|_0 = s = 128$. Both $\Lmodel$ and $\Smodel$ are randomly generated. We tested many combinations of $\Lmodel, \Smodel$ by varying the pixel intensity of $\Lmodel$ so that $\| \Lmodel  \|_2 = \mu \| \Smodel \|_2$, with $\mu>0$ ranging from $1$ to $10^3$. By varying  $\mu$, we unveil different behaviors in the non-convex approach and its convex relaxation. 

The non-convex separation problem reads 
\begin{subequations}
\label{eq:dummy-L+S-l0}
\begin{align}
\{\bs L_{\ell_0}, \bs S_{\ell_0}\} = \argmin_{\bs L, \bs S} &\ts \quad   \frac{1}{2} \|\bs Y - \bs L  -  \bs S\|_2^2\\
  \text{s.t.} & \quad \text{rank}( \bs L)  \leq r, \quad \| \bs S \|_0 \leq s .
\end{align}
\end{subequations}
Let us stress that, whatever the value of $\mu$, the pair $\{\Lmodel, \Smodel\}$ is always a solution of the problem~\eqref{eq:dummy-L+S-l0} (they form a local minimum of its cost), the non-convex rank and $\ell_0$ constraints being independent of the intensities of $\bs L$ and $\bs x$. Provided we can find this solution, we expect no impact from the variation of $\mu$.

A convex relaxation of~\eqref{eq:dummy-L+S-l0} is given by
\begin{subequations}
  \label{eq:dummy-L+S-l1}
  \begin{align}
\{\bs L_{\ell_1}, \bs S_{\ell_1}\} =  \argmin_{\bs L, \bs S} &\ts \quad  \frac{1}{2} \|\bs Y - \bs L  -  \bs S\|_2^2  \\
  \text{s.t.} & \quad \| \bs L \|_* \leq \tau_L, \quad \| \bs S \|_1 \leq \tau_S,
\end{align}
\end{subequations}
where the rank constraint is relaxed to the nuclear norm $\| \cdot \|_*$ and the $\ell_0$ constraint to the $\ell_1$-norm, $\| \cdot \|_1$. These two new convex constraints now depend on the intensity of $\bs L$ and $\bs x$, and thus, $\|\Lmodel\|_*$ increases with $\mu$, while $\|\Smodel\|_1$ remains constant. 

While~\eqref{eq:dummy-L+S-l0} is NP-hard, a decent approximate solution is found by performing a projected gradient descent (PGD) of the $\ell_2$-cost, where the projections amount to hard-thresholding the band $\bs L$ and $\bs S$ in the singular value domain and the pixel domain, respectively (see Sec.~\ref{sec:GreeDS}). This leads to a variant of the iterative hard thresholding (IHT) algorithm \cite{blumensath2008iterative}. The convex problem~\eqref{eq:dummy-L+S-l1} is solved exactly with a PGD where the projections are then equivalent soft-thresholding operators in the same above-mentioned domains \citep{candes2011robust}.

\begin{figure}
  \centering
  {\includegraphics[width=0.23\textwidth]{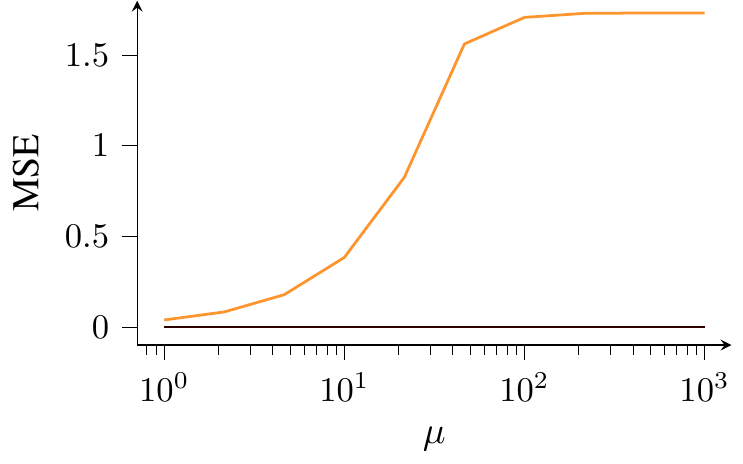}}
  {\includegraphics[width=0.23\textwidth]{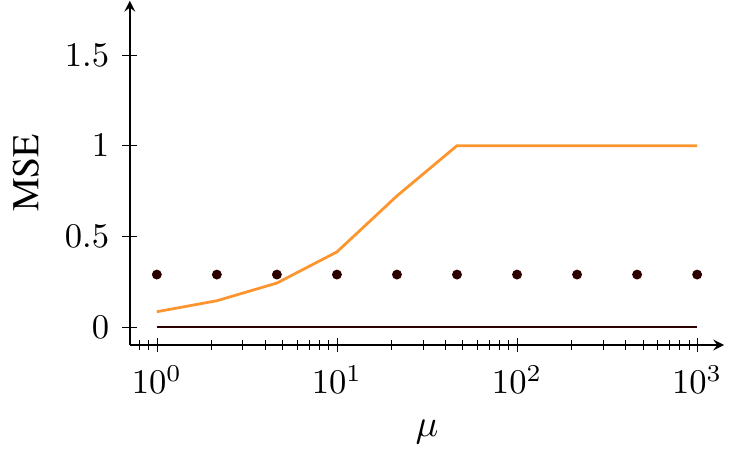}}
\caption{MSE of $\bs S_{\ell_1}$ (brown) and $\bs S_{\ell_0}$ (orange) as $\mu$ increases for $\tau_L = 0.99 \| \Lmodel \|_*$ (left) and $\tau_L = 1.01 \| \Lmodel \|_*$ (right). For the relative squared error of $\bs S_{\ell_0}$, the value of $r$ is the exact value of 1. For completeness, we added the relative squared error of $\bs S_{\ell_0}$ with $r=2$, represented with dots (right). We observe that even the error is more significant than for the nuclear norm, the relative squared error remains at a value of 0.25 and is independent on the scaling factor $\mu$.  }
\label{fig:dummy_convex_low_rank_plus_sparse}
\end{figure}

We display in Fig.~\ref{fig:dummy_convex_low_rank_plus_sparse} the mean square error (MSE) achieved by $\bs S_{\ell_0}$ (obtained with the variant of IHT, and $r$ set to 1 (left) and 2 (right)) and $\bs S_{\ell_1}$ as a function of $\mu$. The displayed MSE is the average for 100 pairs of $\Lmodel$ and $\Smodel$. While the MSE of $\bs S_{\ell_0}$ is constant with $\mu$, the one of $\bs S_{\ell_1}$ is overly sensitive to the value of $\tau_L$ at large contrast, both for $\tau_L = 0.99 \| \Lmodel \|_*$ (left) and $\tau_L = 1.01 \| \Lmodel \|_*$ (right). Although the error in estimating $\tau_L$ is only one percent of the optimal value, the MSE of $\bs S_{\ell_1}$ quickly degrades as $\mu$ increases. In both cases, the support of $\bs S$ is not recovered as soon as $\mu$ reaches a value of $10^2$. 

Concerning the constraints~\eqref{eq:convex_relaxed_objective}, the estimation of $\tau_L$ is a delicate task as we have a significant amount of noise $\NoiseSpeckles$. In practice, a large portion of the speckle field is attributed to the components $\bs x_\inddisk$ and $\bs x_\indpla$, drowning the circumstellar signal in noise at best and at worse, ejecting the circumstellar signal into the noise term. For this reason, we found it was best to convexify the rank constraint of problem~\eqref{eq:final-objective-rank} with projection constraint~\eqref{eq:final-objective-rank-constraint} instead of the nuclear norm constraint~\eqref{eq:nuclear-low-rank-constraint}.

\section{Heuristic to set $\tau_\inddisk$ and $\tau_\indpla$}
\label{app:set-taus}

We propose to set the regularizer parameters $\tau_\inddisk$ and $\tau_\indpla$ using the following heuristic decoupling the inverse problem posed by the observation model from the source separation task.

We first solve~\eqref{eq:convex_relaxed_objective-final_formulation} without the regularizations~\eqref{eq:final-objective-disk-constraint},~\eqref{eq:final-objective-planet-constraint}, and we optimize over the non-regularized vector $\bs x^\text{NR} = \bs x_{\inddisk} + \bs x_{\indpla}$ containing the whole circumstellar signal, summing disks and exoplanets in absence of any morphological regularization. We also switch off the deconvolution ($\cl T$ is set to the identity) that requires there regularizations to work properly.

We thus solve the following problem
\begin{subequations}
\label{eq:convex_relaxed_objective-final_formulation_no_regul}
\begin{align}
\{\bs L^\text{NR}, \bs x^\text{NR}\} = \argmin_{\bs L, \bs x} & \quad  \Hnorm{\maskop(\bs Y - \bs L  -  \rotop[\bs 1_T \bs x^\top ] )},  \\
\text{s.t.} &\quad  \bs L \in \Span( \bs U_{[r]}^{*}), \\
& \quad \bs L,  \bs x \geq 0.
\end{align}
\end{subequations}

The image $\bs x^\text{NR}$ is then used as the observations of a 2-D MCA problem 
\begin{subequations}
\label{eq:objective-MCA-alone}
\begin{align}
\label{eq:objective-MCA-formulation}
\{\hat{\bs x}_\inddisk, \hat{\bs x}_\indpla\} =  \argmin_{ \bs x_\inddisk, \bs x_\indpla} &\ts \quad  \Hnorm{\maskop( \bs x^\text{NR}   - \psf \convOp (\bs x_\inddisk + \bs x_\indpla)^\top  )},  \\
\text{s.t.}&\quad \| \bs \Psi^\top \bs x_\inddisk\|_1 \leq \tau_\inddisk,\\
&\quad \|\bs x_\indpla\|_1 \leq \tau_\indpla \,,\\
& \quad   \bs x_\inddisk, \bs x_\indpla \geq 0.
\end{align}
\end{subequations}

The problem~\eqref{eq:objective-MCA-alone} is significantly faster to solve than problem~\eqref{eq:final-objective} because it only involves 2-D data $T$ times smaller than the initial ADI dataset. Furthermore, solving~\eqref{eq:objective-MCA-alone} does not require the expensive computation involving the rotation operator $\rotop$. We can thus afford to solve it with different values of $\tau_\inddisk$ and $\tau_\indpla$ and then choose the values that produce a satisfying output. Finally, once $\tau_\inddisk$ and $\tau_\indpla$ are selected, we solve~\eqref{eq:final-objective}. 

To further reduce the computational time of this selection, we follow the following strategy that first sets $\tau_\inddisk$ before then $\tau_\indpla$, hence avoiding a costly grid search for these two parameters. We use for this the stopping criteria developed by \cite{almeida2013parameter}. Motivated by the structureless nature of the noise, this procedure aims to select for a deconvolution problem solved by a convex optimization, such as \eqref{eq:objective-MCA-alone}, the regularization parameter that minimizes the \emph{whiteness} of the residual (formed by subtracting from the observation the blurred image estimate). We noted that, given the solution $\{\hat{\bs x}_\inddisk, \hat{\bs x}_\indpla\}$ of~\eqref{eq:objective-MCA-alone} (associated with the parameters $\{\tau_\inddisk, \tau_\indpla\}$) and the residuals
\begin{equation}
\bs r^{\tau_\inddisk}_{\tau_\indpla} = \bs x^\text{NR}   - \psf \convOp (\hat{\bs x}_\inddisk + \hat{\bs x}_\indpla)^\top,
\end{equation}
the whiteness score proposed by~\cite{almeida2013parameter} depends mostly on $\tau_\inddisk$. This is explained by the fact that the disk---an extended structure--- is mostly responsible of the spatial correlations in $\bs x^\text{NR}$. Varying $\tau_\inddisk$ thus quickly changes the whiteness of the residual.

From this observation, we first find the value of $\tau_\inddisk$ that minimizes the whiteness criterion solving~\eqref{eq:objective-MCA-alone} with $\tau_\indpla=0$ and a decreasing value of $\tau_\inddisk$ starting from a large value. As $\tau_\inddisk$ decreases, the whiteness score of the residuals goes to a minimum before to dramatically increase when the value of $\tau_\inddisk$ is too small to allow $\hat{\bs x}_\inddisk$ to account for the disk. Then, once $\tau_\inddisk$ is selected, we solve~\eqref{eq:objective-MCA-alone} with a decreasing value of $\tau_\indpla$ and selection of the satisfying value of $\tau_\indpla$ is done by visual inspection.

% Don't change these lines
\bsp	% typesetting comment
\label{lastpage}
\end{document}